\documentclass[aps,pra,twocolumn,moshowpacs,floatfix,superscriptaddress]{revtex4-1}

\usepackage{dsfont} 
\usepackage{bm} 

\usepackage{bbold} 
\usepackage{xcolor}
\usepackage{times}
\usepackage{physics}
\usepackage{graphicx,braket}
\usepackage{hyperref}
\hypersetup{colorlinks=true,urlcolor=blue,linkcolor=blue,citecolor=blue}
\usepackage{amsmath,amssymb,amsfonts}
\usepackage{wrapfig}
\usepackage{siunitx}

\newcommand{\cov}{\mathrm{cov}}

\usepackage{cleveref}
\crefname{equation}{Eqs.}{Eqs.}
\Crefname{equation}{Equation}{Equations}
\crefrangelabelformat{equation}{(#3#1#4--#5#2#6)}
\crefmultiformat{equation}{Eqs. (#2#1#3}{, #2#1#3)}{#2#1#3}{#2#1#3}
\Crefmultiformat{equation}{Equations (#2#1#3}{, #2#1#3)}{#2#1#3}{#2#1#3}
\begin{document}
\title{Non-stationary dynamics and dissipative freezing in squeezed superradiance}

\author{Carlos S\'anchez Mu\~noz }
\email{carlos.sanchezmunoz@physics.ox.ac.uk}
\affiliation{Clarendon Laboratory, University of Oxford, Parks Road, Oxford OX1 3PU, United Kingdom}
\author{Berislav Bu\v ca}
\affiliation{Clarendon Laboratory, University of Oxford, Parks Road, Oxford OX1 3PU, United Kingdom}
\author{Joseph Tindall}
\affiliation{Clarendon Laboratory, University of Oxford, Parks Road, Oxford OX1 3PU, United Kingdom}
\author{Alejandro Gonz\'alez-Tudela}
\affiliation{Instituto de F\'isica Fundamental IFF-CSIC, Calle Serrano 113b, Madrid 28006, Spain}
\author{Dieter Jaksch}
\affiliation{Clarendon Laboratory, University of Oxford, Parks Road, Oxford OX1 3PU, United Kingdom}
\author{Diego Porras}
\affiliation{Instituto de F\'isica Fundamental IFF-CSIC, Calle Serrano 113b, Madrid 28006, Spain}

\newcommand{\down}{\op{g}{e}}
\newcommand{\up}{\op{e}{g}}
\newcommand{\downd}{\op{+}{-}} 
\newcommand{\upd}{\op{+}{-}}
\newcommand{\app}{a^\dagger}
\newcommand{\ssp}{\sigma^\dagger}
\newcommand*{\Resize}[2]{\resizebox{#1}{!}{$#2$}}%

\begin{abstract}
In this work, we study the driven-dissipative dynamics of a coherently-driven spin ensemble with a squeezed, superradiant decay. This decay consists of a sum of both raising and lowering collective spin operators with a tunable weight.
The  model presents different critical non-equilibrium phases with a gapless Liouvillian that are associated to particular symmetries and that give rise to distinct kinds of non-ergodic dynamics. In Ref.~\cite{arXiv_sanchezmunoz19b} we focus on the case of a strong-symmetry and use this model to introduce and discuss the effect of dissipative freezing, where, regardless of the system size, stochastic quantum trajectories initialized in a superposition of different symmetry sectors always select a single one of them and remain there for the rest of the evolution. Here, we deepen this analysis and study in more detail the other type of non-ergodic physics present in the model, namely, the emergence of non-stationary dynamics in the thermodynamic limit. We complete our description of squeezed superradiance by analysing its metrological properties in terms of spin squeezing and by analysing the features that each of these critical phases imprint on the light emitted by the system.
\end{abstract}
\date{\today} \maketitle

\section{Introduction}
Non-equilibrium systems are present in a wide variety of areas, including physics,  life sciences, sociology and finance. In physics, a typical non-equilibrium situation is realized when driving from an external source is compensated by dissipation to the environment. This is  the case in numerous examples of many-body and cavity QED systems, such as exciton polaritons~\cite{amo09a,rodriguez17a,fink18a}, Rydberg ensembles~\cite{carr13a,melo16a}, superconducting circuits~\cite{fitzpatrick17a}, trapped atoms~\cite{baumann10a,klinder15a,hamsen18a} or in mechanical systems~\cite{teufel11a,kolkowitz12a,pigeau15a}. 
In the ongoing effort to deepen our understanding of out-of-equilibrium phenomena, which typically differ from their equilibrium counterparts~\cite{szymanska06a,roumpos12a,chioccetta13a,altman15a,nitsche14a,caputo18a}, one of the aspects attracting a significant amount of attention are dissipative phase transitions (DPTs)~\cite{kessler12a,minganti18a,carmichael15a,weimer15a,benito16a,
sieberer13a,sanchezmunoz18b,biondi17a,hwang18a,mendoza16a}.  

In non-equilibrium systems, the interplay between driving and losses eventually brings the system into a stationary state defined by a density matrix $\rho_0$. Several non-equilibrium phases associated to different steady states can then exist, and a DPT between these phases is defined as the non-analytical behaviour of a steady-state observable under the change of a system parameter~\cite{kessler12a,minganti18a}. DPTs are less  understood than their classical or quantum counterparts, driven respectively by thermal and quantum fluctuations.  There is an important link between DPTs and the spectral properties of the  Liouvillian superoperator, $\mathcal L$, that governs the dynamics of the density matrix, $\dot\rho=\mathcal L \rho$. Its eigenvalue with largest real part, $\lambda_0$, is exactly zero, and the corresponding eigenvector is the steady state $\rho_0$. In the usual description of DPTs, a transition occurs when the eigenvalue with the second largest real part $\lambda_1$, often called the \emph{asymptotic decay rate} (ADR), tends to zero, which then implies the existence of several degenerate steady states.

A significant amount of research has been devoted to the definition and characterization of  DPTs~\cite{kessler12a,minganti18a}, and to the study of the associated, interrelated phenomena of bistability~\cite{fink18a,carr13a,melo16a,mendoza16a,letscher17a,muppalla18a}, hysteresis~\cite{rodriguez17a,hruby18a}, intermittency~\cite{lee12a,fitzpatrick17a,hruby18a,malossi14a,muppalla18a,ates12a}, multimodality~\cite{letscher17a,malossi14a}, metastability~\cite{macieszczak16a} and symmetry breaking~\cite{manzano14a,hannukainen18a} in open quantum systems. All these phenomena are understood as different manifestations of the coexistence of several non-equilibrium phases. 

One of the problems that hinder our understanding of non-equilibrium systems and DPTs is the enormous computational difficulty typically found when dealing with large quantum open systems. It is thus highly desirable to work with exactly solvable systems or at least computationally tractable models that yield an insight about the physics in the thermodynamic limit.  Unfortunately the existence of tractable many-body non-equilibrium models is still scarce.
In this work, we study one of these models, consisting of a coherently-driven spin ensemble with a collective, squeezed decay. While this model is simple enough to be treated numerically, it displays a variety of dissipative phases with a gapless Liouvillian. More importantly, these non-equilibrium phases are associated with different, non-ergodic behaviours that depart from the usual pictures of phase coexistence mentioned above.

The first of these phenomena is the effect of dissipative freezing, which is related to the existence of a \emph{strong symmetry}~\cite{buca12a}. A strong symmetry consists of an operator $A$ that commutes with the Hamiltonian and all the quantum jump operators of the Liouvillian, implying the existence of several, degenerate steady states and a conservation law for tye symmetry operator, $\dot A = \mathcal L^\dagger A = 0$.   As we discuss in Ref~\cite{arXiv_sanchezmunoz19b}, the effect of dissipative freezing occurs at the level of individual trajectories within the quantum jump formalism, which describes the system in terms of a pure wavefunction undergoing stochastic evolution~\cite{zoller87a,molmer92a,plenio98a}. When the initial state is a superposition involving different symmetry sectors, every quantum trajectory selects only one of these and remains there for the rest of the evolution.
This involves a breakdown of the conservation law for $A$ at the level of individual trajectories. Since such a superposition is not possible in a classical, stochastic system, this is a purely quantum effect. The effect of dissipative freezing is in stark contrast to the notion of intermittency usually reported in bistable systems~\cite{lee12a,fitzpatrick17a,hruby18a,malossi14a,muppalla18a,ates12a}, which is then understood to be a finite-size effect. 

The second non-ergodic phenomenon observed in this model is the appearance of oscillatory, non-stationary dynamics in the long-time limit. This effect is related to the existence of a spectrum of purely imaginary eigenvalues of $\mathcal L$, which needs to be equally spaced in order to prevent eigenstate thermalization. This phenomenon has recently attracted attention in similar systems~\cite{buca19a,iemini18a,tucker18a} and has been linked to the existence of a  \emph{dynamical symmetry} in the system~\cite{buca19a}, e.g. a ladder operator of the Hamiltonian that commutes with all the quantum jump operators of the Liouvillian.

\begin{figure*}[t!]
\begin{center}
\includegraphics[width=0.99\textwidth]{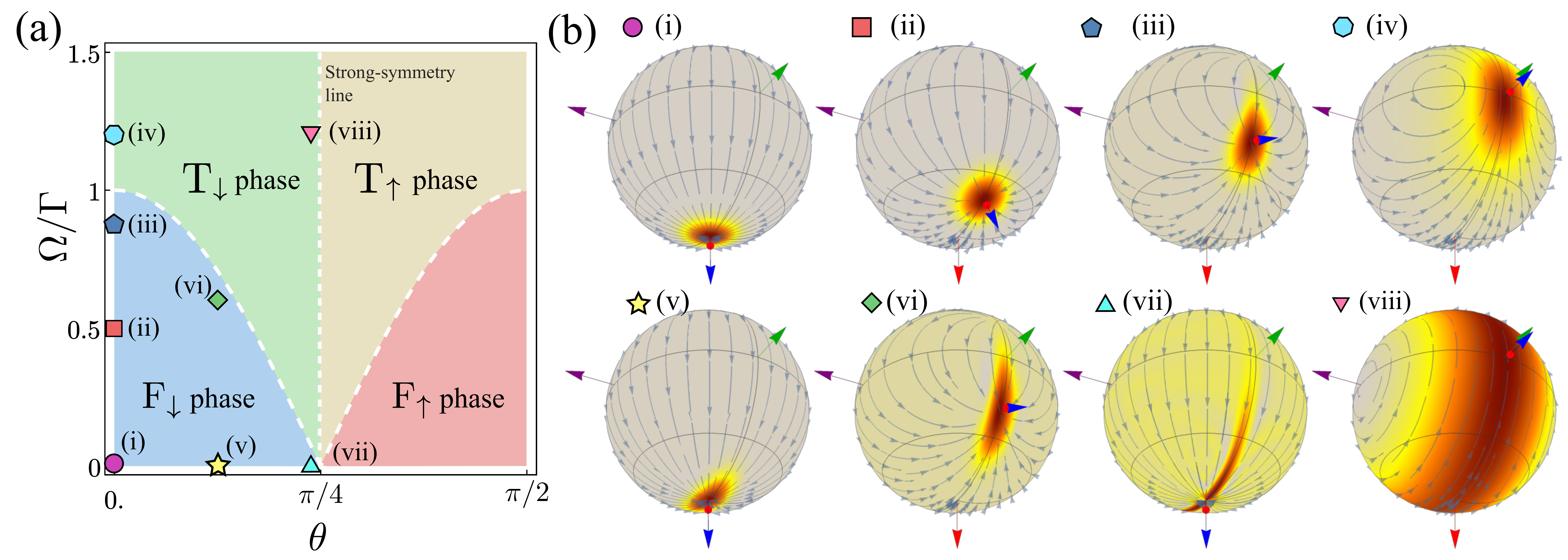}
\end{center}
\caption{Phase diagram and steady state. (a) The phase diagram can be divided into a ferromagnetic ($\text F$) and an thermal ($\text T$) phase, separated by the critical line $\Omega_c(\theta)$ given by Eq.~\eqref{eq:Omega_c} (white, dashed lines). There is spin-up and spin-down version of each of these phases, separated by the strong-symmetry line $\theta=\pi/4$. (b) Spin Wigner functions of the exact steady states of master equation~\eqref{eq:master-equation} for a finite system with $N=50$ at different points  $(\Omega,\theta)$, corresponding to: (i) $(0,0)$, (ii) $(0.5\Gamma,0)$, (iii) $(0.88\Gamma,0)$, (iv) $(1.2\Gamma,0)$, (v) $(0,\pi/8)$, (vi) $(0.6\Gamma, \pi/8)$, (vii) $(0,0.95\pi/4)$, (viii) $(1.2\Gamma,0.95\pi/4)$. Together, we plot the vector field of derivatives described by the mean-field equations \eqref{eq:angles_MF_dot}.}
\label{fig:1}
\end{figure*}

Both non-ergodic phenomena are present in the model of coherently-driven spins with squeezed decay that we discuss here and, interestingly, coexist in similar regions of the phase diagram. Squeezed decay refers to a quantum jump operator that includes both lowering and raising collective spin operators $S_\pm$, with relative weights parametrized by a squeezing angle $\theta$. The amplitude $\Omega$ of the driving field and the squeezing angle $\theta$ are the main tunable parameters. The dissipative phase diagram of the system is hence obtained in the $(\Omega,\theta)$  plane.

The paper is organized as follows. In Sec.~\ref{sec:ii}, we introduce the model of squeezed superradiance, describing the phase diagram and steady-state  of the system. In Sec.~\ref{sec:iii}, we analyse the Liouvillian spectrum of this model and characterize symmetries and regimes of non-ergodicity. In Sec.~\ref{sec:iv}, we describe the novel phenomenon of dissipative freezing, and discuss it in the context of thermodynamics of quantum trajectories and phase transitions. Finally, in Sec.~\ref{sec:v}, we analyse signatures of critical, dissipative dynamics in observables of the light emitted by the system.

\section{Model and phase diagram}
\label{sec:ii}

\subsection{Squeezed superradiance: derivation of the spin master equation}

The model of squeezed superradiance that we consider in this work is given by the following master equation for the reduced density matrix of an ensemble of $N$ spins ($\hbar=1$):
\begin{equation}
\dot{\rho}=-i\Omega[S_x,\rho]+\frac{\Gamma}{2J}\mathcal L_{D_\theta}[\rho],
\label{eq:master-equation}
\end{equation}
where $\mathcal{L}_O[\rho]\equiv 2 O\rho O^\dagger - \{O^\dagger O, \rho\} $ is the usual Lindblad superoperator, and the operator $D_\theta$ describes the quantum jumps undergone by the system
\begin{equation}
D_\theta\equiv \cos(\theta) S_- + \sin(\theta) S_+.
\label{eq:Dtheta}
\end{equation}
In these equations, $\{S_\pm, S_z\}$ are collective spin operators obeying angular momentum commutation relations, $\Omega$ is the driving amplitude, $\Gamma$ is the quantum-jump rate, and $J=N/2$ is the total angular momentum, which is conserved in the dynamics. Notably, $D_\theta$ includes both raising and lowering operators, with a weight that we parametrize by the angle $\theta$.

The dynamics in Eq.~\eqref{eq:master-equation} emerge as the strongly-dissipative limit of the following Hamiltonian%
\begin{equation}
H=\Omega S_x+\frac{g}{\sqrt{N}}\left\{S_+\left[\cos(\theta) a + \sin(\theta)a^\dagger\right] + \mathrm{h.c.}\right\}.
\label{eq:Hamiltonian-with-cavity}
\end{equation}
This Hamiltonian describes a driven spin ensemble coupled to a single cavity mode in the rotating frame of the driving, with $a$  the bosonic annihilation operator of the cavity and $g$ the spin-cavity coupling rate. Since the total angular momentum $J$ is conserved, the spin ensemble can be described as a single big spin; this can be implemented, for instance, with multi-component atomic condensates~\cite{jaksch01a,micheli03a}. The tunable coupling terms in Eq.~\eqref{eq:Hamiltonian-with-cavity} can be achieved via cavity-assisted Raman transitions; this approach has been proposed as a way to implement effective Dicke models~\cite{dimer07a} and used, sucessfully, to observe alternative forms of the superradiant phase transition~\cite{hepp73a,hepp73b,wang73a,emary03a} in atomic condensates~\cite{baumann10a,kessler14a,klinder15a,kroeze18a} and thermal atoms~\cite{baden14a,zhiqiang17a}. 
The great control and versatility provided by these  schemes has motivated research on generalized non-equilibrium Dicke models~\cite{keeling10a,basheen12a}.

In this work, we are focusing on strongly dissipative versions of these systems---where the fast cavity decay yields an effective, collective spin dissipation---which have attracted interest for their applications to the dissipative generation of spin squeezing and entanglement in the steady state~\cite{dallatorre13a,gonzaleztudela13b}. 
By taking into account that the cavity experiences dissipation at a rate $\gamma$, the evolution of the system is described by the master equation~\cite{walls_book94a} $\dot \rho = -i[H,\rho] + \gamma/2 \mathcal L_a[\rho]$. In the limit $\gamma\rightarrow \infty$, the bosonic field tends to a stationary vacuum state, and its adiabatic elimination~\cite{gonzaleztudela13b} yields the effective dynamics for the spins of Eq.~\eqref{eq:master-equation}, with $\Gamma = 2g^2/\gamma$.  It is easy to deduce that the dark state of $D_\theta$ is a spin squeezed state~\cite{dallatorre13a}, which brings us to refer to the dissipative part of Eq.~\eqref{eq:master-equation} as a ``squeezed decay''. Note that, when $\theta=0$, the model corresponds to the standard case of collective resonance flourescence~\cite{puri79a,lawande81a,gonzaleztudela13b}.

\subsection{Phase diagram}

\begin{figure}[t!]
\begin{center}
\includegraphics[width=1\columnwidth]{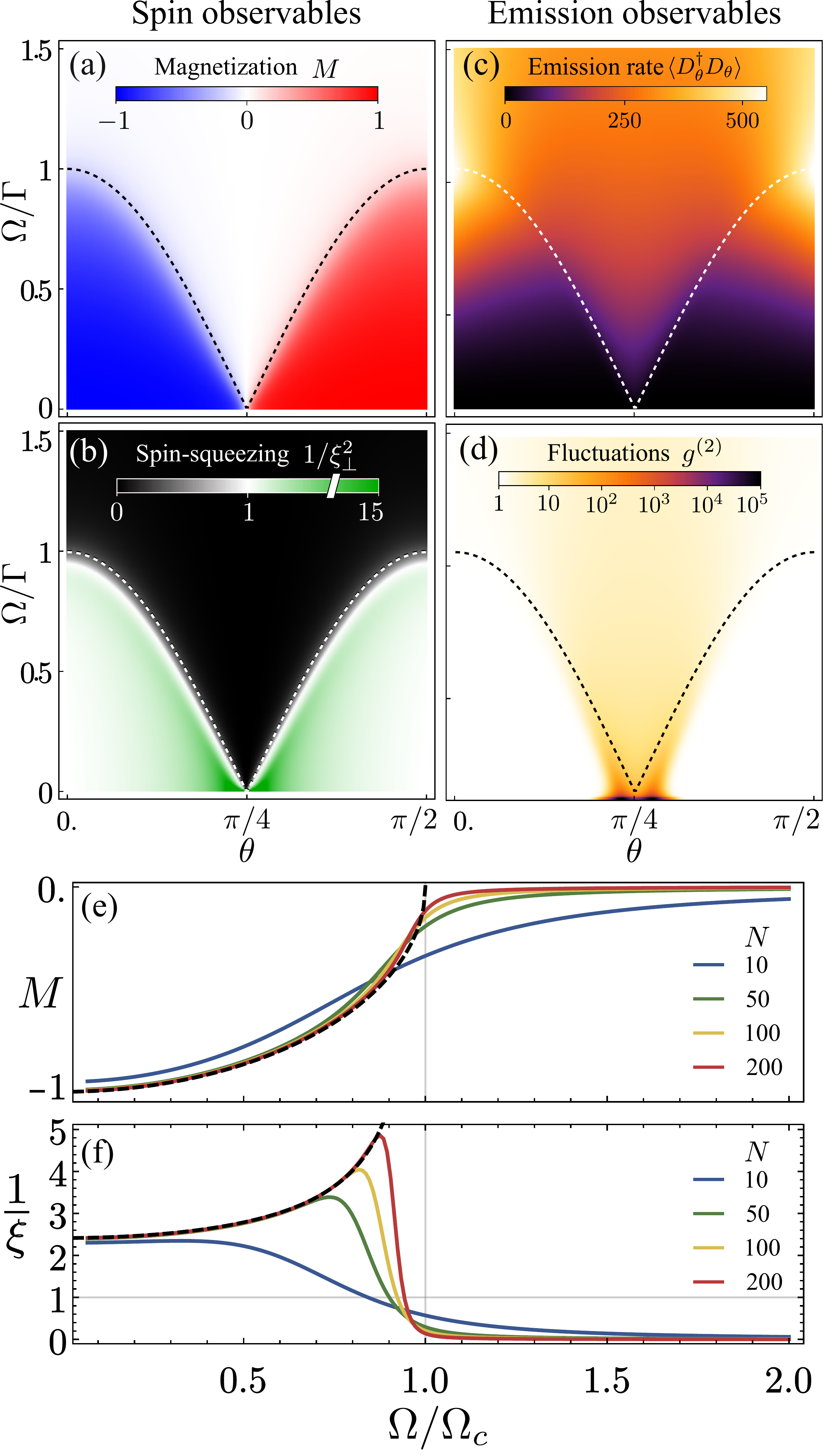}
\end{center}
\caption{(a-d) Steady state observables for a finite system size $N=50$. Dashed lines indicate the critical line, Eq.~\eqref{eq:Omega_c}. 
(e-f) Magnetization (e) and degree of spin squeezing (f) across the phase transition. Black, dashed line is the analytical value in the thermodynamic limit, Eqs.~\eqref{eq:M-mean-field} and \eqref{eq:spin_squeezing}. Solid lines are numerical calculations for finite systems of different sizes.  Calculations were made at $\theta=\pi/8$. }
\label{fig:2}
\end{figure}
The non-equilibrium phases of the system in the $(\Omega,\theta)$ plane are summarized in Fig.~\ref{fig:1}(a), and the corresponding steady-state observables computed exactly for a finite system ($N=50$) are depicted in Fig.~\ref{fig:2}. Fig.~\ref{fig:1}(b) depicts the steady state of a finite system ($N=50$) at several points of the phase diagram using the  spin Wigner function~\cite{agarwal81a,dowling94a}. Additionally, we plot the vector field of derivatives obtained through a mean field approach (see~Appendix~\hyperref[sec:appendix1]{I}). We can divide the phase diagram into two types of phases:

\emph{i)} The \emph{ferromagnetic (F) phase} is characterized by a well-defined magnetization (c.f. Fig.~\ref{fig:2}(a)), a diverging spin-squeezing as we approach the phase transition (Fig.~\ref{fig:2}(b)), small fluctuations in the counting distributions of quantum jumps (described here by the  zero-delay, second-order correlation function of the output field, $g^{(2)}\equiv \langle {D^\dagger_\theta}^2 {D_\theta}^2\rangle/\langle D^\dagger_\theta D_\theta \rangle^2$) (Fig.~\ref{fig:2}(d)), high purity  (not shown) and ergodic dynamics. Any initial state eventually relaxes into a stationary, highly pure gaussian steady-state. In the thermodynamic limit, this phase is well described within a Holstein-Primakoff approximation. 

\emph{ii)} In the \emph{thermal phase} the steady-state is highly mixed, and close to the infinite-temperature state $\rho \propto \mathbb{1}$. This phase is characterized by a mean zero magnetization (Fig.~\ref{fig:2}(a)),  small purity (not shown), large spin fluctuations, high rate of quantum jumps (activity) (Fig.~\ref{fig:2}(c)) and large fluctuations in the output field (Fig.~\ref{fig:2}(d)). As we discuss further below, this phase displays a vanishing asymptotic decay rate (ADR) that leads, in the thermodynamic limit $N\rightarrow \infty$, to a closed gap and non-ergodic dynamics, which manifests itself through closed orbits in the mean-field approach (c.f. point (iv) in Fig.~\ref{fig:1}(b)). 

Both phases have a spin-down ($\downarrow$) and spin-up ($\uparrow$) version at each side of the line $\theta=\pi/4$; each of them being a spin-flipped version of the other.
Therefore, defining
\begin{subequations}
\begin{align}
\Gamma_-&\equiv \Gamma\cos^2\theta\label{eq:gammaplus},\\
\Gamma_+&\equiv \Gamma\sin^2\theta\label{eq:gammaminus},
\end{align}
\end{subequations}
all the results and equations obtained for $\theta \leq\pi/4$ are directly applicable in a spin-flipped basis for $\theta \geq \pi/4$ just by exchanging $\Gamma_- \leftrightarrow \Gamma_+$. Hereafter, the analytical results that we provide refer to the spin-down phases ($\theta \leq \pi/4$).

In Appendix~\hyperref[sec:appendix1]{I} we show that, using a mean-field approach, the transition from the ferromagnetic to the thermal phase occurs at the critical driving:
\begin{equation}
\Omega_c(\theta) = \Gamma_--\Gamma_+=\Gamma(\cos^2\theta-\sin^2\theta).
\label{eq:Omega_c}
\end{equation}

\subsection{Spin observables}
We consider now the expectation values of the normalized spin operators $s_i\equiv S_i/J$, $i\in\{x,y,z \}$ in the steady state.
In the ferromagnetic phase,  these can be obtained by a displaced Holstein-Primakoff (HP) expansion (see~Appendix~\hyperref[sec:appendix2]{II}); the results are the same as the mean-field predictions, with corrections to order $1/J$:
\begin{subequations}
\label{eq:sxsyszHP}
\begin{align}
\langle s_z \rangle  &=M+\mathcal O (1/J) \label{eq:sz},\\
\langle s_x \rangle &=  0  + \mathcal{O}(1/J)\label{eq:sx},\\
\langle s_y \rangle &= \sqrt{1-M^2}+ \mathcal{O}(1/J),\label{eq:sy}
\end{align}
\end{subequations}
where $M$ is the steady-state magnetization that reads:
\begin{equation}
M = -\sqrt{1-\left(\frac{\Omega}{\Gamma_--\Gamma_+}\right)^2}.
\label{eq:M-mean-field}
\end{equation}

The $1/J$ corrections are given by the solution of non-quadratic master equations and therefore analytical expressions are difficult to obtain. It is however possible to get expressions for the spin fluctuations $\Delta s_{z/\pm}^2$ to order $1/J$; this is one of the main advantages of using a HP expansion, since it allows to describe the metrological properties of the spin ensemble~\cite{pezze18a}. In particular, reduced fluctuations along one of the spin directions provides  enhanced phase  sensitivity in atomic interferometers~\cite{gross10a,berrada13a} and greater stability in atomic clocks~\cite{borregaard13a}. States displaying such reduced fluctuations are said to be \emph{spin squeezed}~\cite{wineland94a,ma11a,pezze18a}; the degree of spin squeezing $\xi_\bot$ along any axis $\mathrm u_\bot$ perpendicular to the mean spin direction  is a popular figure of merit, useful as a witness of entanglement~\cite{sorensen01a} and as a direct measure of the phase sensitivity achievable in interferometry protocols. This quantity can be defined as~\cite{wineland94a}:
\begin{equation}
\xi_\bot^2=\frac{N(\Delta S_\bot)^2}{\langle \mathbf{S} \rangle^2}.
\end{equation}
According to this definition, a state is spin-squeezed if a direction $\mathrm{u}_\bot$ exists such that $\xi_\bot^2<1$. In our model, the optimal squeezing direction is always the $\mathrm u _x$ axis (see~Appendix~\hyperref[sec:A-squeezing]{II-D}). Using the HP approximation, we find the following expression for the spin squeezing in the ferromagnetic phase:
\begin{equation}
\xi_\bot^2 = \frac{N(\Delta S_x)^2}{\langle \mathbf S \rangle^2} = (1-M)\left(\frac{1}{2}+\frac{\Gamma_+-\sqrt{\Gamma_-\Gamma_+}}{\Gamma_--\Gamma_+}\right).
\label{eq:spin_squeezing}
\end{equation}
The analytical results in Eq.\eqref{eq:sxsyszHP} and \eqref{eq:spin_squeezing} are shown in Fig.~\ref{fig:2}(e-f), compared with numerical calculations for finite system size. Equation~\eqref{eq:spin_squeezing} shows that, in the thermodynamic limit, spin squeezing diverges (i.e. $\xi_\bot^2\rightarrow 0$) in the vicinity of the critical line, where $M\rightarrow 0$. This implies a greatly enhanced phase sensitivity and the emergence of many-body correlations, which are general properties associated to second-order phase transitions~\cite{kessler12a}.

\begin{figure}
\begin{center}
\includegraphics[width=1\columnwidth]{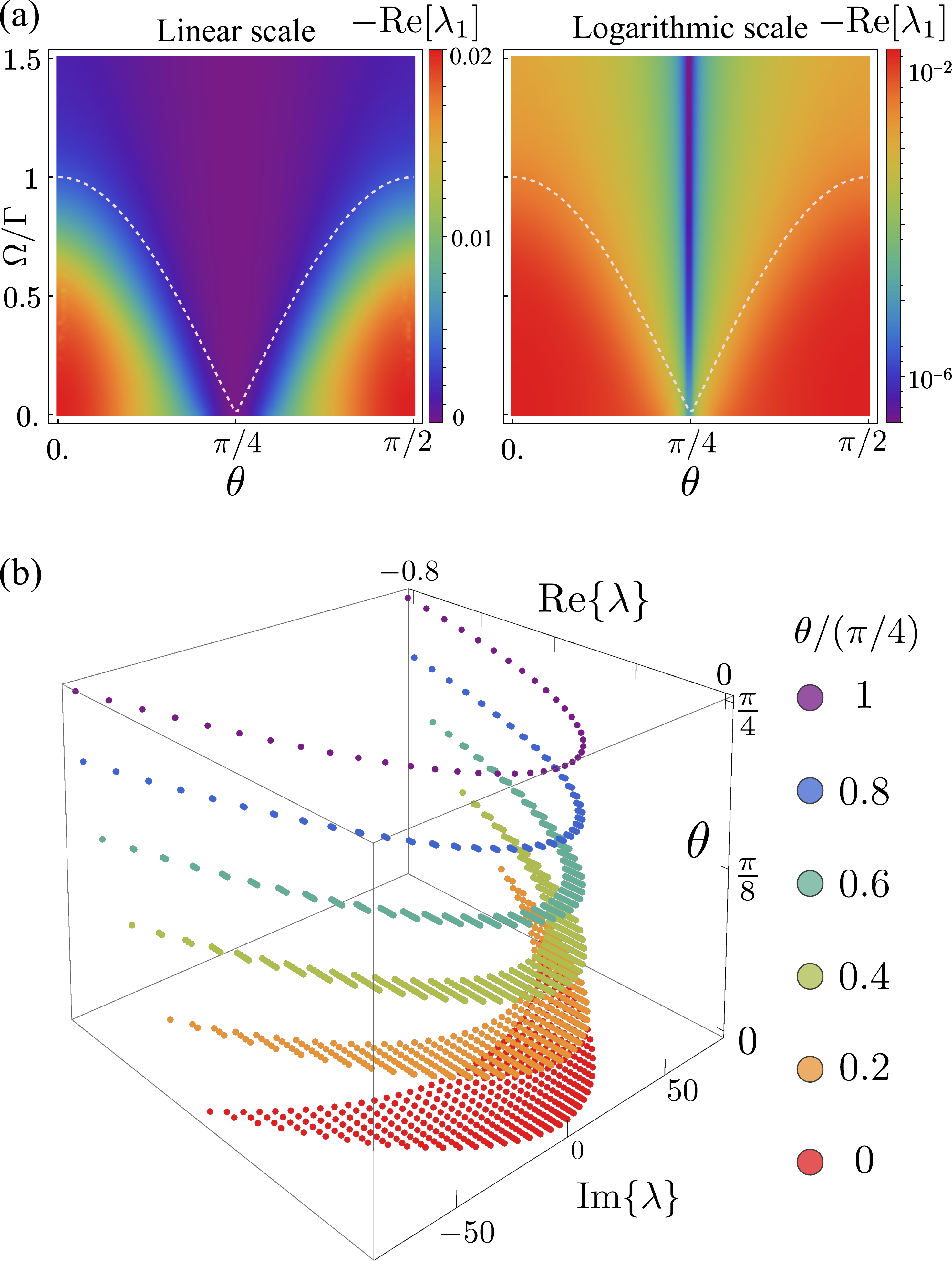}
\end{center}
\caption{(a) Liouvillian gap for $N=100$, $\Omega = 0.4\Gamma$. In the thermodynamic limit, the gap closes at the critical line $\Omega_c(\theta)$ (white, dashed). In logarithmic scale, we observe a closing of the gap for finite $J$ at the point $\theta = \pi/4$ due to the strong symmetry. (b) Liouvillian eigenvalues for a system size $J = 10$ and $\Omega = 200\,\Gamma$.  }
\label{fig:gap-linear-log}
\end{figure}

\section{Spectral properties of the Liouvillian}
\label{sec:iii}
Having characterized the phase diagram of the model, we analyse now the spectral properties of the Liouvillian, which contains essential information about the different dissipative phases and non-ergodic dynamics~\cite{kessler12a,minganti18a,macieszczak16a}.
In the ferromagnetic phase, we can use the Holstein-Primakoff expansion to obtain an expression for the Liouvillian gap in the thermodynamic limit (see~Appendix~\hyperref[sec:appendix2]{II}):
\begin{equation}
\lambda =(\Gamma_--\Gamma_+)M,
\label{eq:gap}
\end{equation}
showing that the gap closes when $M=0$, i.e. at the transition from a ferromagnetic to a thermal phase, in agreement with the usual description of DPTs~\cite{kessler12a,minganti18a}.  Figure~\ref{fig:gap-linear-log}(a) depicts the exact ADR for a finite system, computed by numerical diagonalization. The ADR in the thermal phase features a small but finite value that, as we prove below, scales with system size as $1/J$. Below we focus on this gapless region, which is the most promising in terms of non-ergodic dynamics.

\emph{Strong symmetry.}    Even for a finite system, the ADR closes exactly at the  line that separates the $\mathrm{T}_\downarrow$ and $\mathrm{T}_\uparrow$ phase, $\theta=\pi/4$, as can be seen from the logarithmic-scale plot in Fig.~\ref{fig:gap-linear-log}(a). The reason for this exact closing, that occurs even at finite system size, is the existence of a  \emph{strong symmetry} at $\theta=\pi/4$. For a general Liouvillian given by $\mathcal L \rho = -i[H,\rho]+\sum_\mu  \left(2L_\mu \rho L_\mu^\dagger -\{L_\mu^\dagger L_\mu,\rho\}\right)$, a strong symmetry is defined by a unitary operator $A$ which fulfils
\begin{subequations}
\begin{align}
\left[H,A\right] &= 0 \label{eq:strong-symmetry1},\\
\left[L_\mu,A\right] &= 0\label{eq:strong-symmetry2}.
\end{align}
\end{subequations}
As  demonstrated in Ref.~\cite{buca12a}, the existence of a strong symmetry implies that, if $A$ has $n_A$ distinct eigenvalues, there are at least $n_A$ distinct steady states of $\mathcal L$ with eigenvalue 0. In the particular case $A=H=L$ (with $L\equiv L_1$ being the only quantum-jump operator),  the density matrices $\rho^{(m)}=|m\rangle \langle m |$ are all steady states with $|m\rangle$ being the eigenstates of $A$. In our system, we find $\{H,L\}\propto S_x$ at the strong-symmetry point $\theta=\pi/4$, which means that $S_x$ is a strong symmetry of the Liouvillian and that all its eigenstates are steady-states, explaining the exact closing of the ADR. The existence of a strong symmetry at $\theta=\pi/4$ is key to understand the effect of dissipative freezing that we discuss in the following section.

\emph{Imaginary eigenvalues}
A more general analysis of the Liouvillian spectrum in the large driving limit provides further insight into the different ways in which the gap can be closed well within the thermal phase and reveals the existence of eigenstates with purely imaginary values.   In the limit $\Omega \gg \Gamma/J$, we can remove counter-rotating terms in the master equation and obtain 
\begin{multline}
\dot\rho \approx - i\Omega[ S_x,\rho]+\frac{\Gamma_\theta}{2J}\mathcal L_{S_x}[\rho]+\frac{\chi_{\theta}}{8J}\left(\mathcal L_{S_x^+}[\rho]+\mathcal L_{S_x^-}[\rho]\right),
\label{eq:me-heisenberg}
\end{multline}
where we have defined the ladder operators in the $x$-direction,  $S^{\pm}_x \equiv \frac{1}{2}\left(S_z\pm i S_y\right)$, and $\Gamma_{\theta}\equiv \Gamma(\cos\theta+\sin\theta)^2$, $\chi_{\theta}\equiv \Gamma(\cos\theta-\sin\theta)^2$.
For $\theta \neq \pi/4$, the steady state solution is the infinite-temperature state $\rho_\mathrm{\infty} = \mathbb{1}/(2J)$. One can find an analytical solution of the eigenvalue problem for this Liouvillian~\cite{ribeiro19a}; here, we provide compact, closed form expressions for the eigenvalues and relevant eigenstates. The spectrum of eigenvalues reads:
\begin{equation}
\lambda_{q,k}^\pm=\pm iq\Omega-\frac{\Gamma_\theta}{2J}q^2-\frac{\chi_\theta}{4J}\left[q+k(1+k+2q)\right],
\label{eq:exact-eigenvalues}
\end{equation}
with $q=0,1,\ldots 2J$, $k = 0,1,\ldots 2J-q$. This spectrum is plotted in Fig.~\ref{fig:gap-linear-log}(b) for different values of $\theta$. The corresponding eigenstates can be written in terms of the states:
\begin{equation}
\rho^{(n,m)}\propto (S_x^+)^n \rho_\infty (S_x^-)^m.
\label{eq:eigenstates}
\end{equation}
For a given $q$, the $(2J+1-q)$ eigenstates corresponding to the  eigenvalues $\lambda_{q,k}^i$ can be built from superpositions of different $\rho^{(n,m)}$, with $(n,m)$ fulfilling $q=|n-m|$ and $i = \mathrm{sign}(n-m)$. In particular, the eigenstates with eigenvalue $\lambda^\pm_{q,0}$, which are the slowest-decaying ones among those having the same $q$ (i.e. same imaginary eigenvalue), take the simple form $\rho^{(q,0)}$ and $\rho^{(0,q)}$. In the strong symmetry situation $\theta=\pi/4$, i.e. when $\chi_\theta = 0$, $\rho^{(n,m)}$ are the exact eigenstates themselves.

Equation~\eqref{eq:exact-eigenvalues} clearly shows that, besides the eigenvalue $\lambda_{0,0}=0$, which corresponds to the steady state, other eigenvalues with zero real part can be obtained in two ways: either reaching the thermodynamic limit $J\rightarrow \infty$, or tuning the system into the strong symmetry situation, $\chi_\theta = 0$. For any fixed $q$,  $\lim_{J\rightarrow\infty}\Re[\lambda_{q,k}^\pm]=0$, implying eigenstates with finite, purely imaginary eigenvalues. 
Purely imaginary eigenvalues have as a consequence the absence of stationary states and the emergence of oscillatory dynamics in the long-time limit~\cite{buca19a}, which has recently attracted attention in similar models \cite{iemini18a,tucker18a}.  This can also be observed from a mean-field analysis (see~Appendix~\hyperref[sec:appendix2]{II}), which in the thermal phase yields the closed orbits displayed at points (iv) and (viii) in Figure~\ref{fig:1}(b).

\emph{Dynamical symmetries.}  Recently, it was shown that absence of a stationary state and the presence of long-time oscillatory dynamics in open quantum systems can be directly implied by the existence of a \emph{dynamical symmetry} operator $A$ fulfilling~\cite{buca19a}:
\begin{subequations}
\label{eq:dynamical_symmetry}
\begin{align}
[H,A]&=\Lambda A,\label{eq:aoperator} \\
[L_\mu,A]&=[L_\mu^\dagger,A]=0. 
\end{align}
\end{subequations}
In that case, the matrices $\rho^{(nm)}\equiv A^n \rho_\infty (A^\dagger)^m$, with a form similar to the states that we defined in Eq.~\eqref{eq:eigenstates}, are eigenvectors of the Liouvillian with purely imaginary eigenvalues:
\begin{equation}
\mathcal L \rho^{(nm)} = i(m-n)\Lambda \rho^{(nm)}.
\end{equation}
Despite the similarities, in the particular case of our model, the operator $S_x^-$ does not fulfil the conditions~\eqref{eq:dynamical_symmetry} of a dynamical symmetry. However,
in the $\Omega/\Gamma \gg 1$ limit, where the system is in essence purely Hamiltonian, conditions~\eqref{eq:dynamical_symmetry} are immediately satisfied, yielding  purely imaginary eigenvalues that are integer multiples of $\Omega$. We note that, in general, this will not happen for any arbitrary dissipative system in the purely Hamiltonian limit. Here, the existence of a dynamical symmetry and oscillatory dynamics in the long-time limit is a consequence of having a spin Hamiltonian with equally spaced energy levels, preventing the mechanisms of eigenstate thermalization typical of  closed many-body systems~\cite{deutsch91a,srednicki94a,rigol08a}.

\section{Dissipative freezing of the dynamics}
\label{sec:iv}
Having completely characterized the dissipative phases of the system and the spectral properties of the Liouvillian, we are ready to describe the effect of dissipative freezing. 
Several manifestations of the coexistence of multiple steady states, such as bistability and intermittency, have attracted a great deal of attention in recent years~\cite{fink18a,carr13a,melo16a,mendoza16a,letscher17a,
lee12a,fitzpatrick17a,hruby18a,muppalla18a,ates12a,malossi14a}. 
The timescale $\tau$ associated to this intermittency is related to the inverse of the ADR, which necessarily diverges at a DPT associated with a gapless Liouvillian. These critical phenomena, however, are typically discussed in contexts in which DPTs take place in the thermodynamic limit. Therefore, the long-time limit $\tau$ exists, at least formally, in any real, finite system.

Systems with a strong symmetry differ radically from this situation, since the gap is exactly closed even for a finite system. In these cases, the dynamics is split into several, unconnected symmetry sectors. In this section, we describe the evolution of individual quantum trajectories of the wavefunction, and discuss the particular situation in which the initial state is a superposition involving several of these sectors. We report the emerging phenomenon of dissipative freezing, as we show in Ref.~\cite{arXiv_sanchezmunoz19b}, and discuss in further detail the implications of this effect in several indicators of statistics of the quantum jumps, such as the activity distribution or related quantities that appear naturally in the context of thermodynamics of quantum trajectories~\cite{garrahan10a,garrahan07a,ates12a,flindt13a,hickey14a,carollo18a,garrahan18a}.

\subsection{Freezing in individual trajectories}
\label{sec:freezing}
Dissipative evolution of the system density matrix admits an alternative interpretation in terms of individual, stochastic evolution of pure wavefunctions, the so called quantum-jump or Monte Carlo wavefunction approach~\cite{molmer92a}. The predictions of the master equation are recovered when one takes an ensemble average over a sufficiently high number of trajectories.

The evolution of a single trajectory can be summarized as follows. At every differential time step $dt$, for each element of the type $(\gamma_i/2) \mathcal L_{0_i}[\rho]$ in the master equation, the wavefunction $|\psi(t)\rangle$ can randomly undergo a quantum jump with probability $p_i = \gamma_i \langle\psi(t)|O_i|\psi(t)\rangle dt$ that transforms the system, under proper normalization, as
\begin{equation}
|\psi(t+dt)\rangle \propto O_i|\psi(t)\rangle.
\end{equation}
When no jump occurs, the wavefunction evolves under the action of a non-Hermitian Hamiltonian
\begin{equation}
|\psi(t+dt)\rangle \propto (1-i \tilde H\, dt)|\psi(t)\rangle,
\end{equation}
where $\tilde H\equiv H-i\sum_i (\gamma_i/2)O_i^\dagger O_i$. These trajectories can be physically understood as individual, stochastic realizations of an experiment where quantum jumps are recorded~\footnote{These unravelings are not uniquely defined, since different jump operators can be chosen that yield the same master equation (by changing the Hamiltonian accordingly). These different unravelings would correspond to different detection schemes, such as photon counting or homodyne detection, that differ on the way the system is monitored~\cite{bartolo17a}.}. If the system is ergodic, a time average over a single trajectory also recovers the predictions of the master equation. 

In the presence of a strong symmetry, the system is not ergodic, and multiple degenerate steady states can exist~\cite{buca12a}. The actual steady state of the system is then composed by a particular superposition of these states, fixed by the initial conditions~\cite{macieszczak16a,minganti18a}. However, because the evolution is not ergodic, it is not guaranteed that a single trajectory will switch among these states, which is the main assumption behind the notion of intermittency~\cite{lee12a,fitzpatrick17a,hruby18a,malossi14a,muppalla18a,ates12a}. Another question that poses itself is whether the conservation law associated to the strong symmetry operator, $\dot A = \mathcal L^\dagger A = 0$, will hold at the level of individual trajectories.

\begin{figure}[b!]
\begin{center}
\includegraphics[width=1\columnwidth]{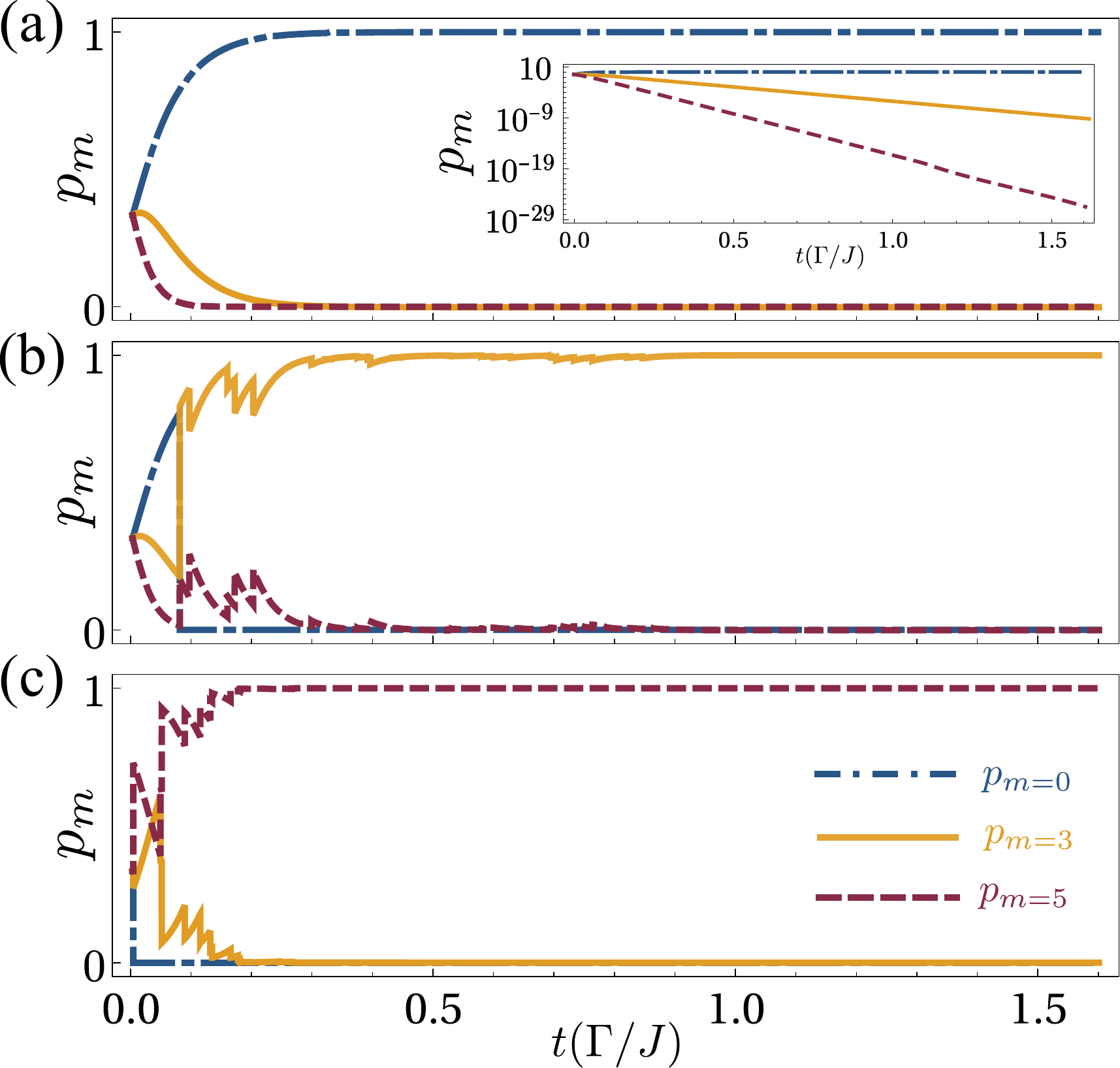}
\end{center}
\caption{Three different quantum trajectories at $\theta=\pi/4$ for the same initial state (a superposition of three eigenstates of $S_x$). Panels (a-c) show the three possible types of trajectories that occur. The inset in (a) shows the exponential decrease of the occupation of non-selected states. Parameters: $J=5$, $\Omega=0.8\Gamma$. }
\label{fig:symmetry}
\end{figure}

In the particular case $\theta=\pi/4$, the model of squeezed superradiance that we study here represents one of the simplest implementations of a strong symmetry, offering a privileged platform to address these questions.
In order to do this, we study the quantum trajectories of states initialized in superpositions of different eigenstates of $S_x$. The evolution of the wavefunction  then features what we term a ``dissipative freezing'' of the dynamics. The phenomenon is depicted on Fig.~\ref{fig:symmetry} (a--c): after initializing the state in a given superposition---in this example, of the $S_x$ eigenstates $|0\rangle$, $|3\rangle$ and $|5\rangle$---the stochastic, dissipative evolution of the wavefunction brings it into one of the eigenstates of $S_x$, with the probability of being in any of the other ones decaying exponentially with time; the evolution is effectively frozen in one eigenstate for an individual realization of the

As we prove in Ref.~\cite{arXiv_sanchezmunoz19b}, an eigenstate of a strong symmetry is invariant under this stochastic evolution, which may suggest that any quantum trajectory could eventually get ``trapped'' into one of them, analogously to a dark-state cooling or population trapping mechanism~\cite{griessner06a,aspect88a}. However, it is not guaranteed that an initial superposition of different eigenstates will always select a single one of these.
We have unambiguously proved~\cite{arXiv_sanchezmunoz19b} that this is indeed what happens in the particular case $\dot\rho = -i\Omega[A,\rho] + \Gamma/(2J)\mathcal L_A\{\rho\}$; i.e. dynamics with a single quantum jump $L$ and a general, Hermitian strong-symmetry $A\propto H \propto L$. In order to prove the emergence of dissipative freezing, we set $t_0 = 0 $ and consider an initial state $|\psi(0)\rangle = \sum_m c_m(0)|m\rangle$, expanded in the basis of eigenstates of $A$, $|m\rangle$, with eigenvalue $m$. For \emph{any} general quantum trajectory that evolves for a time $t$ undergoing $n$ quantum jumps, the probability for the final state to be in an eigenstate of $|m\rangle$ takes the form~(see~Appendix~\hyperref[sec:appendix3]{III}):
\begin{equation}
p(m; t,n) = \frac{1}{\mathcal N} \left(e^{- |m|^2}|m|^{2\alpha}\right)^{t\Gamma/J}|c_m(0)|^2,
\label{eq:S-freezing}
\end{equation}
with ${\alpha = nJ/(t\Gamma)}$ and $\mathcal N$ a normalizing constant. In Ref.~\cite{arXiv_sanchezmunoz19b}, we discuss how this equation gives, in the long time limit $t\Gamma/J\gg 1$, a distribution where only a single eigenspace of $A^\dagger A$ (completely determined by $n$) is occupied. This equation thus encapsulates the dissipative freezing effect. 

\subsection{Activity distribution}
\begin{figure}[b!]
\begin{center}
\includegraphics[width=1\columnwidth]{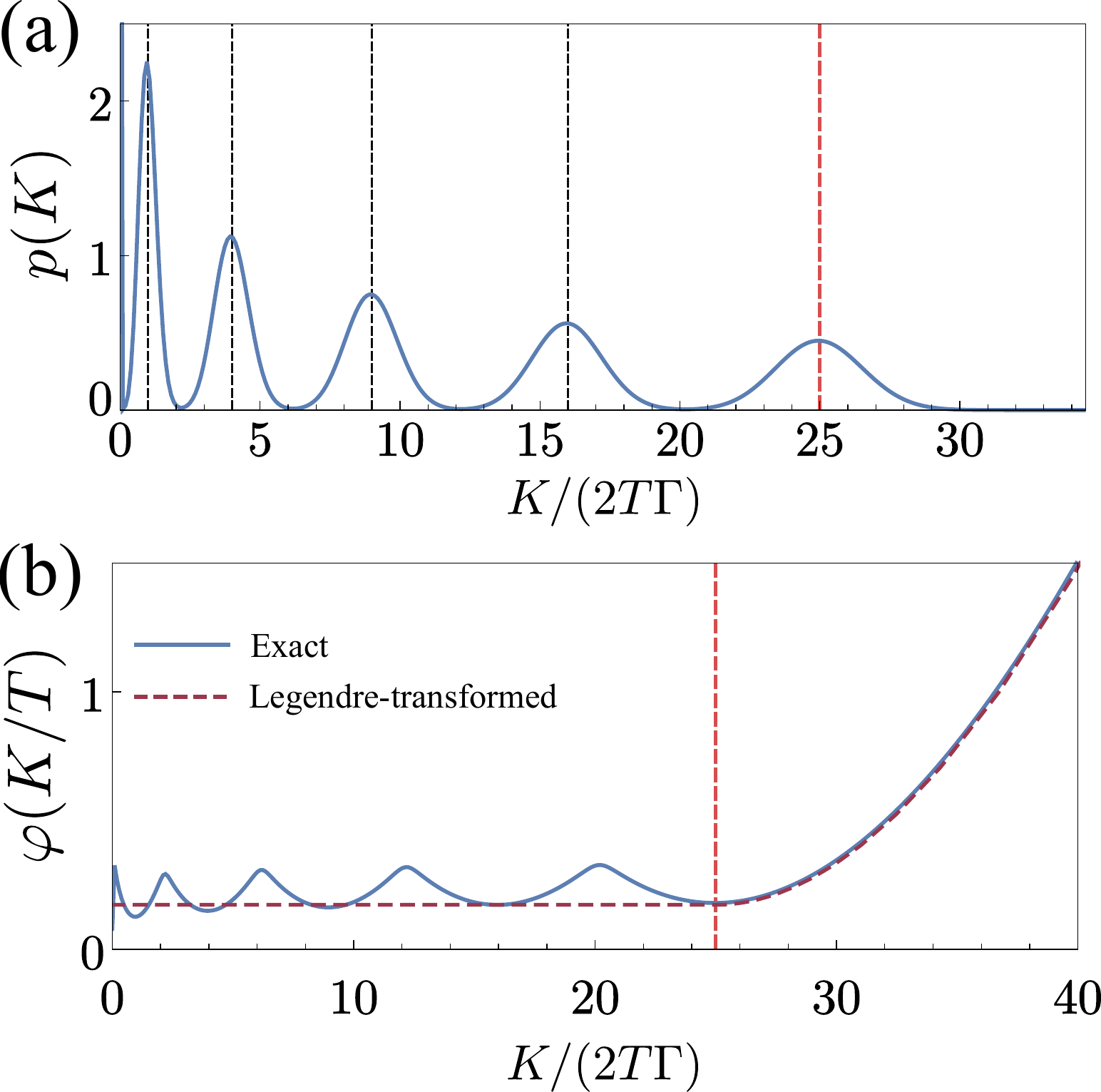}
\end{center}
\caption{(a) Probability of having $K$ quantum jumps in a time $T=3\cdot 10^3/\Gamma$ in the case of our model, where $A=S_x$. $N=20$, $\Omega = 0.8\Gamma$, $\theta = \pi/4$ (strong symmetry point). (b) Rate function  $\varphi(k)$, obtained directly from the logarithm of Eq.~\eqref{eq:pn} (solid, blue) and by an Legendre transformation (dashed,red). The Legendre transformed $\varphi(k)$ is given by Eq.~\eqref{eq:legendre-inverse} plus an additive constant to match the normalization of $p(K)$ for a finite $T$. }
\label{fig:rate}
\end{figure}

Now that we have presented the dissipative freezing effect, it is instructive to analyse it in terms of one of the main observables of interest when discussing multistability; the activity~\cite{garrahan10a,ates12a}. The activity is defined as the mean number of quantum jumps undergone by the system per unit time; this can be defined through the probability distribution $p_T(K)$ of counting $K$ jumps on a time $T$.
Following our previous discussion, we assume the existence of a strong symmetry $A$ with eigenstates $|m\rangle$ and only one quantum jump operator, $L=\sqrt{\Gamma/J}A$. We consider an initial state with the form
\begin{equation}
\rho(0) = \sum_m c_m |m\rangle\langle m|.
\end{equation}
This initial state is a steady-state of the system, meaning that its preparation can always be conceived as the long-time limit of another initial state. Other choices of $\rho(0)$ may involve transient effects that will be irrelevant in the limit $T\rightarrow\infty$.  We can then prove (see~Appendix~\hyperref[sec:appendix4]{IV}) that the photon counting distribution takes the form:
\begin{equation}
p_T(K)= \sum_m \frac{1}{K!}\left(\frac{T\Gamma m^2}{J} \right)^K e^{-\Gamma m^2 T/J}c_m,
\label{eq:pn}
\end{equation}
which is dependent on the initial state. This equation presents the multimodal structure depicted in Fig.~\ref{fig:rate}(a), where we plot it for the particular case of our model, where $A=S_x$. The physical interpretation is simple: to every eigenstate $|m\rangle$ of $A$, there is an associated steady state:
\begin{equation}
\rho_0^{(m)} = |m\rangle \langle m|,
\label{eq:rho0m-ss}
\end{equation}
with a corresponding quantum-jump rate of $\mathrm{Tr}[L^\dagger L \rho]=m^2\Gamma/J $. The set of $\rho^{(m)}_0$ conform a basis, meaning that any combination of these steady-states is a also steady state. The asymptotic state 
\begin{equation}
\rho_\mathrm{ss}=\lim_{t\rightarrow \infty}e^{\mathcal L t}\rho(0) = \sum_m \mathrm{Tr}[\rho_0^{(m)},\rho(0)]\rho_0^{(m)},
\label{eq:rhoss}
\end{equation}
is therefore strongly dependent on the initial state and given by its overlap with each of the $\rho_0^{(m)}$. Those $\rho_0^{(m)}$ having a finite overlap with $\rho(0)$ will manifest as a distinct peak in the counting distribution $p_T(K)$, centered at the value $K_m=Tm^2\Gamma/J $.

Multimodality (as a signature of multistability) has been recently associated with dynamical phase transitions~\cite{ates12a} that feature the coexistence of two phases \emph{in time}, with an stochastic switching between these phases that has been observed experimentally on multiple occasions~\cite{lee12a,fitzpatrick17a,hruby18a,malossi14a,muppalla18a}. While we obtain a clear multimodal structure for the activity distribution, our results on the dissipative freezing do not match this notion of intermittency. Let us therefore put our results in the context of the theory used in Ref~\cite{ates12a}: the thermodynamics of quantum trajectories.

\subsection{Thermodynamics of quantum trajectories: Dynamical phase transition}
\subsubsection{Brief introduction to thermodynamics of quantum trajectories}

Recently, several works ~\cite{garrahan10a,garrahan07a,ates12a,flindt13a,hickey14a,carollo18a,garrahan18a} have approached the questions of multimodality and intermittency from the perspective of the thermodynamics of quantum trajectories. This approach regards the set of quantum trajectories in which the dynamics can be unraveled as an statistical  ensemble that can be analysed using the tools of statistical mechanics.
In the following, we briefly outline this theory (a comprehensive description can be found in Refs.~\cite{garrahan10a,ates12a}) and discuss its implications in systems, such as the one we report here, where dissipative freezing of the dynamics occurs.

Let us consider a system governed by the master equation $\dot \rho =\mathcal L\rho =  -i[H,\rho] + L\rho L^\dagger - \frac{1}{2}\{L^\dagger L ,\rho \} $. The evolution of $\rho$ can be unraveled as a set of quantum trajectories~\cite{zoller87a,molmer92a,plenio98a} by which a conditional density matrix $\rho_K(t)$ can be built from the ensemble average of all the trajectories of duration $t$ having $K$ quantum jumps. The activity distribution is then  given by $p_K(t) = \mathrm{Tr}\rho_K(t)$. We can define a generating function $Z=\langle e^{sK}\rangle$:
\begin{equation}
Z =\sum_{K=0}^\infty e^{sK}p_K(t) = \mathrm{Tr}\sum_{K=0}^\infty e^{sK}\rho_K(t) = \mathrm{Tr}\rho_s(t),
\label{eq:Z}
\end{equation}
where $\rho_s(t)\equiv \sum_{K=0}^\infty e^{sK}\rho_K(t)$ is a Laplace transformed density matrix that evolves according a tilted master equation:
\begin{equation}
\mathcal{W}_s\rho_s = \dot \rho_s = -i[H,\rho_s] + e^{s}L\rho L^\dagger - \frac{1}{2}\{L^\dagger L,\rho \},
\label{eq:me-laplace}
\end{equation}
and the ``counting field'' $s$ is a variable conjugate to $K$. For $s=0$, Eq.~\eqref{eq:me-laplace} corresponds to the normal master equation, $\mathcal W_s =\mathcal L$. For $s\neq 0$,  Eq.~\eqref{eq:me-laplace} is not a physical trace-preserving master equation, and describes a class of dynamics in which the quantum jumps are biased by the factor $e^{s}$. Despite $\mathcal W_s$ being unphysical, its spectral properties contain valuable information about the fluctuations of the ensemble of trajectories. In particular, the partition function acquires, in the long time limit, a large deviation form $Z \asymp e^{t\lambda(s)}$, with $\lambda(s)$  the eigenvalue of $\mathcal W_s$ with the largest real part. This allows us to write the  activity or mean emission rate as:
\begin{equation}
\langle k \rangle = \langle K \rangle / t = \frac{1}{t}\left.\frac{\partial Z}{\partial s}\right|_{s=0}=\left. \frac{\partial \lambda(s)}{\partial_s}\right|_{s=0}.
\label{eq:k}
\end{equation}
This suggest the definition of a $s$-dependent emission rate $\langle k\rangle_s \equiv \partial\lambda/\partial s(s)$. Equivalently, fluctuations in the activity can be described by  Mandel's $Q$ parameter, $Q = (\langle K ^2\rangle -\langle K \rangle ^2)/\langle K \rangle - 1$, given by:
\begin{equation}
Q = \left.\frac{\partial^2\lambda/\partial s^2}{\partial \lambda/\partial s}\right|_{s=0}.
\label{eq:Q}
\end{equation}
To sum up, the behaviour of $\lambda(s)$ around the vicinity of $s=0$ characterizes the fluctuations of the ensemble of quantum trajectories.

The connection to thermodynamics put forward in Ref.~\cite{garrahan10a} can be made by assuming that, in the long-time limit, $p_K(t)$ also acquires a large deviation form
\begin{equation}
p_K(t) \asymp  e^{-t\varphi(K/t)}.
\label{eq:pK-LD}
\end{equation}
If $p_K(t)$ describes the probability distribution of an statistical ensemble, then the \emph{rate function} $\varphi(K/t)=-\ln p_K(t)/t$ plays the role of an entropy density~\cite{touchette09a}. By plugging Eq.~\eqref{eq:pK-LD} into Eq.~\eqref{eq:Z}, we obtain directly that $\varphi(k=K/t)$ and $\lambda(s)$ are related by a Legendre transformation:
\begin{equation}
\lambda(s) = \max_k[ks-\varphi(k)],
\label{eq:legendre}
\end{equation}
meaning that $\lambda(s)$ has the properties of a free energy. The inverse transformation
\begin{equation}
\varphi(k)=\max_s[ks-\lambda(s)]
\label{eq:legendre-inverse}
\end{equation}
is a useful relation that allows us to obtain $\varphi(k)$ from the knowledge of $\lambda(s)$, which can in turn be computed from the eigenvalues of $\mathcal W_s$. However, this relation follows from the G{\"a}rtner-Ellis Theorem~\cite{touchette09a}, that requires $\lambda(s)$ to be differentiable for all $s\in \mathbb{R}$ or, equivalently, $\varphi(k)$ to be concave for all $k\in\mathbb{R}$. These are precisely the conditions that are violated when a  phase transition occurs. 

\subsubsection{Multistability and breaking of the intermittency: connection to known models}
\begin{figure}[t!]
\begin{center}
\includegraphics[width=0.99\columnwidth]{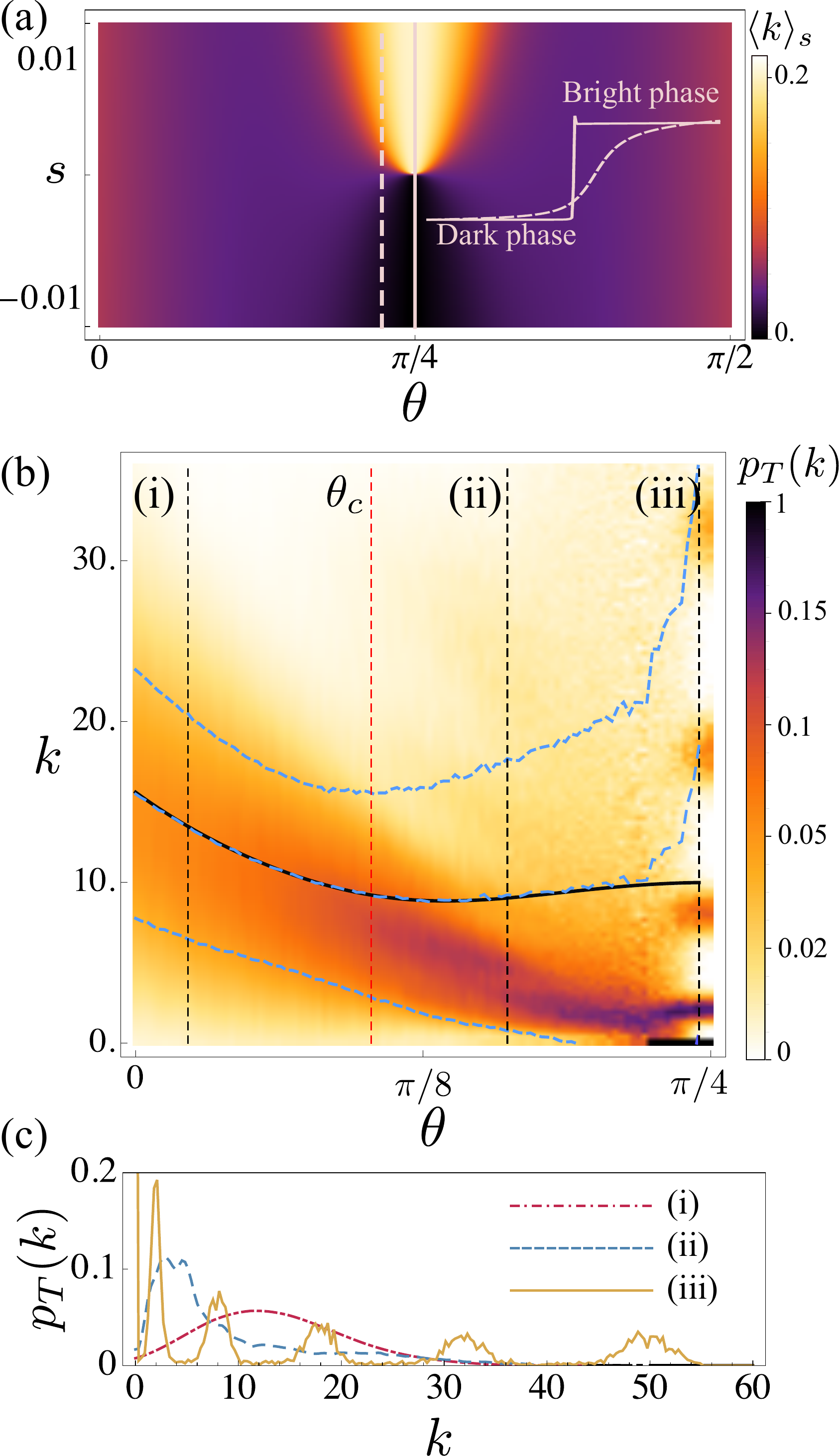}
\end{center}
\caption{(a) $\langle k\rangle_s$ versus $\theta$, featuring the coexistence between a bright and a dark phase in the vicinity of $\theta=\pi/4$.  (b) Probability distribution $p_T(k)$ of the activity versus the squeezing angle $\theta$ for $\Omega = 0.8 \Gamma$. The distribution for each $\theta$ has been computed from 400 Monte Carlo trajectories. The time $T$ has been taken en each case as half the relaxation time, $\tau=1/|\lambda_1|$.  Points (i---iii) indicate the three values of $\theta$ shown in panel (c).  Solid-blue (dashed-black) correspond to $\langle k \rangle$ calculated from the master equation (the Monte Carlo trajectories); dashed, blue lines correspond to the variance $\Delta k^2$ computed from the Monte Carlo trajectories.}
\label{fig:multistability}
\end{figure}

In Ref.~\cite{ates12a}, the coexistence of dynamical phases was linked to a discontinuity in the $s$-dependent order parameter $\langle k\rangle_s$ at the physical point $s=0$, i.e. a first-order phase transition with respect to the counting field. Based on this temporal coexistence between phases, such a first-order phase transition was then referred to as a dynamical phase transition.   As we show in Fig.~\ref{fig:multistability}(a), where we plot a numerical calculation of  $\langle k\rangle_s$ versus $\theta$,  the closing of the gap at the strong symmetry point gives rise to such a discontinuity; the limit $s\rightarrow 0^+$ features a bright phase characterized by a high activity, whereas for $s\rightarrow 0^-$ we find a dark phase with virtually no quantum jumps. The discontinuity turns into a continuous crossover as we depart from the point $\theta=\pi/4$, consistent with a  first-order phase transition smoothed by finite-size effects. Such a crossover is responsible for the phenomenon of intermittency typically observed in finite many-body systems undergoing a DPT~\cite{ates12a}. When the crossover turns into a real discontinuity, here due to appearance of a strong symmetry, intermittency is substituted by the phenomenon of dissipative freezing.

We elaborate this argument by proving first that a strong symmetry implies that $\langle k\rangle_s$ is discontinuous  (a  similar analysis was performed in Ref·~\cite{manzano14a}). Following our previous discussions, we focus on the case where a strong symmetry $A$ is present, and $L=\sqrt{\Gamma/J}A$. We can immediately see that the steady states $\rho_0^{(m)}$ in Eq.~\eqref{eq:rho0m-ss} are also eigenstates of $\mathcal W_s$, with eigenvalues:
\begin{equation}
\lambda^{(m)}(s)=\frac{\Gamma}{J}m^2(e^s-1).
\end{equation}
Since these are the largest eigenvalues for $s=0$, they must also be in the vicinity of that point. Therefore, we can write $\lambda(s)$ around $s=0$ as:
\begin{equation}
\lambda(s)=\begin{cases}
(\Gamma/J) m_{\mathrm{min}}^2(e^s-1) \quad s < 0\\
(\Gamma/J) m_{\mathrm{max}}^2(e^s-1) \quad s > 0,
\end{cases}
\end{equation}
with $m_{\mathrm{min}/\mathrm{max}}$ the minimum/maximum eigenvalues of $A$. If $m_\mathrm{min}\neq m_\mathrm{max}$, it is clear that $\lambda(s)$ shows a singular behaviour at $s=0$, having a discontinuous derivative. Contrary to the situations typically considered, the discontinuity does not become a crossover when the system has a finite size, since its origin is the exact closing of the Liouvillian gap due to the strong symmetry (see Fig.~\ref{fig:gap-linear-log}).
If we were to try to find $\varphi(k)$ by blindly applying Eq.~\eqref{eq:legendre-inverse} with a generic expression for $\lambda(s)=(\Gamma/J)m^2(e^s-1)$, we would find that the value of $s$ that maximizes $ks-\lambda(s)$ is given by:
\begin{equation}
s=
\begin{cases}
\ln\left[Jk/(\Gamma m_\mathrm{min}^2)\right]  \quad k <\frac{\Gamma}{J}m_\mathrm{min}^2,\\
\ln\left[Jk/(\Gamma m_\mathrm{max}^2)\right]  \quad k >\frac{\Gamma}{J}m_\mathrm{max}^2,\\
0 \quad \quad \quad \quad   \frac{\Gamma}{J}m_\mathrm{min}^2 < k <\frac{\Gamma}{J}m_\mathrm{max}^2,
\end{cases}
\label{eq:lambda_discontinuity}
\end{equation} 
This yields the rate function shown in dashed-red in Fig.~\ref{fig:rate}(b): the non-concave regions of $\varphi(K/t)$ associated to multimodality translate into a nonphysical flat plateau when one tries to use the inverse Legendre transformation in Eq.~\eqref{eq:legendre-inverse}. This result connects back to standard thermodynamics, where phase transitions are associated with non-concavities in the underlying fundamental equations for the thermodynamic potentials. 
A multimodal distribution $p_T(k)$ as we obtained in Eq.~\eqref{eq:pn} will always yield a discontinuous $\lambda(s)$ and will therefore be linked to a first-order phase transition.

To summarize, we have discussed the notions of dissipative freezing~\eqref{eq:S-freezing}, multimodal activity distributions~\eqref{eq:pn} and first order phase transitions at the trajectory level \eqref{eq:lambda_discontinuity}. We conclude that these phenomena are linked, since all of them emerge from the existence of a strong symmetry that yields a perfect closing of the Liouvillian gap for any system size.  Intermittency is therefore a consequence of the finite system size; it implies a smoothing of the phase transition that allows to make use of Eq.~\eqref{eq:legendre-inverse}, but that gives in turn a unimodal probability distribution: i.e. in the long time limit, intermittency destroys multimodality~\cite{macieszczak16b}. Dissipative freezing can therefore be alternatively described as the survival of multimodality in the long-time limit. In quantum metrology, this has strong implications for the scaling in time of the Fisher information~\cite{macieszczak16b}.

These ideas are further supported by numerical calculations in Fig.~\ref{fig:multistability}(b-c), where we show $p_T(k)$ computed from sets of quantum trajectories, for time windows approximately twice the inverse Liouvillian gap, $T \approx (2\Re\{\lambda_2\})^{-1}$.  The value of $\Omega = 0.8\Gamma$ is such that we can observe the transition from the ferromagnetic to the thermal phase at $\theta_c\approx 0.4$. When this transition is crossed,  fluctuations start increasing with $\theta$---see dashed blue lines in panel (b)---and the unimodal distribution is strongly distorted. This characteristic of the thermal phase is the consequence of the increased asymmetry on $\langle k\rangle_s$ at $s=0$---see panel (a)---which is associated to the closing of the Liouvillian gap. As we get close to $\pi=\pi/4$, where $\Re\{\lambda_2\} =  0 $, it  becomes impossible to simulate times of the order of $\Re\{\lambda_2\}^{-1}$. In the plot, this is identified by the emergence of several peaks in $p_T(k)$: the crossover in $\langle k \rangle_s$ gives rise to a  multi-peaked structure that would merge into a single peak were $T$  long enough. Since this multimodality does not correspond to the long-time limit, the large-deviations approach is unable to describe it; this is the situation in which intermittency occurs. On the other hand, the strong symmetry point features a multimodal $p_T(k)$ for any $T$; that survival of the multimodal structure is the signature of dissipative freezing of the dynamics.

\begin{figure}[t!]
\begin{center}
\includegraphics[width=1\columnwidth]{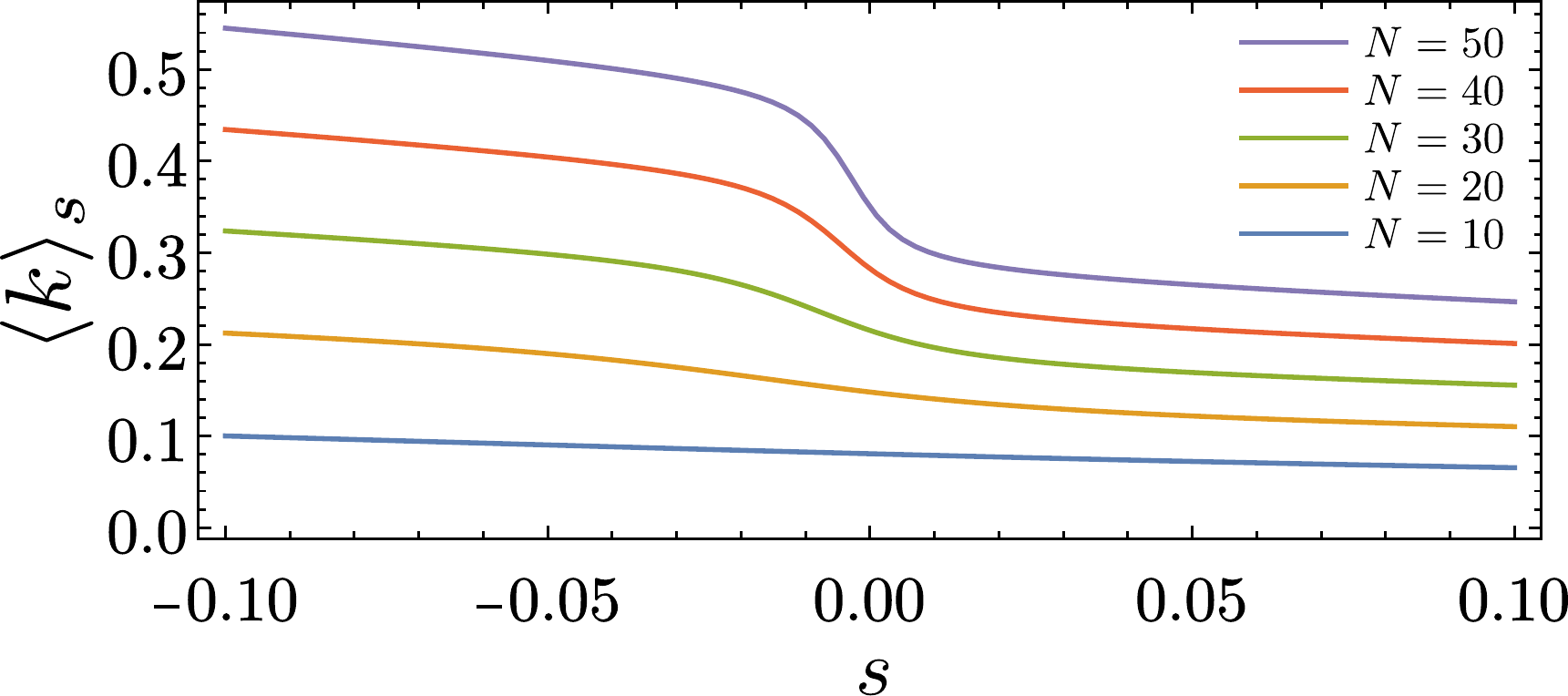}
\end{center}
\caption{Emergence of a crossover in the $s$-dependent activity parameter with increasing system size at $\theta=0$. $\Omega = 4\Gamma$. }
\label{fig:kink-N}
\end{figure}

Finally, we note that the closing of the Liouvillian gap in the thermodynamic limit of the thermal phase---c.f. Eq.~\eqref{eq:exact-eigenvalues}---also yields a crossover in $\langle k \rangle_s$ (see Fig.~\ref{fig:kink-N}). Since this closing is of a different nature (associated with eigenvalues with imaginary part), it offers the interesting prospect of studying multistability and intermittency between phases displaying coherent, oscillatory dynamics in the long-time limit. This could be done, for instance, by analysing the time correlations between the spectral features of the different phases, as we discuss in the following section.

\begin{figure*}[t!]
\begin{center}
\includegraphics[width=0.97\textwidth]{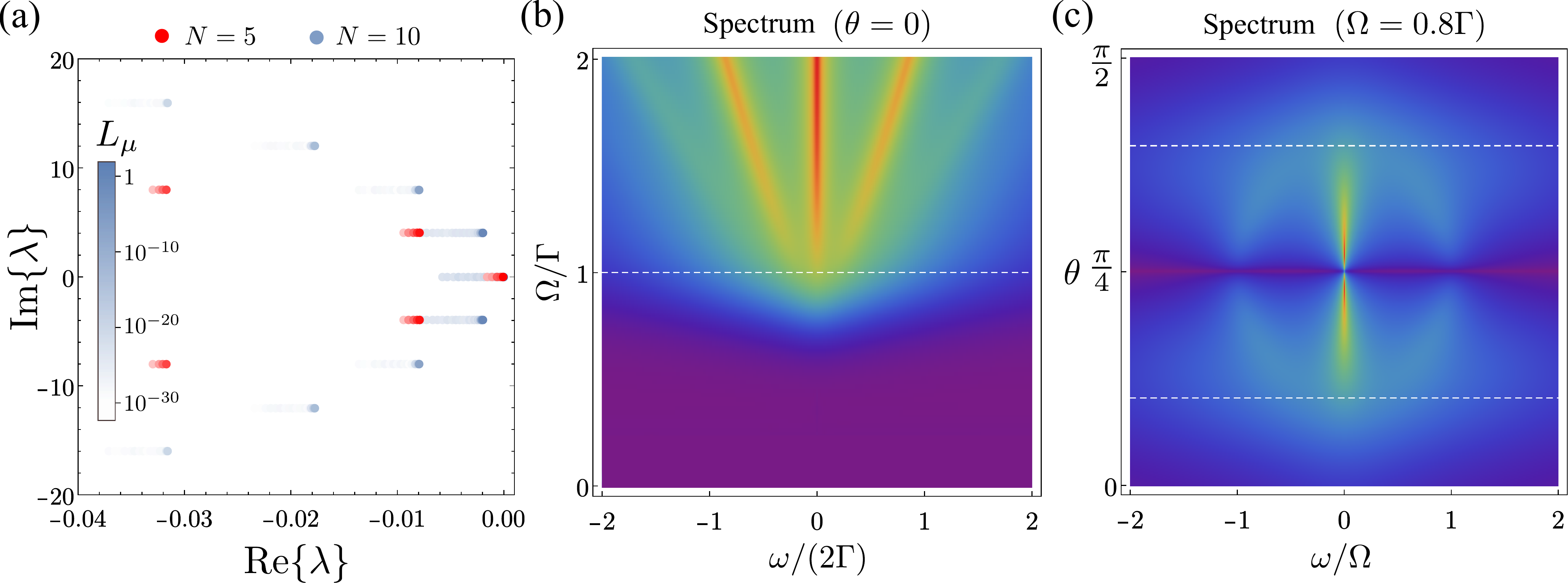}
\end{center}
\caption{(a) Liouvillian eigenvalues $\lambda_\mu$ weighted by $L_\mu$. This illustrates the set of eigenvalues that are experimentally accessible by the measurement of the spectrum of emission. (b) Spectrum of emission versus $\Omega$, for $\theta=0$. (c) Spectrum versus $\theta$, for $\Omega = 0.8 \Gamma$. At the strong symmetry point, the gap closes exactly, and the spectrum features an extreme line-narrowing.  White, dashed lines indicate where a phase transition occurs; the signature of the phase transition is the emergence of sideband peaks. }
\label{fig:spectrum}
\end{figure*}

\begin{figure}[b!]
\begin{center}
\includegraphics[width=0.97\columnwidth]{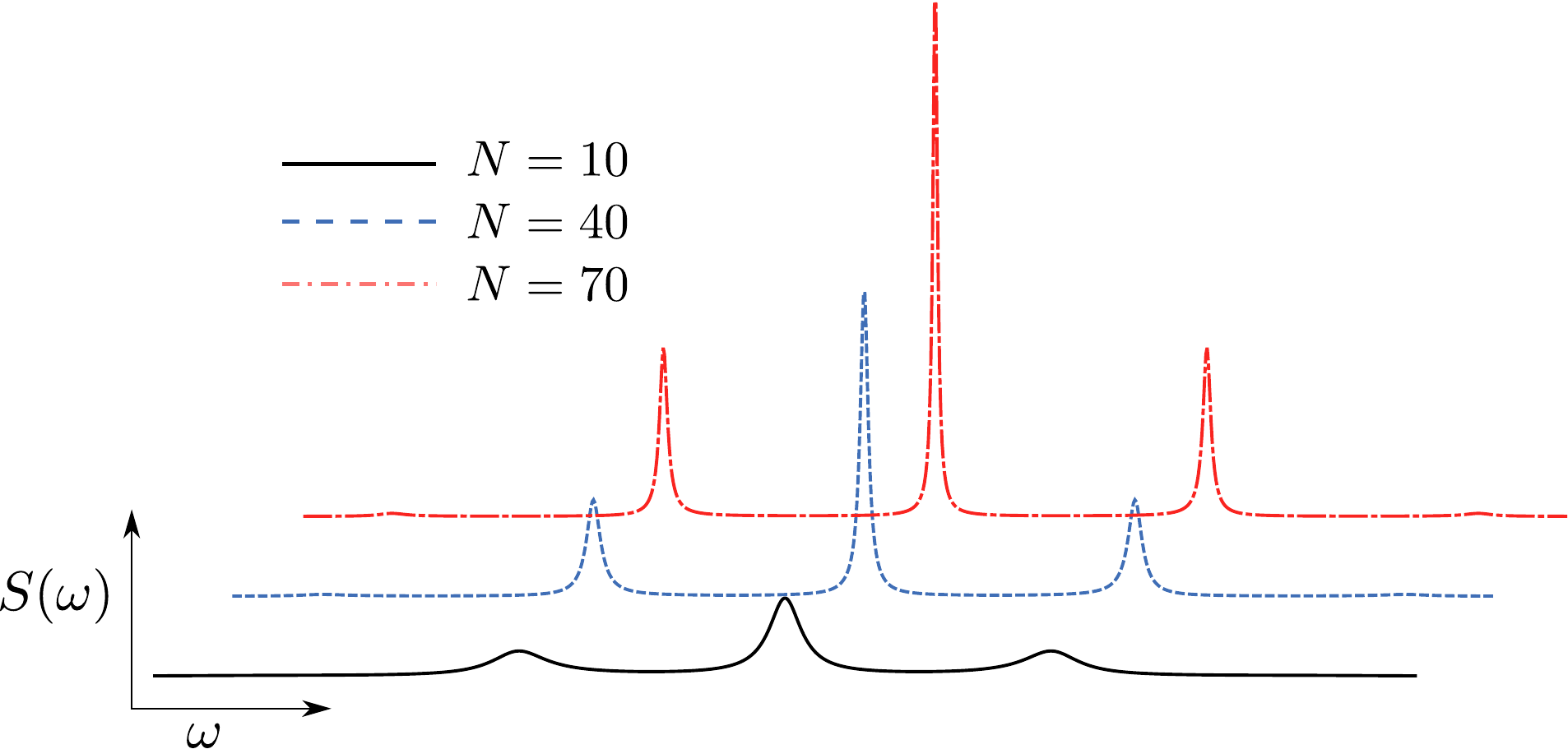}
\end{center}
\caption{Spectrum of emission, $ S(\omega)$,  in the thermal phase, for different values of $N$. The fact that the real part of the highest eigenvalues goes to zero as $1/N$ can be measured as a narrowing of the spectral peaks. Parameters: $\theta=0$, $\Omega = 2\Gamma$. }
\label{fig:spectrumN}
\end{figure}

\section{Signatures of critical dynamics in the emitted light}
\label{sec:v}
In this section, we discuss the possibility of probing some of the essential features of the low-energy spectrum of the Liouvillian by analysing the light emitted by the system. 
Many of the essential features of critical dissipative dynamics are encoded in the spectral properties of the Liouvillian. Stationary observables of the form $\langle O \rangle = \mathrm{Tr}[O \rho_0]$  contain a limited amount of information about these properties, since they depend only on the lowest eigenvalue of $\mathcal L$. However,  observables involving two-time correlators of the form $\langle O (t) O(t+\tau)\rangle$  require a knowledge not only of $\rho_0$, but also of the Liouvillian $\mathcal L$. Consequently, they carry information about the dynamics of the system that is not present $\rho_0$, and can provide valuable data about $\mathcal L$, such as its spectral properties, in an experimentally accessible way. To illustrate this point, we focus here on the case of the spectrum of emission, providing a closed-form expression in terms of the Liouvillian eigenvalues and right and left eigenstates.

We define the (unnormalized) spectrum of emission as 
\begin{equation}
S(\omega) = \lim_{t\rightarrow \infty}\frac{1}{\pi} \Re\int_0^\infty d\tau\, e^{i\omega \tau} \langle a^\dagger (t) a(t+\tau) \rangle,
\end{equation}
where, generally, $a$ is some system operator linked to the bath output operator by input-output relations (in our case, $a=D_\theta$).  By applying the quantum regression theorem~\cite{gardiner_book00a}, we obtain:
\begin{equation}
S(\omega) = \lim_{t\rightarrow \infty}\frac{1}{\pi} \Re\int_0^\infty d\tau\, e^{i\omega \tau} \mathrm{Tr}\left[a e^{\mathcal L \tau}(\rho(t) a^\dagger) \right].
\label{eq:spectrum-int}
\end{equation}
Note that, typically, the limit $t\rightarrow \infty$ will  imply that $\rho(t)$ is simply $\rho_\mathrm{0}$, the steady state of the system. In most systems this steady state is unique, but here we want to take into account the possibility of multiple steady states (i.e. multiple eigenstates of $\mathcal L$ with eigenvalues with zero real part), meaning that $\rho(t)$ can be any superposition of these steady states, defined by the initial state. Therefore, we take the limit $t\rightarrow \infty$ and substitute $\rho(t)$ by an arbitrary superposition of steady states, $\rho_\mathrm{ss}$, determined by the initial state. We can perform a spectral decomposition of the Liouvillian to write, for any $\rho$:
\begin{equation}
e^{\mathcal L t}\rho = \sum_\mu e^{\lambda_\mu t}\mathrm{Tr}[\rho_{ \mathrm L, \mu}\rho]\rho_{ \mathrm R,\mu},
\end{equation}
where $\rho_{L/R,\mu}$ is the left/right eigenstate of $\mathcal L$ with eigenvalue $\lambda_\mu$. This allows us to write
\begin{multline}
S(\omega) =\frac{1}{\pi}\Re\int_0^\infty d\tau \, \sum_ {\mu}e^{(i\omega+\lambda_\mu)\tau}\\
\times \mathrm{Tr}[a\rho_{\mathrm{R},\mu}]   \mathrm{Tr}[ a^\dagger\rho_{\mathrm L , \mu}\rho_\mathrm{ss}] .
\label{eq:spectrum-formal}
\end{multline}
By defining
\begin{subequations}
\begin{eqnarray}
\omega_\mu &\equiv & \Im{\lambda_\mu}\\
\gamma_\mu/2 &\equiv & -\Re{\lambda_\mu}\\
L_\mu & \equiv & \Re{\mathrm{Tr}[a\rho_{\mathrm{R},\mu}]\mathrm{Tr}[a^\dagger\rho_{\mathrm L , \mu}\rho_\mathrm{ss}]}\\
K_\mu & \equiv & \Im{\mathrm{Tr}[a\rho_{\mathrm{R},\mu}]\mathrm{Tr}[a^\dagger\rho_{\mathrm L , \mu}\rho_\mathrm{ss} ]},
\end{eqnarray}
\end{subequations}
we can formally integrate Eq.~\eqref{eq:spectrum-formal} and obtain
\begin{multline}
S(\omega) = \frac{1}{\pi}\sum_ {\mu, \Re{\lambda_\mu}\neq 0}\frac{(\gamma_\mu/2) L_\mu -(\omega+\omega_\mu)K_\mu}{(\gamma_\mu/2)^2+(\omega+\omega_\mu)^2}\\
+\sum_ {\mu, \Re{\lambda_\mu}= 0}\left[ L_\mu \delta(\omega+\omega_\mu)+\frac{K_\mu}{\pi}\mathrm{ P. V.}\left(\frac{1}{\omega+\omega_\mu}\right)\right].
\label{eq:spectrum-final}
\end{multline}
Note that terms with $\Re{\lambda_\mu}=0$ give rise to a series of $\delta$-peaks in the spectrum,  positioned at frequencies that are given by the imaginary part of the eigenvalues with zero real part. The last term means that the principal value integral of $1/(\omega+\omega_\mu)$ should be computed when integrating that distribution. That term never appears in the case of a unique steady state ($\mu=0$), since in that case $\rho_{L,0}= \mathbb{1}$ and $K_0 = \Im{|\langle a \rangle_\mathrm{ss}|^2}=0$. All the terms proportional to $K_\mu$ in Eq.~\eqref{eq:spectrum-final} are \emph{dispersive} lineshapes that break the symmetry of the corresponding Lorentzians (proportional to $L_\mu$). Although they may appear unphysical (since they can yield negative values) they give a physical result once the sum is performed.

Equation~\eqref{eq:spectrum-final} tells us that the spectrum of emission can be used to probe the Liouvillian spectrum and  also infer, indirectly, information about the right and left eigenvectors. Similar formal integrations of Eq.~\eqref{eq:spectrum-int} have been presented before~\cite{delvalle09b,delvalle12a,ridolfo13a}; ours differ from these in that they make explicit use of the left and right eigenvectors of $\mathcal L$. In particular, we see that the existence of eigenvalues with zero real part and finite imaginary part translates into the presence of measurable $\delta$-peaks in the spectrum. These turn into peaks with a finite width when the linewidth of the detectors and/or other unavoidable losses to different channels are included in the description. Figure~\ref{fig:spectrum} illustrates the information about the Liouvillian eigenvalues provided by the spectrum in the model discussed in this work: panel (a) shows distribution of eigenvalues for $N=(5,10)$, weighted by their value of $L_\mu$. This way, features like the emergence of imaginary eigenvalues with vanishing real part in the thermodynamic limit can be directly measured in the laboratory. We show this in panel (b), where the ferromagnetic-thermal DPT is shown to be accompanied by the emergence of sideband peaks in the fluorescence spectrum; this is the well known generalization of the Mollow triplet to the case of collective resonance fluorescence~\cite{carmichael77a}. The result that the Liouvillian gap closes in this phase as $1/N$ can be confirmed experimentally: as shown in Fig.~\ref{fig:spectrumN}, it can be measured directly as a decrease in the linewidth of the spectral peaks. Finally, Fig.~\ref{fig:spectrum}(c) shows the emergence of sideband peaks when $\theta$ is varied so as to enter in the thermal phase, and the observation of extreme line-narrowing as the gap is closed exactly at the strong-symmetry point.

These results open the intriguing possibility of exploring the notions of ergodicity, intermittency and dissipative freezing in systems with Liouvillian eigenvalues with vanishing real part and finite imaginary part by studying temporal correlations between different spectral windows~\cite{cohentannoudji79a,schrama92a,nienhuis93a,ulhaq12a,%
delvalle12a,silva16a,peiris15a,bounouar17a}. This will be a topic of study for future works.

\section{Conclusions}
We have studied the model of squeezed superradiance an analysed the different types of non-ergodic dynamics emerging in dissipative phases with a gapless Liouvillian. In order to identify the relevant regimes of non-ergodic dynamics, we have completely characterized the phase diagram of the system, its metrological properties, and its Liouvillian spectrum. 
We have shown the existence of non-stationary dynamics linked to Liouvillian eigenvalues with a finite imaginary part and a vanishing real part in the thermodynamic limit, and we have reported the phenomenon of dissipative freezing that appears when the Liouvillian has a strong symmetry. We have connected the phenomenon of dissipative freezing with the theory of thermodynamics of quantum trajectories, showing that it is linked to a real discontinuity in the associated first-order phase transition with respect to the counting field. Intermittency is, on the other hand, linked to the smoothing of such first-order phase transition into a crossover due to finite-size effects. Notably, the model studied here allows to explore all this phenomenology with a finite size system that can be treated numerically. Our work sheds new light on the critical behaviour of open systems with finite system size, and might provide new routes in the development of sensors based on the critical behaviour of driven-dissipative quantum systems~\cite{macieszczak16b,fernandezlorenzo17a,gammelmark14a,kiilerich14a,kiilerich16a}.

\section*{ACKNOWLEDGEMENTS}
C.S.M. kindly acknowledges F. Minganti for fruitful and insightful discussions. B.B and C.S.M are grateful to Juan P. Garrahan for very useful comments and insightful discussions. 
C.S.M. is funded by the Marie Sklodowska-Curie Fellowship QUSON (Project  No. 752180). B.B., J.T. and D.J. acknowledge support from the EPSRC grants No. EP/P009565/1 and EP/K038311/1, and the European Research Council under the European Union’s Seventh Framework Programme (FP7/2007-2013)/ERC Grant Agreement No. 319286 Q-MAC. AGT and DP acknowledge support from CSIC Research Platform on Quantum Technologies PTI-001 and from Spanish project PGC2018-094792-B-100 (MCIU/AEI/FEDER, EU).

\addcontentsline{toc}{chapter}{Bibliography}
\bibliographystyle{mybibstyle}
\bibliography{Sci,arXiv,books}

\begin{thebibliography}{107}%
\makeatletter
\providecommand \@ifxundefined [1]{%
 \@ifx{#1\undefined}
}%
\providecommand \@ifnum [1]{%
 \ifnum #1\expandafter \@firstoftwo
 \else \expandafter \@secondoftwo
 \fi
}%
\providecommand \@ifx [1]{%
 \ifx #1\expandafter \@firstoftwo
 \else \expandafter \@secondoftwo
 \fi
}%
\providecommand \natexlab [1]{#1}%
\providecommand \emph  [1]{``#1''}%
\providecommand \bibnamefont  [1]{#1}%
\providecommand \bibfnamefont [1]{#1}%
\providecommand \citenamefont [1]{#1}%
\providecommand \href@noop [0]{\@secondoftwo}%
\providecommand \href [0]{\begingroup \@sanitize@url \@href}%
\providecommand \@href[1]{\@@startlink{#1}\@@href}%
\providecommand \@@href[1]{\endgroup#1\@@endlink}%
\providecommand \@sanitize@url [0]{\catcode `\\12\catcode `\$12\catcode
  `\&12\catcode `\#12\catcode `\^12\catcode `\_12\catcode `\%12\relax}%
\providecommand \@@startlink[1]{}%
\providecommand \@@endlink[0]{}%
\providecommand \url  [0]{\begingroup\@sanitize@url \@url }%
\providecommand \@url [1]{\endgroup\@href {#1}{\urlprefix }}%
\providecommand \urlprefix  [0]{URL }%
\providecommand \Eprint [0]{\href }%
\providecommand \doibase [0]{http://dx.doi.org/}%
\providecommand \selectlanguage [0]{\@gobble}%
\providecommand \bibinfo  [0]{\@secondoftwo}%
\providecommand \bibfield  [0]{\@secondoftwo}%
\providecommand \translation [1]{[#1]}%
\providecommand \BibitemOpen [0]{}%
\providecommand \bibitemStop [0]{}%
\providecommand \bibitemNoStop [0]{.\EOS\space}%
\providecommand \EOS [0]{\spacefactor3000\relax}%
\providecommand \BibitemShut  [1]{\csname bibitem#1\endcsname}%
\let\auto@bib@innerbib\@empty
\bibitem [{\citenamefont {S{\'a}nchez~Mu{\~n}oz}\ \emph
  {et~al.}(2019)\citenamefont {S{\'a}nchez~Mu{\~n}oz}, \citenamefont
  {Bu{\v{c}}a}, \citenamefont {Tindall}, \citenamefont {Gonz{\'a}lez-Tudela},
  \citenamefont {Jaksch},\ and\ \citenamefont
  {Porras}}]{arXiv_sanchezmunoz19b}%
  \BibitemOpen
  \bibfield  {author} {\bibinfo {author} {\bibfnamefont {C.}~\bibnamefont
  {S{\'a}nchez~Mu{\~n}oz}}, \bibinfo {author} {\bibfnamefont {B.}~\bibnamefont
  {Bu{\v{c}}a}}, \bibinfo {author} {\bibfnamefont {J.}~\bibnamefont {Tindall}},
  \bibinfo {author} {\bibfnamefont {A.}~\bibnamefont {Gonz{\'a}lez-Tudela}},
  \bibinfo {author} {\bibfnamefont {D.}~\bibnamefont {Jaksch}}, \ and\ \bibinfo
  {author} {\bibfnamefont {D.}~\bibnamefont {Porras}},\ }\bibfield  {title}
  {\emph {\bibinfo {title} {Symmetries and Conservation Laws in Quantum
  Trajectories: Dissipative Freezing},}\ }\href@noop {} {\bibfield  {journal}
  {\bibinfo  {journal} {arXiv:1908.11862}\ } (\bibinfo {year}
  {2019})}\BibitemShut {NoStop}%
\bibitem [{\citenamefont {Amo}\ \emph {et~al.}(2009)\citenamefont {Amo},
  \citenamefont {Sanvitto}, \citenamefont {Laussy}, \citenamefont {Ballarini},
  \citenamefont {del Valle}, \citenamefont {Martin}, \citenamefont
  {Lema\^itre}, \citenamefont {Bloch}, \citenamefont {Krizhanovskii},
  \citenamefont {Skolnick}, \citenamefont {Tejedor},\ and\ \citenamefont
  {Vi{\~n}a}}]{amo09a}%
  \BibitemOpen
  \bibfield  {author} {\bibinfo {author} {\bibfnamefont {A.}~\bibnamefont
  {Amo}}, \bibinfo {author} {\bibfnamefont {D.}~\bibnamefont {Sanvitto}},
  \bibinfo {author} {\bibfnamefont {F.~P.}\ \bibnamefont {Laussy}}, \bibinfo
  {author} {\bibfnamefont {D.}~\bibnamefont {Ballarini}}, \bibinfo {author}
  {\bibfnamefont {E.}~\bibnamefont {del Valle}}, \bibinfo {author}
  {\bibfnamefont {M.~D.}\ \bibnamefont {Martin}}, \bibinfo {author}
  {\bibfnamefont {A.}~\bibnamefont {Lema\^itre}}, \bibinfo {author}
  {\bibfnamefont {J.}~\bibnamefont {Bloch}}, \bibinfo {author} {\bibfnamefont
  {D.~N.}\ \bibnamefont {Krizhanovskii}}, \bibinfo {author} {\bibfnamefont
  {M.~S.}\ \bibnamefont {Skolnick}}, \bibinfo {author} {\bibfnamefont
  {C.}~\bibnamefont {Tejedor}}, \ and\ \bibinfo {author} {\bibfnamefont
  {L.}~\bibnamefont {Vi{\~n}a}},\ }\bibfield  {title} {\emph {\bibinfo {title}
  {Collective fluid dynamics of a polariton condensate in a semiconductor
  microcavity},}\ }\href@noop {} {\bibfield  {journal} {\bibinfo  {journal}
  {Nature}\ }\textbf {\bibinfo {volume} {457}},\ \bibinfo {pages} {291}
  (\bibinfo {year} {2009})}\BibitemShut {NoStop}%
\bibitem [{\citenamefont {Rodriguez}\ \emph {et~al.}(2017)\citenamefont
  {Rodriguez}, \citenamefont {Casteels}, \citenamefont {Storme}, \citenamefont
  {Carlon~Zambon}, \citenamefont {Sagnes}, \citenamefont {Le~Gratiet},
  \citenamefont {Galopin}, \citenamefont {Lema\^{\i}tre}, \citenamefont {Amo},
  \citenamefont {Ciuti},\ and\ \citenamefont {Bloch}}]{rodriguez17a}%
  \BibitemOpen
  \bibfield  {author} {\bibinfo {author} {\bibfnamefont {S.~R.~K.}\
  \bibnamefont {Rodriguez}}, \bibinfo {author} {\bibfnamefont {W.}~\bibnamefont
  {Casteels}}, \bibinfo {author} {\bibfnamefont {F.}~\bibnamefont {Storme}},
  \bibinfo {author} {\bibfnamefont {N.}~\bibnamefont {Carlon~Zambon}}, \bibinfo
  {author} {\bibfnamefont {I.}~\bibnamefont {Sagnes}}, \bibinfo {author}
  {\bibfnamefont {L.}~\bibnamefont {Le~Gratiet}}, \bibinfo {author}
  {\bibfnamefont {E.}~\bibnamefont {Galopin}}, \bibinfo {author} {\bibfnamefont
  {A.}~\bibnamefont {Lema\^{\i}tre}}, \bibinfo {author} {\bibfnamefont
  {A.}~\bibnamefont {Amo}}, \bibinfo {author} {\bibfnamefont {C.}~\bibnamefont
  {Ciuti}}, \ and\ \bibinfo {author} {\bibfnamefont {J.}~\bibnamefont
  {Bloch}},\ }\bibfield  {title} {\emph {\bibinfo {title} {Probing a
  Dissipative Phase Transition via Dynamical Optical Hysteresis},}\ }\href@noop
  {} {\bibfield  {journal} {\bibinfo  {journal} {Phys. Rev. Lett.}\ }\textbf
  {\bibinfo {volume} {118}},\ \bibinfo {pages} {247402} (\bibinfo {year}
  {2017})}\BibitemShut {NoStop}%
\bibitem [{\citenamefont {Fink}\ \emph {et~al.}(2018)\citenamefont {Fink},
  \citenamefont {Schade}, \citenamefont {H{\"o}fling}, \citenamefont
  {Schneider},\ and\ \citenamefont {Imamoglu}}]{fink18a}%
  \BibitemOpen
  \bibfield  {author} {\bibinfo {author} {\bibfnamefont {T.}~\bibnamefont
  {Fink}}, \bibinfo {author} {\bibfnamefont {A.}~\bibnamefont {Schade}},
  \bibinfo {author} {\bibfnamefont {S.}~\bibnamefont {H{\"o}fling}}, \bibinfo
  {author} {\bibfnamefont {C.}~\bibnamefont {Schneider}}, \ and\ \bibinfo
  {author} {\bibfnamefont {A.}~\bibnamefont {Imamoglu}},\ }\bibfield  {title}
  {\emph {\bibinfo {title} {Signatures of a dissipative phase transition in
  photon correlation measurements},}\ }\href@noop {} {\bibfield  {journal}
  {\bibinfo  {journal} {Nat. Phys.}\ }\textbf {\bibinfo {volume} {14}},\
  \bibinfo {pages} {365} (\bibinfo {year} {2018})}\BibitemShut {NoStop}%
\bibitem [{\citenamefont {Carr}\ \emph {et~al.}(2013)\citenamefont {Carr},
  \citenamefont {Ritter}, \citenamefont {Wade}, \citenamefont {Adams},\ and\
  \citenamefont {Weatherill}}]{carr13a}%
  \BibitemOpen
  \bibfield  {author} {\bibinfo {author} {\bibfnamefont {C.}~\bibnamefont
  {Carr}}, \bibinfo {author} {\bibfnamefont {R.}~\bibnamefont {Ritter}},
  \bibinfo {author} {\bibfnamefont {C.}~\bibnamefont {Wade}}, \bibinfo {author}
  {\bibfnamefont {C.~S.}\ \bibnamefont {Adams}}, \ and\ \bibinfo {author}
  {\bibfnamefont {K.~J.}\ \bibnamefont {Weatherill}},\ }\bibfield  {title}
  {\emph {\bibinfo {title} {Nonequilibrium phase transition in a dilute Rydberg
  ensemble},}\ }\href@noop {} {\bibfield  {journal} {\bibinfo  {journal} {Phys.
  Rev. Lett.}\ }\textbf {\bibinfo {volume} {111}},\ \bibinfo {pages} {113901}
  (\bibinfo {year} {2013})}\BibitemShut {NoStop}%
\bibitem [{\citenamefont {de~Melo}\ \emph {et~al.}(2016)\citenamefont
  {de~Melo}, \citenamefont {Wade}, \citenamefont {{\v{S}}ibali{\'c}},
  \citenamefont {Kondo}, \citenamefont {Adams},\ and\ \citenamefont
  {Weatherill}}]{melo16a}%
  \BibitemOpen
  \bibfield  {author} {\bibinfo {author} {\bibfnamefont {N.~R.}\ \bibnamefont
  {de~Melo}}, \bibinfo {author} {\bibfnamefont {C.~G.}\ \bibnamefont {Wade}},
  \bibinfo {author} {\bibfnamefont {N.}~\bibnamefont {{\v{S}}ibali{\'c}}},
  \bibinfo {author} {\bibfnamefont {J.~M.}\ \bibnamefont {Kondo}}, \bibinfo
  {author} {\bibfnamefont {C.~S.}\ \bibnamefont {Adams}}, \ and\ \bibinfo
  {author} {\bibfnamefont {K.~J.}\ \bibnamefont {Weatherill}},\ }\bibfield
  {title} {\emph {\bibinfo {title} {Intrinsic optical bistability in a strongly
  driven Rydberg ensemble},}\ }\href@noop {} {\bibfield  {journal} {\bibinfo
  {journal} {Phys. Rev. A}\ }\textbf {\bibinfo {volume} {93}},\ \bibinfo
  {pages} {063863} (\bibinfo {year} {2016})}\BibitemShut {NoStop}%
\bibitem [{\citenamefont {Fitzpatrick}\ \emph {et~al.}(2017)\citenamefont
  {Fitzpatrick}, \citenamefont {Sundaresan}, \citenamefont {Li}, \citenamefont
  {Koch},\ and\ \citenamefont {Houck}}]{fitzpatrick17a}%
  \BibitemOpen
  \bibfield  {author} {\bibinfo {author} {\bibfnamefont {M.}~\bibnamefont
  {Fitzpatrick}}, \bibinfo {author} {\bibfnamefont {N.~M.}\ \bibnamefont
  {Sundaresan}}, \bibinfo {author} {\bibfnamefont {A.~C.~Y.}\ \bibnamefont
  {Li}}, \bibinfo {author} {\bibfnamefont {J.}~\bibnamefont {Koch}}, \ and\
  \bibinfo {author} {\bibfnamefont {A.~A.}\ \bibnamefont {Houck}},\ }\bibfield
  {title} {\emph {\bibinfo {title} {Observation of a Dissipative Phase
  Transition in a One-Dimensional Circuit QED Lattice},}\ }\href@noop {}
  {\bibfield  {journal} {\bibinfo  {journal} {Phys. Rev. X}\ }\textbf {\bibinfo
  {volume} {7}},\ \bibinfo {pages} {011016} (\bibinfo {year}
  {2017})}\BibitemShut {NoStop}%
\bibitem [{\citenamefont {Baumann}\ \emph {et~al.}(2010)\citenamefont
  {Baumann}, \citenamefont {Guerlin}, \citenamefont {Brennecke},\ and\
  \citenamefont {Esslinger}}]{baumann10a}%
  \BibitemOpen
  \bibfield  {author} {\bibinfo {author} {\bibfnamefont {K.}~\bibnamefont
  {Baumann}}, \bibinfo {author} {\bibfnamefont {C.}~\bibnamefont {Guerlin}},
  \bibinfo {author} {\bibfnamefont {F.}~\bibnamefont {Brennecke}}, \ and\
  \bibinfo {author} {\bibfnamefont {T.}~\bibnamefont {Esslinger}},\ }\bibfield
  {title} {\emph {\bibinfo {title} {Dicke quantum phase transition with a
  superfluid gas in an optical cavity},}\ }\href@noop {} {\bibfield  {journal}
  {\bibinfo  {journal} {Nature}\ }\textbf {\bibinfo {volume} {464}},\ \bibinfo
  {pages} {1301} (\bibinfo {year} {2010})}\BibitemShut {NoStop}%
\bibitem [{\citenamefont {Klinder}\ \emph {et~al.}(2015)\citenamefont
  {Klinder}, \citenamefont {Ke{\ss}ler}, \citenamefont {Wolke}, \citenamefont
  {Mathey},\ and\ \citenamefont {Hemmerich}}]{klinder15a}%
  \BibitemOpen
  \bibfield  {author} {\bibinfo {author} {\bibfnamefont {J.}~\bibnamefont
  {Klinder}}, \bibinfo {author} {\bibfnamefont {H.}~\bibnamefont {Ke{\ss}ler}},
  \bibinfo {author} {\bibfnamefont {M.}~\bibnamefont {Wolke}}, \bibinfo
  {author} {\bibfnamefont {L.}~\bibnamefont {Mathey}}, \ and\ \bibinfo {author}
  {\bibfnamefont {A.}~\bibnamefont {Hemmerich}},\ }\bibfield  {title} {\emph
  {\bibinfo {title} {Dynamical phase transition in the open Dicke model},}\
  }\href@noop {} {\bibfield  {journal} {\bibinfo  {journal} {Proc. Natl. Acad.
  Sci.}\ }\textbf {\bibinfo {volume} {112}},\ \bibinfo {pages} {3290} (\bibinfo
  {year} {2015})}\BibitemShut {NoStop}%
\bibitem [{\citenamefont {Hamsen}\ \emph {et~al.}(2018)\citenamefont {Hamsen},
  \citenamefont {Tolazzi}, \citenamefont {Wilk},\ and\ \citenamefont
  {Rempe}}]{hamsen18a}%
  \BibitemOpen
  \bibfield  {author} {\bibinfo {author} {\bibfnamefont {C.}~\bibnamefont
  {Hamsen}}, \bibinfo {author} {\bibfnamefont {K.~N.}\ \bibnamefont {Tolazzi}},
  \bibinfo {author} {\bibfnamefont {T.}~\bibnamefont {Wilk}}, \ and\ \bibinfo
  {author} {\bibfnamefont {G.}~\bibnamefont {Rempe}},\ }\bibfield  {title}
  {\emph {\bibinfo {title} {Strong coupling between photons of two light fields
  mediated by one atom},}\ }\href@noop {} {\bibfield  {journal} {\bibinfo
  {journal} {Nat. Phys.}\ }\textbf {\bibinfo {volume} {14}},\ \bibinfo {pages}
  {885} (\bibinfo {year} {2018})}\BibitemShut {NoStop}%
\bibitem [{\citenamefont {Teufel}\ \emph {et~al.}(2011)\citenamefont {Teufel},
  \citenamefont {Donner}, \citenamefont {Li}, \citenamefont {Harlow},
  \citenamefont {Allman}, \citenamefont {Cicak}, \citenamefont {Sirois},
  \citenamefont {Whittaker}, \citenamefont {Lehnert},\ and\ \citenamefont
  {Simmonds}}]{teufel11a}%
  \BibitemOpen
  \bibfield  {author} {\bibinfo {author} {\bibfnamefont {J.}~\bibnamefont
  {Teufel}}, \bibinfo {author} {\bibfnamefont {T.}~\bibnamefont {Donner}},
  \bibinfo {author} {\bibfnamefont {D.}~\bibnamefont {Li}}, \bibinfo {author}
  {\bibfnamefont {J.}~\bibnamefont {Harlow}}, \bibinfo {author} {\bibfnamefont
  {M.}~\bibnamefont {Allman}}, \bibinfo {author} {\bibfnamefont
  {K.}~\bibnamefont {Cicak}}, \bibinfo {author} {\bibfnamefont
  {A.}~\bibnamefont {Sirois}}, \bibinfo {author} {\bibfnamefont {J.~D.}\
  \bibnamefont {Whittaker}}, \bibinfo {author} {\bibfnamefont {K.}~\bibnamefont
  {Lehnert}}, \ and\ \bibinfo {author} {\bibfnamefont {R.~W.}\ \bibnamefont
  {Simmonds}},\ }\bibfield  {title} {\emph {\bibinfo {title} {Sideband cooling
  of micromechanical motion to the quantum ground state},}\ }\href@noop {}
  {\bibfield  {journal} {\bibinfo  {journal} {Nature}\ }\textbf {\bibinfo
  {volume} {475}},\ \bibinfo {pages} {359} (\bibinfo {year}
  {2011})}\BibitemShut {NoStop}%
\bibitem [{\citenamefont {Kolkowitz}\ \emph {et~al.}(2012)\citenamefont
  {Kolkowitz}, \citenamefont {Jayich}, \citenamefont {Unterreithmeier},
  \citenamefont {Bennett}, \citenamefont {Rabl}, \citenamefont {Harris},\ and\
  \citenamefont {Lukin}}]{kolkowitz12a}%
  \BibitemOpen
  \bibfield  {author} {\bibinfo {author} {\bibfnamefont {S.}~\bibnamefont
  {Kolkowitz}}, \bibinfo {author} {\bibfnamefont {A.~C.~B.}\ \bibnamefont
  {Jayich}}, \bibinfo {author} {\bibfnamefont {Q.~P.}\ \bibnamefont
  {Unterreithmeier}}, \bibinfo {author} {\bibfnamefont {S.~D.}\ \bibnamefont
  {Bennett}}, \bibinfo {author} {\bibfnamefont {P.}~\bibnamefont {Rabl}},
  \bibinfo {author} {\bibfnamefont {J.}~\bibnamefont {Harris}}, \ and\ \bibinfo
  {author} {\bibfnamefont {M.~D.}\ \bibnamefont {Lukin}},\ }\bibfield  {title}
  {\emph {\bibinfo {title} {Coherent sensing of a mechanical resonator with a
  single-spin qubit},}\ }\href@noop {} {\bibfield  {journal} {\bibinfo
  {journal} {Science}\ }\textbf {\bibinfo {volume} {335}},\ \bibinfo {pages}
  {1603} (\bibinfo {year} {2012})}\BibitemShut {NoStop}%
\bibitem [{\citenamefont {Pigeau}\ \emph {et~al.}(2015)\citenamefont {Pigeau},
  \citenamefont {Rohr}, \citenamefont {De~Lépinay}, \citenamefont {Gloppe},
  \citenamefont {Jacques},\ and\ \citenamefont {Arcizet}}]{pigeau15a}%
  \BibitemOpen
  \bibfield  {author} {\bibinfo {author} {\bibfnamefont {B.}~\bibnamefont
  {Pigeau}}, \bibinfo {author} {\bibfnamefont {S.}~\bibnamefont {Rohr}},
  \bibinfo {author} {\bibfnamefont {L.~M.}\ \bibnamefont {De~Lépinay}},
  \bibinfo {author} {\bibfnamefont {A.}~\bibnamefont {Gloppe}}, \bibinfo
  {author} {\bibfnamefont {V.}~\bibnamefont {Jacques}}, \ and\ \bibinfo
  {author} {\bibfnamefont {O.}~\bibnamefont {Arcizet}},\ }\bibfield  {title}
  {\emph {\bibinfo {title} {Observation of a phononic {Mollow} triplet in a
  multimode hybrid spin-nanomechanical system},}\ }\href@noop {} {\bibfield
  {journal} {\bibinfo  {journal} {Nat. Comm.}\ }\textbf {\bibinfo {volume} {6}}
  (\bibinfo {year} {2015})}\BibitemShut {NoStop}%
\bibitem [{\citenamefont {Szymanska}\ \emph {et~al.}(2006)\citenamefont
  {Szymanska}, \citenamefont {Keeling},\ and\ \citenamefont
  {Littlewood}}]{szymanska06a}%
  \BibitemOpen
  \bibfield  {author} {\bibinfo {author} {\bibfnamefont {M.~H.}\ \bibnamefont
  {Szymanska}}, \bibinfo {author} {\bibfnamefont {J.}~\bibnamefont {Keeling}},
  \ and\ \bibinfo {author} {\bibfnamefont {P.~B.}\ \bibnamefont {Littlewood}},\
  }\bibfield  {title} {\emph {\bibinfo {title} {Nonequilibrium Quantum
  Condensation in an Incoherently Pumped Dissipative System},}\ }\href@noop {}
  {\bibfield  {journal} {\bibinfo  {journal} {Phys. Rev. Lett.}\ }\textbf
  {\bibinfo {volume} {96}},\ \bibinfo {pages} {230602} (\bibinfo {year}
  {2006})}\BibitemShut {NoStop}%
\bibitem [{\citenamefont {Roumpos}\ \emph {et~al.}(2012)\citenamefont
  {Roumpos}, \citenamefont {Lohse}, \citenamefont {Nitsche}, \citenamefont
  {Keeling}, \citenamefont {Szyma{\'n}ska}, \citenamefont {Littlewood},
  \citenamefont {L{\"o}ffler}, \citenamefont {H{\"o}fling}, \citenamefont
  {Worschech}, \citenamefont {Forchel},\ and\ \citenamefont
  {Yamamoto}}]{roumpos12a}%
  \BibitemOpen
  \bibfield  {author} {\bibinfo {author} {\bibfnamefont {G.}~\bibnamefont
  {Roumpos}}, \bibinfo {author} {\bibfnamefont {M.}~\bibnamefont {Lohse}},
  \bibinfo {author} {\bibfnamefont {W.~H.}\ \bibnamefont {Nitsche}}, \bibinfo
  {author} {\bibfnamefont {J.}~\bibnamefont {Keeling}}, \bibinfo {author}
  {\bibfnamefont {M.~H.}\ \bibnamefont {Szyma{\'n}ska}}, \bibinfo {author}
  {\bibfnamefont {P.~B.}\ \bibnamefont {Littlewood}}, \bibinfo {author}
  {\bibfnamefont {A.}~\bibnamefont {L{\"o}ffler}}, \bibinfo {author}
  {\bibfnamefont {S.}~\bibnamefont {H{\"o}fling}}, \bibinfo {author}
  {\bibfnamefont {L.}~\bibnamefont {Worschech}}, \bibinfo {author}
  {\bibfnamefont {A.}~\bibnamefont {Forchel}}, \ and\ \bibinfo {author}
  {\bibfnamefont {Y.}~\bibnamefont {Yamamoto}},\ }\bibfield  {title} {\emph
  {\bibinfo {title} {Power-law decay of the spatial correlation function in
  exciton-polariton condensates},}\ }\href@noop {} {\bibfield  {journal}
  {\bibinfo  {journal} {Proc. Natl. Acad. Sci.}\ }\textbf {\bibinfo {volume}
  {109}},\ \bibinfo {pages} {6467} (\bibinfo {year} {2012})}\BibitemShut
  {NoStop}%
\bibitem [{\citenamefont {Chiocchetta}\ and\ \citenamefont
  {Carusotto}(2013)}]{chioccetta13a}%
  \BibitemOpen
  \bibfield  {author} {\bibinfo {author} {\bibfnamefont {A.}~\bibnamefont
  {Chiocchetta}}\ and\ \bibinfo {author} {\bibfnamefont {I.}~\bibnamefont
  {Carusotto}},\ }\bibfield  {title} {\emph {\bibinfo {title} {Non-equilibrium
  quasi-condensates in reduced dimensions},}\ }\href@noop {} {\bibfield
  {journal} {\bibinfo  {journal} {Europhys. Lett.}\ }\textbf {\bibinfo {volume}
  {102}},\ \bibinfo {pages} {67007} (\bibinfo {year} {2013})}\BibitemShut
  {NoStop}%
\bibitem [{\citenamefont {Altman}\ \emph {et~al.}(2015)\citenamefont {Altman},
  \citenamefont {Sieberer}, \citenamefont {Chen}, \citenamefont {Diehl},\ and\
  \citenamefont {Toner}}]{altman15a}%
  \BibitemOpen
  \bibfield  {author} {\bibinfo {author} {\bibfnamefont {E.}~\bibnamefont
  {Altman}}, \bibinfo {author} {\bibfnamefont {L.~M.}\ \bibnamefont
  {Sieberer}}, \bibinfo {author} {\bibfnamefont {L.}~\bibnamefont {Chen}},
  \bibinfo {author} {\bibfnamefont {S.}~\bibnamefont {Diehl}}, \ and\ \bibinfo
  {author} {\bibfnamefont {J.}~\bibnamefont {Toner}},\ }\bibfield  {title}
  {\emph {\bibinfo {title} {Two-Dimensional Superfluidity of Exciton Polaritons
  Requires Strong Anisotropy},}\ }\href@noop {} {\bibfield  {journal} {\bibinfo
   {journal} {Phys. Rev. X}\ }\textbf {\bibinfo {volume} {5}},\ \bibinfo
  {pages} {011017} (\bibinfo {year} {2015})}\BibitemShut {NoStop}%
\bibitem [{\citenamefont {Nitsche}\ \emph {et~al.}(2014)\citenamefont
  {Nitsche}, \citenamefont {Kim}, \citenamefont {Roumpos}, \citenamefont
  {Schneider}, \citenamefont {Kamp}, \citenamefont {H\"ofling}, \citenamefont
  {Forchel},\ and\ \citenamefont {Yamamoto}}]{nitsche14a}%
  \BibitemOpen
  \bibfield  {author} {\bibinfo {author} {\bibfnamefont {W.~H.}\ \bibnamefont
  {Nitsche}}, \bibinfo {author} {\bibfnamefont {N.~Y.}\ \bibnamefont {Kim}},
  \bibinfo {author} {\bibfnamefont {G.}~\bibnamefont {Roumpos}}, \bibinfo
  {author} {\bibfnamefont {C.}~\bibnamefont {Schneider}}, \bibinfo {author}
  {\bibfnamefont {M.}~\bibnamefont {Kamp}}, \bibinfo {author} {\bibfnamefont
  {S.}~\bibnamefont {H\"ofling}}, \bibinfo {author} {\bibfnamefont
  {A.}~\bibnamefont {Forchel}}, \ and\ \bibinfo {author} {\bibfnamefont
  {Y.}~\bibnamefont {Yamamoto}},\ }\bibfield  {title} {\emph {\bibinfo {title}
  {Algebraic order and the Berezinskii-Kosterlitz-Thouless transition in an
  exciton-polariton gas},}\ }\href@noop {} {\bibfield  {journal} {\bibinfo
  {journal} {Phys. Rev. B}\ }\textbf {\bibinfo {volume} {90}},\ \bibinfo
  {pages} {205430} (\bibinfo {year} {2014})}\BibitemShut {NoStop}%
\bibitem [{\citenamefont {Caputo}\ \emph {et~al.}(2018)\citenamefont {Caputo},
  \citenamefont {Ballarini}, \citenamefont {Dagvadorj}, \citenamefont {{
  S{\'a}nchez Mu{\~n}oz}}, \citenamefont {De~Giorgi}, \citenamefont {Dominici},
  \citenamefont {West}, \citenamefont {Pfeiffer}, \citenamefont {Gigli},
  \citenamefont {Laussy} \emph {et~al.}}]{caputo18a}%
  \BibitemOpen
  \bibfield  {author} {\bibinfo {author} {\bibfnamefont {D.}~\bibnamefont
  {Caputo}}, \bibinfo {author} {\bibfnamefont {D.}~\bibnamefont {Ballarini}},
  \bibinfo {author} {\bibfnamefont {G.}~\bibnamefont {Dagvadorj}}, \bibinfo
  {author} {\bibfnamefont {C.}~\bibnamefont {{ S{\'a}nchez Mu{\~n}oz}}},
  \bibinfo {author} {\bibfnamefont {M.}~\bibnamefont {De~Giorgi}}, \bibinfo
  {author} {\bibfnamefont {L.}~\bibnamefont {Dominici}}, \bibinfo {author}
  {\bibfnamefont {K.}~\bibnamefont {West}}, \bibinfo {author} {\bibfnamefont
  {L.~N.}\ \bibnamefont {Pfeiffer}}, \bibinfo {author} {\bibfnamefont
  {G.}~\bibnamefont {Gigli}}, \bibinfo {author} {\bibfnamefont {F.~P.}\
  \bibnamefont {Laussy}},  \emph {et~al.},\ }\bibfield  {title} {\emph
  {\bibinfo {title} {Topological order and thermal equilibrium in polariton
  condensates},}\ }\href@noop {} {\bibfield  {journal} {\bibinfo  {journal}
  {Nat. Mater.}\ }\textbf {\bibinfo {volume} {17}},\ \bibinfo {pages} {145}
  (\bibinfo {year} {2018})}\BibitemShut {NoStop}%
\bibitem [{\citenamefont {Kessler}\ \emph {et~al.}(2012)\citenamefont
  {Kessler}, \citenamefont {Giedke}, \citenamefont {Imamoglu}, \citenamefont
  {Yelin}, \citenamefont {Lukin},\ and\ \citenamefont {Cirac}}]{kessler12a}%
  \BibitemOpen
  \bibfield  {author} {\bibinfo {author} {\bibfnamefont {E.~M.}\ \bibnamefont
  {Kessler}}, \bibinfo {author} {\bibfnamefont {G.}~\bibnamefont {Giedke}},
  \bibinfo {author} {\bibfnamefont {A.}~\bibnamefont {Imamoglu}}, \bibinfo
  {author} {\bibfnamefont {S.~F.}\ \bibnamefont {Yelin}}, \bibinfo {author}
  {\bibfnamefont {M.~D.}\ \bibnamefont {Lukin}}, \ and\ \bibinfo {author}
  {\bibfnamefont {J.~I.}\ \bibnamefont {Cirac}},\ }\bibfield  {title} {\emph
  {\bibinfo {title} {Dissipative phase transition in a central spin system},}\
  }\href@noop {} {\bibfield  {journal} {\bibinfo  {journal} {Phys. Rev. A}\
  }\textbf {\bibinfo {volume} {86}},\ \bibinfo {pages} {012116} (\bibinfo
  {year} {2012})}\BibitemShut {NoStop}%
\bibitem [{\citenamefont {Minganti}\ \emph {et~al.}(2018)\citenamefont
  {Minganti}, \citenamefont {Biella}, \citenamefont {Bartolo},\ and\
  \citenamefont {Ciuti}}]{minganti18a}%
  \BibitemOpen
  \bibfield  {author} {\bibinfo {author} {\bibfnamefont {F.}~\bibnamefont
  {Minganti}}, \bibinfo {author} {\bibfnamefont {A.}~\bibnamefont {Biella}},
  \bibinfo {author} {\bibfnamefont {N.}~\bibnamefont {Bartolo}}, \ and\
  \bibinfo {author} {\bibfnamefont {C.}~\bibnamefont {Ciuti}},\ }\bibfield
  {title} {\emph {\bibinfo {title} {Spectral theory of Liouvillians for
  dissipative phase transitions},}\ }\href@noop {} {\bibfield  {journal}
  {\bibinfo  {journal} {Phys. Rev. A}\ }\textbf {\bibinfo {volume} {98}},\
  \bibinfo {pages} {042118} (\bibinfo {year} {2018})}\BibitemShut {NoStop}%
\bibitem [{\citenamefont {Carmichael}(2015)}]{carmichael15a}%
  \BibitemOpen
  \bibfield  {author} {\bibinfo {author} {\bibfnamefont {H.~J.}\ \bibnamefont
  {Carmichael}},\ }\bibfield  {title} {\emph {\bibinfo {title} {Breakdown of
  Photon Blockade: A Dissipative Quantum Phase Transition in Zero
  Dimensions},}\ }\href@noop {} {\bibfield  {journal} {\bibinfo  {journal}
  {Phys. Rev. X}\ }\textbf {\bibinfo {volume} {5}},\ \bibinfo {pages} {031028}
  (\bibinfo {year} {2015})}\BibitemShut {NoStop}%
\bibitem [{\citenamefont {Weimer}(2015)}]{weimer15a}%
  \BibitemOpen
  \bibfield  {author} {\bibinfo {author} {\bibfnamefont {H.}~\bibnamefont
  {Weimer}},\ }\bibfield  {title} {\emph {\bibinfo {title} {Variational
  Principle for Steady States of Dissipative Quantum Many-Body Systems},}\
  }\href@noop {} {\bibfield  {journal} {\bibinfo  {journal} {Phys. Rev. Lett.}\
  }\textbf {\bibinfo {volume} {114}},\ \bibinfo {pages} {040402} (\bibinfo
  {year} {2015})}\BibitemShut {NoStop}%
\bibitem [{\citenamefont {Benito}\ \emph {et~al.}(2016)\citenamefont {Benito},
  \citenamefont {{S{\'a}nchez~Mu{\~n}oz}},\ and\ \citenamefont
  {Navarrete-Benlloch}}]{benito16a}%
  \BibitemOpen
  \bibfield  {author} {\bibinfo {author} {\bibfnamefont {M.}~\bibnamefont
  {Benito}}, \bibinfo {author} {\bibfnamefont {C.}~\bibnamefont
  {{S{\'a}nchez~Mu{\~n}oz}}}, \ and\ \bibinfo {author} {\bibfnamefont
  {C.}~\bibnamefont {Navarrete-Benlloch}},\ }\bibfield  {title} {\emph
  {\bibinfo {title} {Degenerate parametric oscillation in quantum membrane
  optomechanics},}\ }\href@noop {} {\bibfield  {journal} {\bibinfo  {journal}
  {Phys. Rev. A}\ }\textbf {\bibinfo {volume} {93}},\ \bibinfo {pages} {023846}
  (\bibinfo {year} {2016})}\BibitemShut {NoStop}%
\bibitem [{\citenamefont {Sieberer}\ \emph {et~al.}(2013)\citenamefont
  {Sieberer}, \citenamefont {Huber}, \citenamefont {Altman},\ and\
  \citenamefont {Diehl}}]{sieberer13a}%
  \BibitemOpen
  \bibfield  {author} {\bibinfo {author} {\bibfnamefont {L.~M.}\ \bibnamefont
  {Sieberer}}, \bibinfo {author} {\bibfnamefont {S.~D.}\ \bibnamefont {Huber}},
  \bibinfo {author} {\bibfnamefont {E.}~\bibnamefont {Altman}}, \ and\ \bibinfo
  {author} {\bibfnamefont {S.}~\bibnamefont {Diehl}},\ }\bibfield  {title}
  {\emph {\bibinfo {title} {Dynamical Critical Phenomena in Driven-Dissipative
  Systems},}\ }\href@noop {} {\bibfield  {journal} {\bibinfo  {journal} {Phys.
  Rev. Lett.}\ }\textbf {\bibinfo {volume} {110}},\ \bibinfo {pages} {195301}
  (\bibinfo {year} {2013})}\BibitemShut {NoStop}%
\bibitem [{\citenamefont {S\'anchez Mu\~noz}\ \emph {et~al.}(2018)\citenamefont
  {S\'anchez Mu\~noz}, \citenamefont {Lara}, \citenamefont {Puebla},\ and\
  \citenamefont {Nori}}]{sanchezmunoz18b}%
  \BibitemOpen
  \bibfield  {author} {\bibinfo {author} {\bibfnamefont {C.}~\bibnamefont
  {S\'anchez Mu\~noz}}, \bibinfo {author} {\bibfnamefont {A.}~\bibnamefont
  {Lara}}, \bibinfo {author} {\bibfnamefont {J.}~\bibnamefont {Puebla}}, \ and\
  \bibinfo {author} {\bibfnamefont {F.}~\bibnamefont {Nori}},\ }\bibfield
  {title} {\emph {\bibinfo {title} {Hybrid Systems for the Generation of
  Nonclassical Mechanical States via Quadratic Interactions},}\ }\href@noop {}
  {\bibfield  {journal} {\bibinfo  {journal} {Phys. Rev. Lett.}\ }\textbf
  {\bibinfo {volume} {121}},\ \bibinfo {pages} {123604} (\bibinfo {year}
  {2018})}\BibitemShut {NoStop}%
\bibitem [{\citenamefont {Biondi}\ \emph {et~al.}(2017)\citenamefont {Biondi},
  \citenamefont {Blatter}, \citenamefont {T\"ureci},\ and\ \citenamefont
  {Schmidt}}]{biondi17a}%
  \BibitemOpen
  \bibfield  {author} {\bibinfo {author} {\bibfnamefont {M.}~\bibnamefont
  {Biondi}}, \bibinfo {author} {\bibfnamefont {G.}~\bibnamefont {Blatter}},
  \bibinfo {author} {\bibfnamefont {H.~E.}\ \bibnamefont {T\"ureci}}, \ and\
  \bibinfo {author} {\bibfnamefont {S.}~\bibnamefont {Schmidt}},\ }\bibfield
  {title} {\emph {\bibinfo {title} {Nonequilibrium gas-liquid transition in the
  driven-dissipative photonic lattice},}\ }\href@noop {} {\bibfield  {journal}
  {\bibinfo  {journal} {Phys. Rev. A}\ }\textbf {\bibinfo {volume} {96}},\
  \bibinfo {pages} {043809} (\bibinfo {year} {2017})}\BibitemShut {NoStop}%
\bibitem [{\citenamefont {Hwang}\ \emph {et~al.}(2018)\citenamefont {Hwang},
  \citenamefont {Rabl},\ and\ \citenamefont {Plenio}}]{hwang18a}%
  \BibitemOpen
  \bibfield  {author} {\bibinfo {author} {\bibfnamefont {M.-J.}\ \bibnamefont
  {Hwang}}, \bibinfo {author} {\bibfnamefont {P.}~\bibnamefont {Rabl}}, \ and\
  \bibinfo {author} {\bibfnamefont {M.~B.}\ \bibnamefont {Plenio}},\ }\bibfield
   {title} {\emph {\bibinfo {title} {Dissipative phase transition in the open
  quantum Rabi model},}\ }\href@noop {} {\bibfield  {journal} {\bibinfo
  {journal} {Phys. Rev. A}\ }\textbf {\bibinfo {volume} {97}},\ \bibinfo
  {pages} {013825} (\bibinfo {year} {2018})}\BibitemShut {NoStop}%
\bibitem [{\citenamefont {Mendoza-Arenas}\ \emph {et~al.}(2016)\citenamefont
  {Mendoza-Arenas}, \citenamefont {Clark}, \citenamefont {Felicetti},
  \citenamefont {Romero}, \citenamefont {Solano}, \citenamefont {Angelakis},\
  and\ \citenamefont {Jaksch}}]{mendoza16a}%
  \BibitemOpen
  \bibfield  {author} {\bibinfo {author} {\bibfnamefont {J.~J.}\ \bibnamefont
  {Mendoza-Arenas}}, \bibinfo {author} {\bibfnamefont {S.~R.}\ \bibnamefont
  {Clark}}, \bibinfo {author} {\bibfnamefont {S.}~\bibnamefont {Felicetti}},
  \bibinfo {author} {\bibfnamefont {G.}~\bibnamefont {Romero}}, \bibinfo
  {author} {\bibfnamefont {E.}~\bibnamefont {Solano}}, \bibinfo {author}
  {\bibfnamefont {D.~G.}\ \bibnamefont {Angelakis}}, \ and\ \bibinfo {author}
  {\bibfnamefont {D.}~\bibnamefont {Jaksch}},\ }\bibfield  {title} {\emph
  {\bibinfo {title} {Beyond mean-field bistability in driven-dissipative
  lattices: Bunching-antibunching transition and quantum simulation},}\
  }\href@noop {} {\bibfield  {journal} {\bibinfo  {journal} {Phys. Rev. A}\
  }\textbf {\bibinfo {volume} {93}},\ \bibinfo {pages} {023821} (\bibinfo
  {year} {2016})}\BibitemShut {NoStop}%
\bibitem [{\citenamefont {Letscher}\ \emph {et~al.}(2017)\citenamefont
  {Letscher}, \citenamefont {Thomas}, \citenamefont {Niederpr{\"u}m},
  \citenamefont {Fleischhauer},\ and\ \citenamefont {Ott}}]{letscher17a}%
  \BibitemOpen
  \bibfield  {author} {\bibinfo {author} {\bibfnamefont {F.}~\bibnamefont
  {Letscher}}, \bibinfo {author} {\bibfnamefont {O.}~\bibnamefont {Thomas}},
  \bibinfo {author} {\bibfnamefont {T.}~\bibnamefont {Niederpr{\"u}m}},
  \bibinfo {author} {\bibfnamefont {M.}~\bibnamefont {Fleischhauer}}, \ and\
  \bibinfo {author} {\bibfnamefont {H.}~\bibnamefont {Ott}},\ }\bibfield
  {title} {\emph {\bibinfo {title} {Bistability versus metastability in driven
  dissipative Rydberg gases},}\ }\href@noop {} {\bibfield  {journal} {\bibinfo
  {journal} {Phys. Rev. X}\ }\textbf {\bibinfo {volume} {7}},\ \bibinfo {pages}
  {021020} (\bibinfo {year} {2017})}\BibitemShut {NoStop}%
\bibitem [{\citenamefont {Muppalla}\ \emph {et~al.}(2018)\citenamefont
  {Muppalla}, \citenamefont {Gargiulo}, \citenamefont {Mirzaei}, \citenamefont
  {Venkatesh}, \citenamefont {Juan}, \citenamefont {Gr{\"u}nhaupt},
  \citenamefont {Pop},\ and\ \citenamefont {Kirchmair}}]{muppalla18a}%
  \BibitemOpen
  \bibfield  {author} {\bibinfo {author} {\bibfnamefont {P.}~\bibnamefont
  {Muppalla}}, \bibinfo {author} {\bibfnamefont {O.}~\bibnamefont {Gargiulo}},
  \bibinfo {author} {\bibfnamefont {S.}~\bibnamefont {Mirzaei}}, \bibinfo
  {author} {\bibfnamefont {B.~P.}\ \bibnamefont {Venkatesh}}, \bibinfo {author}
  {\bibfnamefont {M.}~\bibnamefont {Juan}}, \bibinfo {author} {\bibfnamefont
  {L.}~\bibnamefont {Gr{\"u}nhaupt}}, \bibinfo {author} {\bibfnamefont
  {I.}~\bibnamefont {Pop}}, \ and\ \bibinfo {author} {\bibfnamefont
  {G.}~\bibnamefont {Kirchmair}},\ }\bibfield  {title} {\emph {\bibinfo {title}
  {Bistability in a mesoscopic Josephson junction array resonator},}\
  }\href@noop {} {\bibfield  {journal} {\bibinfo  {journal} {Phys. Rev. B}\
  }\textbf {\bibinfo {volume} {97}},\ \bibinfo {pages} {024518} (\bibinfo
  {year} {2018})}\BibitemShut {NoStop}%
\bibitem [{\citenamefont {Hruby}\ \emph {et~al.}(2018)\citenamefont {Hruby},
  \citenamefont {Dogra}, \citenamefont {Landini}, \citenamefont {Donner},\ and\
  \citenamefont {Esslinger}}]{hruby18a}%
  \BibitemOpen
  \bibfield  {author} {\bibinfo {author} {\bibfnamefont {L.}~\bibnamefont
  {Hruby}}, \bibinfo {author} {\bibfnamefont {N.}~\bibnamefont {Dogra}},
  \bibinfo {author} {\bibfnamefont {M.}~\bibnamefont {Landini}}, \bibinfo
  {author} {\bibfnamefont {T.}~\bibnamefont {Donner}}, \ and\ \bibinfo {author}
  {\bibfnamefont {T.}~\bibnamefont {Esslinger}},\ }\bibfield  {title} {\emph
  {\bibinfo {title} {Metastability and avalanche dynamics in strongly
  correlated gases with long-range interactions},}\ }\href@noop {} {\bibfield
  {journal} {\bibinfo  {journal} {Proc. Natl. Acad. Sci.}\ }\textbf {\bibinfo
  {volume} {115}},\ \bibinfo {pages} {3279} (\bibinfo {year}
  {2018})}\BibitemShut {NoStop}%
\bibitem [{\citenamefont {Lee}\ \emph {et~al.}(2012)\citenamefont {Lee},
  \citenamefont {Haeffner},\ and\ \citenamefont {Cross}}]{lee12a}%
  \BibitemOpen
  \bibfield  {author} {\bibinfo {author} {\bibfnamefont {T.~E.}\ \bibnamefont
  {Lee}}, \bibinfo {author} {\bibfnamefont {H.}~\bibnamefont {Haeffner}}, \
  and\ \bibinfo {author} {\bibfnamefont {M.}~\bibnamefont {Cross}},\ }\bibfield
   {title} {\emph {\bibinfo {title} {Collective quantum jumps of Rydberg
  atoms},}\ }\href@noop {} {\bibfield  {journal} {\bibinfo  {journal} {Phys.
  Rev. Lett.}\ }\textbf {\bibinfo {volume} {108}},\ \bibinfo {pages} {023602}
  (\bibinfo {year} {2012})}\BibitemShut {NoStop}%
\bibitem [{\citenamefont {Malossi}\ \emph {et~al.}(2014)\citenamefont
  {Malossi}, \citenamefont {Valado}, \citenamefont {Scotto}, \citenamefont
  {Huillery}, \citenamefont {Pillet}, \citenamefont {Ciampini}, \citenamefont
  {Arimondo},\ and\ \citenamefont {Morsch}}]{malossi14a}%
  \BibitemOpen
  \bibfield  {author} {\bibinfo {author} {\bibfnamefont {N.}~\bibnamefont
  {Malossi}}, \bibinfo {author} {\bibfnamefont {M.}~\bibnamefont {Valado}},
  \bibinfo {author} {\bibfnamefont {S.}~\bibnamefont {Scotto}}, \bibinfo
  {author} {\bibfnamefont {P.}~\bibnamefont {Huillery}}, \bibinfo {author}
  {\bibfnamefont {P.}~\bibnamefont {Pillet}}, \bibinfo {author} {\bibfnamefont
  {D.}~\bibnamefont {Ciampini}}, \bibinfo {author} {\bibfnamefont
  {E.}~\bibnamefont {Arimondo}}, \ and\ \bibinfo {author} {\bibfnamefont
  {O.}~\bibnamefont {Morsch}},\ }\bibfield  {title} {\emph {\bibinfo {title}
  {Full counting statistics and phase diagram of a dissipative Rydberg gas},}\
  }\href@noop {} {\bibfield  {journal} {\bibinfo  {journal} {Phys. Rev. Lett.}\
  }\textbf {\bibinfo {volume} {113}},\ \bibinfo {pages} {023006} (\bibinfo
  {year} {2014})}\BibitemShut {NoStop}%
\bibitem [{\citenamefont {Ates}\ \emph {et~al.}(2012)\citenamefont {Ates},
  \citenamefont {Olmos}, \citenamefont {Garrahan},\ and\ \citenamefont
  {Lesanovsky}}]{ates12a}%
  \BibitemOpen
  \bibfield  {author} {\bibinfo {author} {\bibfnamefont {C.}~\bibnamefont
  {Ates}}, \bibinfo {author} {\bibfnamefont {B.}~\bibnamefont {Olmos}},
  \bibinfo {author} {\bibfnamefont {J.~P.}\ \bibnamefont {Garrahan}}, \ and\
  \bibinfo {author} {\bibfnamefont {I.}~\bibnamefont {Lesanovsky}},\ }\bibfield
   {title} {\emph {\bibinfo {title} {Dynamical phases and intermittency of the
  dissipative quantum Ising model},}\ }\href@noop {} {\bibfield  {journal}
  {\bibinfo  {journal} {Phys. Rev. A}\ }\textbf {\bibinfo {volume} {85}},\
  \bibinfo {pages} {043620} (\bibinfo {year} {2012})}\BibitemShut {NoStop}%
\bibitem [{\citenamefont {Macieszczak}\ \emph
  {et~al.}(2016{\natexlab{a}})\citenamefont {Macieszczak}, \citenamefont
  {Gu{\c{t}}{\u{a}}}, \citenamefont {Lesanovsky},\ and\ \citenamefont
  {Garrahan}}]{macieszczak16a}%
  \BibitemOpen
  \bibfield  {author} {\bibinfo {author} {\bibfnamefont {K.}~\bibnamefont
  {Macieszczak}}, \bibinfo {author} {\bibfnamefont {M.}~\bibnamefont
  {Gu{\c{t}}{\u{a}}}}, \bibinfo {author} {\bibfnamefont {I.}~\bibnamefont
  {Lesanovsky}}, \ and\ \bibinfo {author} {\bibfnamefont {J.~P.}\ \bibnamefont
  {Garrahan}},\ }\bibfield  {title} {\emph {\bibinfo {title} {Towards a theory
  of metastability in open quantum dynamics},}\ }\href@noop {} {\bibfield
  {journal} {\bibinfo  {journal} {Phys. Rev. Lett.}\ }\textbf {\bibinfo
  {volume} {116}},\ \bibinfo {pages} {240404} (\bibinfo {year}
  {2016}{\natexlab{a}})}\BibitemShut {NoStop}%
\bibitem [{\citenamefont {Manzano}\ and\ \citenamefont
  {Hurtado}(2014)}]{manzano14a}%
  \BibitemOpen
  \bibfield  {author} {\bibinfo {author} {\bibfnamefont {D.}~\bibnamefont
  {Manzano}}\ and\ \bibinfo {author} {\bibfnamefont {P.~I.}\ \bibnamefont
  {Hurtado}},\ }\bibfield  {title} {\emph {\bibinfo {title} {Symmetry and the
  thermodynamics of currents in open quantum systems},}\ }\href@noop {}
  {\bibfield  {journal} {\bibinfo  {journal} {Phys. Rev. B}\ }\textbf {\bibinfo
  {volume} {90}},\ \bibinfo {pages} {125138} (\bibinfo {year}
  {2014})}\BibitemShut {NoStop}%
\bibitem [{\citenamefont {Hannukainen}\ and\ \citenamefont
  {Larson}()}]{hannukainen18a}%
  \BibitemOpen
  \bibfield  {author} {\bibinfo {author} {\bibfnamefont {J.}~\bibnamefont
  {Hannukainen}}\ and\ \bibinfo {author} {\bibfnamefont {J.}~\bibnamefont
  {Larson}},\ }\bibfield  {title} {\emph {\bibinfo {title} {Dissipation-driven
  quantum phase transitions and symmetry breaking},}\ }\href@noop {} {\bibinfo
  {journal} {Phys. Rev. A}\ }\BibitemShut {NoStop}%
\bibitem [{\citenamefont {Bu{\v{c}}a}\ and\ \citenamefont
  {Prosen}(2012)}]{buca12a}%
  \BibitemOpen
\bibfield  {journal} {  }\bibfield  {author} {\bibinfo {author} {\bibfnamefont
  {B.}~\bibnamefont {Bu{\v{c}}a}}\ and\ \bibinfo {author} {\bibfnamefont
  {T.}~\bibnamefont {Prosen}},\ }\bibfield  {title} {\emph {\bibinfo {title} {A
  note on symmetry reductions of the Lindblad equation: transport in
  constrained open spin chains},}\ }\href@noop {} {\bibfield  {journal}
  {\bibinfo  {journal} {New J. Phys.}\ }\textbf {\bibinfo {volume} {14}},\
  \bibinfo {pages} {073007} (\bibinfo {year} {2012})}\BibitemShut {NoStop}%
\bibitem [{\citenamefont {Zoller}\ \emph {et~al.}(1987)\citenamefont {Zoller},
  \citenamefont {Marte},\ and\ \citenamefont {Walls}}]{zoller87a}%
  \BibitemOpen
  \bibfield  {author} {\bibinfo {author} {\bibfnamefont {P.}~\bibnamefont
  {Zoller}}, \bibinfo {author} {\bibfnamefont {M.}~\bibnamefont {Marte}}, \
  and\ \bibinfo {author} {\bibfnamefont {D.}~\bibnamefont {Walls}},\ }\bibfield
   {title} {\emph {\bibinfo {title} {Quantum jumps in atomic systems},}\
  }\href@noop {} {\bibfield  {journal} {\bibinfo  {journal} {Phys. Rev. A}\
  }\textbf {\bibinfo {volume} {35}},\ \bibinfo {pages} {198} (\bibinfo {year}
  {1987})}\BibitemShut {NoStop}%
\bibitem [{\citenamefont {Dalibard}\ \emph {et~al.}(1992)\citenamefont
  {Dalibard}, \citenamefont {Castin},\ and\ \citenamefont
  {M{\o}lmer}}]{molmer92a}%
  \BibitemOpen
  \bibfield  {author} {\bibinfo {author} {\bibfnamefont {J.}~\bibnamefont
  {Dalibard}}, \bibinfo {author} {\bibfnamefont {Y.}~\bibnamefont {Castin}}, \
  and\ \bibinfo {author} {\bibfnamefont {K.}~\bibnamefont {M{\o}lmer}},\
  }\bibfield  {title} {\emph {\bibinfo {title} {Wave-function approach to
  dissipative processes in quantum optics},}\ }\href@noop {} {\bibfield
  {journal} {\bibinfo  {journal} {Phys. Rev. Lett.}\ }\textbf {\bibinfo
  {volume} {68}},\ \bibinfo {pages} {580} (\bibinfo {year} {1992})}\BibitemShut
  {NoStop}%
\bibitem [{\citenamefont {Plenio}\ and\ \citenamefont
  {Knight}(1998)}]{plenio98a}%
  \BibitemOpen
  \bibfield  {author} {\bibinfo {author} {\bibfnamefont {M.~B.}\ \bibnamefont
  {Plenio}}\ and\ \bibinfo {author} {\bibfnamefont {P.~L.}\ \bibnamefont
  {Knight}},\ }\bibfield  {title} {\emph {\bibinfo {title} {The quantum-jump
  approach to dissipative dynamics in quantum optics},}\ }\href@noop {}
  {\bibfield  {journal} {\bibinfo  {journal} {Rev. Mod. Phys.}\ }\textbf
  {\bibinfo {volume} {70}},\ \bibinfo {pages} {101} (\bibinfo {year}
  {1998})}\BibitemShut {NoStop}%
\bibitem [{\citenamefont {Bu{\v{c}}a}\ \emph {et~al.}(2019)\citenamefont
  {Bu{\v{c}}a}, \citenamefont {Tindall},\ and\ \citenamefont
  {Jaksch}}]{buca19a}%
  \BibitemOpen
  \bibfield  {author} {\bibinfo {author} {\bibfnamefont {B.}~\bibnamefont
  {Bu{\v{c}}a}}, \bibinfo {author} {\bibfnamefont {J.}~\bibnamefont {Tindall}},
  \ and\ \bibinfo {author} {\bibfnamefont {D.}~\bibnamefont {Jaksch}},\
  }\bibfield  {title} {\emph {\bibinfo {title} {Non-stationary coherent quantum
  many-body dynamics through dissipation},}\ }\href@noop {} {\bibfield
  {journal} {\bibinfo  {journal} {Nat. Comm.}\ }\textbf {\bibinfo {volume}
  {10}},\ \bibinfo {pages} {1730} (\bibinfo {year} {2019})}\BibitemShut
  {NoStop}%
\bibitem [{\citenamefont {Iemini}\ \emph {et~al.}(2018)\citenamefont {Iemini},
  \citenamefont {Russomanno}, \citenamefont {Keeling}, \citenamefont
  {Schir\`o}, \citenamefont {Dalmonte},\ and\ \citenamefont
  {Fazio}}]{iemini18a}%
  \BibitemOpen
  \bibfield  {author} {\bibinfo {author} {\bibfnamefont {F.}~\bibnamefont
  {Iemini}}, \bibinfo {author} {\bibfnamefont {A.}~\bibnamefont {Russomanno}},
  \bibinfo {author} {\bibfnamefont {J.}~\bibnamefont {Keeling}}, \bibinfo
  {author} {\bibfnamefont {M.}~\bibnamefont {Schir\`o}}, \bibinfo {author}
  {\bibfnamefont {M.}~\bibnamefont {Dalmonte}}, \ and\ \bibinfo {author}
  {\bibfnamefont {R.}~\bibnamefont {Fazio}},\ }\bibfield  {title} {\emph
  {\bibinfo {title} {Boundary Time Crystals},}\ }\href@noop {} {\bibfield
  {journal} {\bibinfo  {journal} {Phys. Rev. Lett.}\ }\textbf {\bibinfo
  {volume} {121}},\ \bibinfo {pages} {035301} (\bibinfo {year}
  {2018})}\BibitemShut {NoStop}%
\bibitem [{\citenamefont {Tucker}\ \emph {et~al.}(2018)\citenamefont {Tucker},
  \citenamefont {Zhu}, \citenamefont {Lewis-Swan}, \citenamefont {Marino},
  \citenamefont {Jimenez}, \citenamefont {Restrepo},\ and\ \citenamefont
  {Rey}}]{tucker18a}%
  \BibitemOpen
  \bibfield  {author} {\bibinfo {author} {\bibfnamefont {K.}~\bibnamefont
  {Tucker}}, \bibinfo {author} {\bibfnamefont {B.}~\bibnamefont {Zhu}},
  \bibinfo {author} {\bibfnamefont {R.}~\bibnamefont {Lewis-Swan}}, \bibinfo
  {author} {\bibfnamefont {J.}~\bibnamefont {Marino}}, \bibinfo {author}
  {\bibfnamefont {F.}~\bibnamefont {Jimenez}}, \bibinfo {author} {\bibfnamefont
  {J.}~\bibnamefont {Restrepo}}, \ and\ \bibinfo {author} {\bibfnamefont
  {A.~M.}\ \bibnamefont {Rey}},\ }\bibfield  {title} {\emph {\bibinfo {title}
  {Shattered time: can a dissipative time crystal survive many-body
  correlations?}}\ }\href@noop {} {\bibfield  {journal} {\bibinfo  {journal}
  {New J. Phys.}\ }\textbf {\bibinfo {volume} {20}},\ \bibinfo {pages} {123003}
  (\bibinfo {year} {2018})}\BibitemShut {NoStop}%
\bibitem [{\citenamefont {Jaksch}\ \emph {et~al.}(2001)\citenamefont {Jaksch},
  \citenamefont {Gardiner}, \citenamefont {Schulze}, \citenamefont {Cirac},\
  and\ \citenamefont {Zoller}}]{jaksch01a}%
  \BibitemOpen
  \bibfield  {author} {\bibinfo {author} {\bibfnamefont {D.}~\bibnamefont
  {Jaksch}}, \bibinfo {author} {\bibfnamefont {S.}~\bibnamefont {Gardiner}},
  \bibinfo {author} {\bibfnamefont {K.}~\bibnamefont {Schulze}}, \bibinfo
  {author} {\bibfnamefont {J.}~\bibnamefont {Cirac}}, \ and\ \bibinfo {author}
  {\bibfnamefont {P.}~\bibnamefont {Zoller}},\ }\bibfield  {title} {\emph
  {\bibinfo {title} {Uniting Bose-Einstein condensates in optical
  resonators},}\ }\href@noop {} {\ \textbf {\bibinfo {volume} {86}},\ \bibinfo
  {pages} {4733} (\bibinfo {year} {2001})}\BibitemShut {NoStop}%
\bibitem [{\citenamefont {Micheli}\ \emph {et~al.}(2003)\citenamefont
  {Micheli}, \citenamefont {Jaksch}, \citenamefont {Cirac},\ and\ \citenamefont
  {Zoller}}]{micheli03a}%
  \BibitemOpen
  \bibfield  {author} {\bibinfo {author} {\bibfnamefont {A.}~\bibnamefont
  {Micheli}}, \bibinfo {author} {\bibfnamefont {D.}~\bibnamefont {Jaksch}},
  \bibinfo {author} {\bibfnamefont {J.~I.}\ \bibnamefont {Cirac}}, \ and\
  \bibinfo {author} {\bibfnamefont {P.}~\bibnamefont {Zoller}},\ }\bibfield
  {title} {\emph {\bibinfo {title} {Many-particle entanglement in two-component
  Bose-Einstein condensates},}\ }\href@noop {} {\bibfield  {journal} {\bibinfo
  {journal} {Phys. Rev. A}\ }\textbf {\bibinfo {volume} {67}},\ \bibinfo
  {pages} {013607} (\bibinfo {year} {2003})}\BibitemShut {NoStop}%
\bibitem [{\citenamefont {Dimer}\ \emph {et~al.}(2007)\citenamefont {Dimer},
  \citenamefont {Estienne}, \citenamefont {Parkins},\ and\ \citenamefont
  {Carmichael}}]{dimer07a}%
  \BibitemOpen
  \bibfield  {author} {\bibinfo {author} {\bibfnamefont {F.}~\bibnamefont
  {Dimer}}, \bibinfo {author} {\bibfnamefont {B.}~\bibnamefont {Estienne}},
  \bibinfo {author} {\bibfnamefont {A.~S.}\ \bibnamefont {Parkins}}, \ and\
  \bibinfo {author} {\bibfnamefont {H.~J.}\ \bibnamefont {Carmichael}},\
  }\bibfield  {title} {\emph {\bibinfo {title} {Proposed realization of the
  Dicke-model quantum phase transition in an optical cavity QED system},}\
  }\href@noop {} {\bibfield  {journal} {\bibinfo  {journal} {Phys. Rev. A}\
  }\textbf {\bibinfo {volume} {75}},\ \bibinfo {pages} {013804} (\bibinfo
  {year} {2007})}\BibitemShut {NoStop}%
\bibitem [{\citenamefont {Hepp}\ and\ \citenamefont
  {Lieb}(1973{\natexlab{a}})}]{hepp73a}%
  \BibitemOpen
  \bibfield  {author} {\bibinfo {author} {\bibfnamefont {K.}~\bibnamefont
  {Hepp}}\ and\ \bibinfo {author} {\bibfnamefont {E.~H.}\ \bibnamefont
  {Lieb}},\ }\bibfield  {title} {\emph {\bibinfo {title} {On the superradiant
  phase transition for molecules in a quantized radiation field: the Dicke
  maser model},}\ }\href@noop {} {\bibfield  {journal} {\bibinfo  {journal}
  {Annals of Physics}\ }\textbf {\bibinfo {volume} {76}},\ \bibinfo {pages}
  {360} (\bibinfo {year} {1973}{\natexlab{a}})}\BibitemShut {NoStop}%
\bibitem [{\citenamefont {Hepp}\ and\ \citenamefont
  {Lieb}(1973{\natexlab{b}})}]{hepp73b}%
  \BibitemOpen
  \bibfield  {author} {\bibinfo {author} {\bibfnamefont {K.}~\bibnamefont
  {Hepp}}\ and\ \bibinfo {author} {\bibfnamefont {E.~H.}\ \bibnamefont
  {Lieb}},\ }\bibfield  {title} {\emph {\bibinfo {title} {Equilibrium
  Statistical Mechanics of Matter Interacting with the Quantized Radiation
  Field},}\ }\href@noop {} {\bibfield  {journal} {\bibinfo  {journal} {Phys.
  Rev. A}\ }\textbf {\bibinfo {volume} {8}},\ \bibinfo {pages} {2517} (\bibinfo
  {year} {1973}{\natexlab{b}})}\BibitemShut {NoStop}%
\bibitem [{\citenamefont {Wang}\ and\ \citenamefont {Hioe}(1973)}]{wang73a}%
  \BibitemOpen
  \bibfield  {author} {\bibinfo {author} {\bibfnamefont {Y.~K.}\ \bibnamefont
  {Wang}}\ and\ \bibinfo {author} {\bibfnamefont {F.~T.}\ \bibnamefont
  {Hioe}},\ }\bibfield  {title} {\emph {\bibinfo {title} {Phase Transition in
  the Dicke Model of Superradiance},}\ }\href@noop {} {\bibfield  {journal}
  {\bibinfo  {journal} {Phys. Rev. A}\ }\textbf {\bibinfo {volume} {7}},\
  \bibinfo {pages} {831} (\bibinfo {year} {1973})}\BibitemShut {NoStop}%
\bibitem [{\citenamefont {Emary}\ and\ \citenamefont
  {Brandes}(2003)}]{emary03a}%
  \BibitemOpen
  \bibfield  {author} {\bibinfo {author} {\bibfnamefont {C.}~\bibnamefont
  {Emary}}\ and\ \bibinfo {author} {\bibfnamefont {T.}~\bibnamefont
  {Brandes}},\ }\bibfield  {title} {\emph {\bibinfo {title} {Quantum Chaos
  Triggered by Precursors of a Quantum Phase Transition: The Dicke Model},}\
  }\href@noop {} {\bibfield  {journal} {\bibinfo  {journal} {Phys. Rev. Lett.}\
  }\textbf {\bibinfo {volume} {90}},\ \bibinfo {pages} {044101} (\bibinfo
  {year} {2003})}\BibitemShut {NoStop}%
\bibitem [{\citenamefont {Ke{\ss}ler}\ \emph {et~al.}(2014)\citenamefont
  {Ke{\ss}ler}, \citenamefont {Klinder}, \citenamefont {Wolke},\ and\
  \citenamefont {Hemmerich}}]{kessler14a}%
  \BibitemOpen
  \bibfield  {author} {\bibinfo {author} {\bibfnamefont {H.}~\bibnamefont
  {Ke{\ss}ler}}, \bibinfo {author} {\bibfnamefont {J.}~\bibnamefont {Klinder}},
  \bibinfo {author} {\bibfnamefont {M.}~\bibnamefont {Wolke}}, \ and\ \bibinfo
  {author} {\bibfnamefont {A.}~\bibnamefont {Hemmerich}},\ }\bibfield  {title}
  {\emph {\bibinfo {title} {Steering matter wave superradiance with an
  ultranarrow-band optical cavity},}\ }\href@noop {} {\bibfield  {journal}
  {\bibinfo  {journal} {Phys. Rev. Lett.}\ }\textbf {\bibinfo {volume} {113}},\
  \bibinfo {pages} {070404} (\bibinfo {year} {2014})}\BibitemShut {NoStop}%
\bibitem [{\citenamefont {Kroeze}\ \emph {et~al.}(2018)\citenamefont {Kroeze},
  \citenamefont {Guo}, \citenamefont {Vaidya}, \citenamefont {Keeling},\ and\
  \citenamefont {Lev}}]{kroeze18a}%
  \BibitemOpen
  \bibfield  {author} {\bibinfo {author} {\bibfnamefont {R.~M.}\ \bibnamefont
  {Kroeze}}, \bibinfo {author} {\bibfnamefont {Y.}~\bibnamefont {Guo}},
  \bibinfo {author} {\bibfnamefont {V.~D.}\ \bibnamefont {Vaidya}}, \bibinfo
  {author} {\bibfnamefont {J.}~\bibnamefont {Keeling}}, \ and\ \bibinfo
  {author} {\bibfnamefont {B.~L.}\ \bibnamefont {Lev}},\ }\bibfield  {title}
  {\emph {\bibinfo {title} {Spinor Self-Ordering of a Quantum Gas in a
  Cavity},}\ }\href@noop {} {\bibfield  {journal} {\bibinfo  {journal} {Phys.
  Rev. Lett.}\ }\textbf {\bibinfo {volume} {121}},\ \bibinfo {pages} {163601}
  (\bibinfo {year} {2018})}\BibitemShut {NoStop}%
\bibitem [{\citenamefont {Baden}\ \emph {et~al.}(2014)\citenamefont {Baden},
  \citenamefont {Arnold}, \citenamefont {Grimsmo}, \citenamefont {Parkins},\
  and\ \citenamefont {Barrett}}]{baden14a}%
  \BibitemOpen
  \bibfield  {author} {\bibinfo {author} {\bibfnamefont {M.~P.}\ \bibnamefont
  {Baden}}, \bibinfo {author} {\bibfnamefont {K.~J.}\ \bibnamefont {Arnold}},
  \bibinfo {author} {\bibfnamefont {A.~L.}\ \bibnamefont {Grimsmo}}, \bibinfo
  {author} {\bibfnamefont {S.}~\bibnamefont {Parkins}}, \ and\ \bibinfo
  {author} {\bibfnamefont {M.~D.}\ \bibnamefont {Barrett}},\ }\bibfield
  {title} {\emph {\bibinfo {title} {Realization of the Dicke Model Using
  Cavity-Assisted Raman Transitions},}\ }\href@noop {} {\bibfield  {journal}
  {\bibinfo  {journal} {Phys. Rev. Lett.}\ }\textbf {\bibinfo {volume} {113}},\
  \bibinfo {pages} {020408} (\bibinfo {year} {2014})}\BibitemShut {NoStop}%
\bibitem [{\citenamefont {Zhiqiang}\ \emph {et~al.}(2017)\citenamefont
  {Zhiqiang}, \citenamefont {Lee}, \citenamefont {Kumar}, \citenamefont
  {Arnold}, \citenamefont {Masson}, \citenamefont {Parkins},\ and\
  \citenamefont {Barrett}}]{zhiqiang17a}%
  \BibitemOpen
  \bibfield  {author} {\bibinfo {author} {\bibfnamefont {Z.}~\bibnamefont
  {Zhiqiang}}, \bibinfo {author} {\bibfnamefont {C.~H.}\ \bibnamefont {Lee}},
  \bibinfo {author} {\bibfnamefont {R.}~\bibnamefont {Kumar}}, \bibinfo
  {author} {\bibfnamefont {K.}~\bibnamefont {Arnold}}, \bibinfo {author}
  {\bibfnamefont {S.~J.}\ \bibnamefont {Masson}}, \bibinfo {author}
  {\bibfnamefont {A.}~\bibnamefont {Parkins}}, \ and\ \bibinfo {author}
  {\bibfnamefont {M.}~\bibnamefont {Barrett}},\ }\bibfield  {title} {\emph
  {\bibinfo {title} {Nonequilibrium phase transition in a spin-1 Dicke
  model},}\ }\href@noop {} {\bibfield  {journal} {\bibinfo  {journal} {Optica}\
  }\textbf {\bibinfo {volume} {4}},\ \bibinfo {pages} {424} (\bibinfo {year}
  {2017})}\BibitemShut {NoStop}%
\bibitem [{\citenamefont {Keeling}\ \emph {et~al.}(2010)\citenamefont
  {Keeling}, \citenamefont {Bhaseen},\ and\ \citenamefont
  {Simons}}]{keeling10a}%
  \BibitemOpen
  \bibfield  {author} {\bibinfo {author} {\bibfnamefont {J.}~\bibnamefont
  {Keeling}}, \bibinfo {author} {\bibfnamefont {M.}~\bibnamefont {Bhaseen}}, \
  and\ \bibinfo {author} {\bibfnamefont {B.}~\bibnamefont {Simons}},\
  }\bibfield  {title} {\emph {\bibinfo {title} {Collective dynamics of
  Bose-Einstein condensates in optical cavities},}\ }\href@noop {} {\bibfield
  {journal} {\bibinfo  {journal} {Phys. Rev. Lett.}\ }\textbf {\bibinfo
  {volume} {105}},\ \bibinfo {pages} {043001} (\bibinfo {year}
  {2010})}\BibitemShut {NoStop}%
\bibitem [{\citenamefont {Bhaseen}\ \emph {et~al.}(2012)\citenamefont
  {Bhaseen}, \citenamefont {Mayoh}, \citenamefont {Simons},\ and\ \citenamefont
  {Keeling}}]{basheen12a}%
  \BibitemOpen
  \bibfield  {author} {\bibinfo {author} {\bibfnamefont {M.~J.}\ \bibnamefont
  {Bhaseen}}, \bibinfo {author} {\bibfnamefont {J.}~\bibnamefont {Mayoh}},
  \bibinfo {author} {\bibfnamefont {B.~D.}\ \bibnamefont {Simons}}, \ and\
  \bibinfo {author} {\bibfnamefont {J.}~\bibnamefont {Keeling}},\ }\bibfield
  {title} {\emph {\bibinfo {title} {Dynamics of nonequilibrium Dicke models},}\
  }\href@noop {} {\bibfield  {journal} {\bibinfo  {journal} {Phys. Rev. A}\
  }\textbf {\bibinfo {volume} {85}},\ \bibinfo {pages} {013817} (\bibinfo
  {year} {2012})}\BibitemShut {NoStop}%
\bibitem [{\citenamefont {Dalla~Torre}\ \emph {et~al.}(2013)\citenamefont
  {Dalla~Torre}, \citenamefont {Otterbach}, \citenamefont {Demler},
  \citenamefont {Vuletic},\ and\ \citenamefont {Lukin}}]{dallatorre13a}%
  \BibitemOpen
  \bibfield  {author} {\bibinfo {author} {\bibfnamefont {E.~G.}\ \bibnamefont
  {Dalla~Torre}}, \bibinfo {author} {\bibfnamefont {J.}~\bibnamefont
  {Otterbach}}, \bibinfo {author} {\bibfnamefont {E.}~\bibnamefont {Demler}},
  \bibinfo {author} {\bibfnamefont {V.}~\bibnamefont {Vuletic}}, \ and\
  \bibinfo {author} {\bibfnamefont {M.~D.}\ \bibnamefont {Lukin}},\ }\bibfield
  {title} {\emph {\bibinfo {title} {Dissipative preparation of spin squeezed
  atomic ensembles in a steady state},}\ }\href@noop {} {\bibfield  {journal}
  {\bibinfo  {journal} {Phys. Rev. Lett.}\ }\textbf {\bibinfo {volume} {110}},\
  \bibinfo {pages} {120402} (\bibinfo {year} {2013})}\BibitemShut {NoStop}%
\bibitem [{\citenamefont {Gonzalez-Tudela}\ and\ \citenamefont
  {Porras}(2013)}]{gonzaleztudela13b}%
  \BibitemOpen
  \bibfield  {author} {\bibinfo {author} {\bibfnamefont {A.}~\bibnamefont
  {Gonzalez-Tudela}}\ and\ \bibinfo {author} {\bibfnamefont {D.}~\bibnamefont
  {Porras}},\ }\bibfield  {title} {\emph {\bibinfo {title} {Mesoscopic
  Entanglement Induced by Spontaneous Emission in Solid-State Quantum
  Optics},}\ }\href@noop {} {\bibfield  {journal} {\bibinfo  {journal} {Phys.
  Rev. Lett.}\ }\textbf {\bibinfo {volume} {110}},\ \bibinfo {pages} {080502}
  (\bibinfo {year} {2013})}\BibitemShut {NoStop}%
\bibitem [{\citenamefont {Walls}\ and\ \citenamefont
  {Milburn}(1994)}]{walls_book94a}%
  \BibitemOpen
  \bibfield  {author} {\bibinfo {author} {\bibfnamefont {D.~F.}\ \bibnamefont
  {Walls}}\ and\ \bibinfo {author} {\bibfnamefont {G.~J.}\ \bibnamefont
  {Milburn}},\ }\href@noop {} {\emph {\bibinfo {title} {Quantum Optics}}}\
  (\bibinfo  {publisher} {Springer-Verlag},\ \bibinfo {year}
  {1994})\BibitemShut {NoStop}%
\bibitem [{\citenamefont {Puri}\ and\ \citenamefont {Lawande}(1979)}]{puri79a}%
  \BibitemOpen
  \bibfield  {author} {\bibinfo {author} {\bibfnamefont {R.}~\bibnamefont
  {Puri}}\ and\ \bibinfo {author} {\bibfnamefont {S.}~\bibnamefont {Lawande}},\
  }\bibfield  {title} {\emph {\bibinfo {title} {Exact steady-state density
  operator for a collective atomic system in an external field},}\ }\href@noop
  {} {\bibfield  {journal} {\bibinfo  {journal} {Phys. Lett. A}\ }\textbf
  {\bibinfo {volume} {72}},\ \bibinfo {pages} {200} (\bibinfo {year}
  {1979})}\BibitemShut {NoStop}%
\bibitem [{\citenamefont {Lawande}\ \emph {et~al.}(1981)\citenamefont
  {Lawande}, \citenamefont {Puri},\ and\ \citenamefont {Hassan}}]{lawande81a}%
  \BibitemOpen
  \bibfield  {author} {\bibinfo {author} {\bibfnamefont {S.}~\bibnamefont
  {Lawande}}, \bibinfo {author} {\bibfnamefont {R.}~\bibnamefont {Puri}}, \
  and\ \bibinfo {author} {\bibfnamefont {S.}~\bibnamefont {Hassan}},\
  }\bibfield  {title} {\emph {\bibinfo {title} {Non-resonant effects in the
  fluorescent Dicke model. I. Exact steady state analysis},}\ }\href@noop {}
  {\bibfield  {journal} {\bibinfo  {journal} {J. Phys. B.: At. Mol. Phys.}\
  }\textbf {\bibinfo {volume} {14}},\ \bibinfo {pages} {4171} (\bibinfo {year}
  {1981})}\BibitemShut {NoStop}%
\bibitem [{\citenamefont {Agarwal}(1981)}]{agarwal81a}%
  \BibitemOpen
  \bibfield  {author} {\bibinfo {author} {\bibfnamefont {G.~S.}\ \bibnamefont
  {Agarwal}},\ }\bibfield  {title} {\emph {\bibinfo {title} {Relation between
  atomic coherent-state representation, state multipoles, and generalized
  phase-space distributions},}\ }\href@noop {} {\bibfield  {journal} {\bibinfo
  {journal} {Phys. Rev. A}\ }\textbf {\bibinfo {volume} {24}},\ \bibinfo
  {pages} {2889} (\bibinfo {year} {1981})}\BibitemShut {NoStop}%
\bibitem [{\citenamefont {Dowling}\ \emph {et~al.}(1994)\citenamefont
  {Dowling}, \citenamefont {Agarwal},\ and\ \citenamefont
  {Schleich}}]{dowling94a}%
  \BibitemOpen
  \bibfield  {author} {\bibinfo {author} {\bibfnamefont {J.~P.}\ \bibnamefont
  {Dowling}}, \bibinfo {author} {\bibfnamefont {G.~S.}\ \bibnamefont
  {Agarwal}}, \ and\ \bibinfo {author} {\bibfnamefont {W.~P.}\ \bibnamefont
  {Schleich}},\ }\bibfield  {title} {\emph {\bibinfo {title} {Wigner
  distribution of a general angular-momentum state: Applications to a
  collection of two-level atoms},}\ }\href@noop {} {\bibfield  {journal}
  {\bibinfo  {journal} {Phys. Rev. A}\ }\textbf {\bibinfo {volume} {49}},\
  \bibinfo {pages} {4101} (\bibinfo {year} {1994})}\BibitemShut {NoStop}%
\bibitem [{\citenamefont {Pezz{\`e}}\ \emph {et~al.}(2018)\citenamefont
  {Pezz{\`e}}, \citenamefont {Smerzi}, \citenamefont {Oberthaler},
  \citenamefont {Schmied},\ and\ \citenamefont {Treutlein}}]{pezze18a}%
  \BibitemOpen
  \bibfield  {author} {\bibinfo {author} {\bibfnamefont {L.}~\bibnamefont
  {Pezz{\`e}}}, \bibinfo {author} {\bibfnamefont {A.}~\bibnamefont {Smerzi}},
  \bibinfo {author} {\bibfnamefont {M.~K.}\ \bibnamefont {Oberthaler}},
  \bibinfo {author} {\bibfnamefont {R.}~\bibnamefont {Schmied}}, \ and\
  \bibinfo {author} {\bibfnamefont {P.}~\bibnamefont {Treutlein}},\ }\bibfield
  {title} {\emph {\bibinfo {title} {Quantum metrology with nonclassical states
  of atomic ensembles},}\ }\href@noop {} {\bibfield  {journal} {\bibinfo
  {journal} {Rev. Mod. Phys.}\ }\textbf {\bibinfo {volume} {90}},\ \bibinfo
  {pages} {035005} (\bibinfo {year} {2018})}\BibitemShut {NoStop}%
\bibitem [{\citenamefont {Gross}\ \emph {et~al.}(2010)\citenamefont {Gross},
  \citenamefont {Zibold}, \citenamefont {Nicklas}, \citenamefont {Esteve},\
  and\ \citenamefont {Oberthaler}}]{gross10a}%
  \BibitemOpen
  \bibfield  {author} {\bibinfo {author} {\bibfnamefont {C.}~\bibnamefont
  {Gross}}, \bibinfo {author} {\bibfnamefont {T.}~\bibnamefont {Zibold}},
  \bibinfo {author} {\bibfnamefont {E.}~\bibnamefont {Nicklas}}, \bibinfo
  {author} {\bibfnamefont {J.}~\bibnamefont {Esteve}}, \ and\ \bibinfo {author}
  {\bibfnamefont {M.~K.}\ \bibnamefont {Oberthaler}},\ }\bibfield  {title}
  {\emph {\bibinfo {title} {Nonlinear atom interferometer surpasses classical
  precision limit},}\ }\href@noop {} {\bibfield  {journal} {\bibinfo  {journal}
  {Nature}\ }\textbf {\bibinfo {volume} {464}},\ \bibinfo {pages} {1165}
  (\bibinfo {year} {2010})}\BibitemShut {NoStop}%
\bibitem [{\citenamefont {Berrada}\ \emph {et~al.}(2013)\citenamefont
  {Berrada}, \citenamefont {van Frank}, \citenamefont {B{\"u}cker},
  \citenamefont {Schumm}, \citenamefont {Schaff},\ and\ \citenamefont
  {Schmiedmayer}}]{berrada13a}%
  \BibitemOpen
  \bibfield  {author} {\bibinfo {author} {\bibfnamefont {T.}~\bibnamefont
  {Berrada}}, \bibinfo {author} {\bibfnamefont {S.}~\bibnamefont {van Frank}},
  \bibinfo {author} {\bibfnamefont {R.}~\bibnamefont {B{\"u}cker}}, \bibinfo
  {author} {\bibfnamefont {T.}~\bibnamefont {Schumm}}, \bibinfo {author}
  {\bibfnamefont {J.-F.}\ \bibnamefont {Schaff}}, \ and\ \bibinfo {author}
  {\bibfnamefont {J.}~\bibnamefont {Schmiedmayer}},\ }\bibfield  {title} {\emph
  {\bibinfo {title} {Integrated mach--zehnder interferometer for bose--einstein
  condensates},}\ }\href@noop {} {\ \textbf {\bibinfo {volume} {4}},\ \bibinfo
  {pages} {2077} (\bibinfo {year} {2013})}\BibitemShut {NoStop}%
\bibitem [{\citenamefont {Borregaard}\ and\ \citenamefont
  {S{\o}rensen}(2013)}]{borregaard13a}%
  \BibitemOpen
  \bibfield  {author} {\bibinfo {author} {\bibfnamefont {J.}~\bibnamefont
  {Borregaard}}\ and\ \bibinfo {author} {\bibfnamefont {A.~S.}\ \bibnamefont
  {S{\o}rensen}},\ }\bibfield  {title} {\emph {\bibinfo {title}
  {Near-Heisenberg-limited atomic clocks in the presence of decoherence},}\
  }\href@noop {} {\bibfield  {journal} {\bibinfo  {journal} {Phys. Rev. Lett.}\
  }\textbf {\bibinfo {volume} {111}},\ \bibinfo {pages} {090801} (\bibinfo
  {year} {2013})}\BibitemShut {NoStop}%
\bibitem [{\citenamefont {Wineland}\ \emph {et~al.}(1994)\citenamefont
  {Wineland}, \citenamefont {Bollinger},\ and\ \citenamefont
  {M.}}]{wineland94a}%
  \BibitemOpen
  \bibfield  {author} {\bibinfo {author} {\bibfnamefont {D.~J.}\ \bibnamefont
  {Wineland}}, \bibinfo {author} {\bibfnamefont {J.~J.}\ \bibnamefont
  {Bollinger}}, \ and\ \bibinfo {author} {\bibfnamefont {I.~W.}\ \bibnamefont
  {M.}},\ }\bibfield  {title} {\emph {\bibinfo {title} {{Squeezed atomic states
  and projection noise in spectroscopy}},}\ }\href@noop {} {\bibfield
  {journal} {\bibinfo  {journal} {Phys. Rev. A}\ }\textbf {\bibinfo {volume}
  {50}},\ \bibinfo {pages} {67} (\bibinfo {year} {1994})}\BibitemShut {NoStop}%
\bibitem [{\citenamefont {Ma}\ \emph {et~al.}(2011)\citenamefont {Ma},
  \citenamefont {Wang}, \citenamefont {Sun},\ and\ \citenamefont
  {Nori}}]{ma11a}%
  \BibitemOpen
  \bibfield  {author} {\bibinfo {author} {\bibfnamefont {J.}~\bibnamefont
  {Ma}}, \bibinfo {author} {\bibfnamefont {X.}~\bibnamefont {Wang}}, \bibinfo
  {author} {\bibfnamefont {C.-P.}\ \bibnamefont {Sun}}, \ and\ \bibinfo
  {author} {\bibfnamefont {F.}~\bibnamefont {Nori}},\ }\bibfield  {title}
  {\emph {\bibinfo {title} {Quantum spin squeezing},}\ }\href@noop {}
  {\bibfield  {journal} {\bibinfo  {journal} {Phys. Rep.}\ }\textbf {\bibinfo
  {volume} {509}},\ \bibinfo {pages} {89} (\bibinfo {year} {2011})}\BibitemShut
  {NoStop}%
\bibitem [{\citenamefont {S{\o}rensen}\ \emph {et~al.}()\citenamefont
  {S{\o}rensen}, \citenamefont {Duan}, \citenamefont {Cirac},\ and\
  \citenamefont {Zoller}}]{sorensen01a}%
  \BibitemOpen
  \bibfield  {author} {\bibinfo {author} {\bibfnamefont {A.}~\bibnamefont
  {S{\o}rensen}}, \bibinfo {author} {\bibfnamefont {L.-M.}\ \bibnamefont
  {Duan}}, \bibinfo {author} {\bibfnamefont {J.}~\bibnamefont {Cirac}}, \ and\
  \bibinfo {author} {\bibfnamefont {P.}~\bibnamefont {Zoller}},\ }\bibfield
  {title} {\emph {\bibinfo {title} {Many-particle entanglement with
  Bose--Einstein condensates},}\ }\href@noop {} {\bibfield  {journal} {\bibinfo
   {journal} {Nature}\ }\textbf {\bibinfo {volume} {409}}}\BibitemShut
  {NoStop}%
\bibitem [{\citenamefont {Ribeiro}\ and\ \citenamefont
  {Prosen}(2019)}]{ribeiro19a}%
  \BibitemOpen
  \bibfield  {author} {\bibinfo {author} {\bibfnamefont {P.}~\bibnamefont
  {Ribeiro}}\ and\ \bibinfo {author} {\bibfnamefont {T.~c.~v.}\ \bibnamefont
  {Prosen}},\ }\bibfield  {title} {\emph {\bibinfo {title} {Integrable Quantum
  Dynamics of Open Collective Spin Models},}\ }\href@noop {} {\bibfield
  {journal} {\bibinfo  {journal} {Phys. Rev. Lett.}\ }\textbf {\bibinfo
  {volume} {122}},\ \bibinfo {pages} {010401} (\bibinfo {year}
  {2019})}\BibitemShut {NoStop}%
\bibitem [{\citenamefont {Deutsch}(1991)}]{deutsch91a}%
  \BibitemOpen
  \bibfield  {author} {\bibinfo {author} {\bibfnamefont {J.~M.}\ \bibnamefont
  {Deutsch}},\ }\bibfield  {title} {\emph {\bibinfo {title} {Quantum
  statistical mechanics in a closed system},}\ }\href@noop {} {\bibfield
  {journal} {\bibinfo  {journal} {Phys. Rev. A}\ }\textbf {\bibinfo {volume}
  {43}},\ \bibinfo {pages} {2046} (\bibinfo {year} {1991})}\BibitemShut
  {NoStop}%
\bibitem [{\citenamefont {Srednicki}(1994)}]{srednicki94a}%
  \BibitemOpen
  \bibfield  {author} {\bibinfo {author} {\bibfnamefont {M.}~\bibnamefont
  {Srednicki}},\ }\bibfield  {title} {\emph {\bibinfo {title} {Chaos and
  quantum thermalization},}\ }\href@noop {} {\bibfield  {journal} {\bibinfo
  {journal} {Phys. Rev. E}\ }\textbf {\bibinfo {volume} {50}},\ \bibinfo
  {pages} {888} (\bibinfo {year} {1994})}\BibitemShut {NoStop}%
\bibitem [{\citenamefont {Rigol}\ \emph {et~al.}(2008)\citenamefont {Rigol},
  \citenamefont {Dunjko},\ and\ \citenamefont {Olshanii}}]{rigol08a}%
  \BibitemOpen
  \bibfield  {author} {\bibinfo {author} {\bibfnamefont {M.}~\bibnamefont
  {Rigol}}, \bibinfo {author} {\bibfnamefont {V.}~\bibnamefont {Dunjko}}, \
  and\ \bibinfo {author} {\bibfnamefont {M.}~\bibnamefont {Olshanii}},\
  }\bibfield  {title} {\emph {\bibinfo {title} {Thermalization and its
  mechanism for generic isolated quantum systems},}\ }\href@noop {} {\bibfield
  {journal} {\bibinfo  {journal} {Nature}\ }\textbf {\bibinfo {volume} {452}},\
  \bibinfo {pages} {854} (\bibinfo {year} {2008})}\BibitemShut {NoStop}%
\bibitem [{\citenamefont {Garrahan}\ and\ \citenamefont
  {Lesanovsky}(2010)}]{garrahan10a}%
  \BibitemOpen
  \bibfield  {author} {\bibinfo {author} {\bibfnamefont {J.~P.}\ \bibnamefont
  {Garrahan}}\ and\ \bibinfo {author} {\bibfnamefont {I.}~\bibnamefont
  {Lesanovsky}},\ }\bibfield  {title} {\emph {\bibinfo {title} {Thermodynamics
  of Quantum Jump Trajectories},}\ }\href@noop {} {\bibfield  {journal}
  {\bibinfo  {journal} {Phys. Rev. Lett.}\ }\textbf {\bibinfo {volume} {104}},\
  \bibinfo {pages} {160601} (\bibinfo {year} {2010})}\BibitemShut {NoStop}%
\bibitem [{\citenamefont {Garrahan}\ \emph {et~al.}(2007)\citenamefont
  {Garrahan}, \citenamefont {Jack}, \citenamefont {Lecomte}, \citenamefont
  {Pitard}, \citenamefont {van Duijvendijk},\ and\ \citenamefont {van
  Wijland}}]{garrahan07a}%
  \BibitemOpen
  \bibfield  {author} {\bibinfo {author} {\bibfnamefont {J.~P.}\ \bibnamefont
  {Garrahan}}, \bibinfo {author} {\bibfnamefont {R.~L.}\ \bibnamefont {Jack}},
  \bibinfo {author} {\bibfnamefont {V.}~\bibnamefont {Lecomte}}, \bibinfo
  {author} {\bibfnamefont {E.}~\bibnamefont {Pitard}}, \bibinfo {author}
  {\bibfnamefont {K.}~\bibnamefont {van Duijvendijk}}, \ and\ \bibinfo {author}
  {\bibfnamefont {F.}~\bibnamefont {van Wijland}},\ }\bibfield  {title} {\emph
  {\bibinfo {title} {Dynamical first-order phase transition in kinetically
  constrained models of glasses},}\ }\href@noop {} {\bibfield  {journal}
  {\bibinfo  {journal} {Phys. Rev. Lett.}\ }\textbf {\bibinfo {volume} {98}},\
  \bibinfo {pages} {195702} (\bibinfo {year} {2007})}\BibitemShut {NoStop}%
\bibitem [{\citenamefont {Flindt}\ and\ \citenamefont
  {Garrahan}(2013)}]{flindt13a}%
  \BibitemOpen
  \bibfield  {author} {\bibinfo {author} {\bibfnamefont {C.}~\bibnamefont
  {Flindt}}\ and\ \bibinfo {author} {\bibfnamefont {J.~P.}\ \bibnamefont
  {Garrahan}},\ }\bibfield  {title} {\emph {\bibinfo {title} {Trajectory phase
  transitions, Lee-Yang zeros, and high-order cumulants in full counting
  statistics},}\ }\href@noop {} {\bibfield  {journal} {\bibinfo  {journal}
  {Phys. Rev. Lett.}\ }\textbf {\bibinfo {volume} {110}},\ \bibinfo {pages}
  {050601} (\bibinfo {year} {2013})}\BibitemShut {NoStop}%
\bibitem [{\citenamefont {Hickey}\ \emph {et~al.}(2014)\citenamefont {Hickey},
  \citenamefont {Flindt},\ and\ \citenamefont {Garrahan}}]{hickey14a}%
  \BibitemOpen
  \bibfield  {author} {\bibinfo {author} {\bibfnamefont {J.~M.}\ \bibnamefont
  {Hickey}}, \bibinfo {author} {\bibfnamefont {C.}~\bibnamefont {Flindt}}, \
  and\ \bibinfo {author} {\bibfnamefont {J.~P.}\ \bibnamefont {Garrahan}},\
  }\bibfield  {title} {\emph {\bibinfo {title} {Intermittency and dynamical
  Lee-Yang zeros of open quantum systems},}\ }\href@noop {} {\bibfield
  {journal} {\bibinfo  {journal} {Phys. Rev. E}\ }\textbf {\bibinfo {volume}
  {90}},\ \bibinfo {pages} {062128} (\bibinfo {year} {2014})}\BibitemShut
  {NoStop}%
\bibitem [{\citenamefont {Carollo}\ \emph {et~al.}(2018)\citenamefont
  {Carollo}, \citenamefont {Garrahan}, \citenamefont {Lesanovsky},\ and\
  \citenamefont {P\'erez-Espigares}}]{carollo18a}%
  \BibitemOpen
  \bibfield  {author} {\bibinfo {author} {\bibfnamefont {F.}~\bibnamefont
  {Carollo}}, \bibinfo {author} {\bibfnamefont {J.~P.}\ \bibnamefont
  {Garrahan}}, \bibinfo {author} {\bibfnamefont {I.}~\bibnamefont
  {Lesanovsky}}, \ and\ \bibinfo {author} {\bibfnamefont {C.}~\bibnamefont
  {P\'erez-Espigares}},\ }\bibfield  {title} {\emph {\bibinfo {title} {Making
  rare events typical in Markovian open quantum systems},}\ }\href@noop {}
  {\bibfield  {journal} {\bibinfo  {journal} {Phys. Rev. A}\ }\textbf {\bibinfo
  {volume} {98}},\ \bibinfo {pages} {010103} (\bibinfo {year}
  {2018})}\BibitemShut {NoStop}%
\bibitem [{\citenamefont {Garrahan}\ and\ \citenamefont
  {Gu{\c{t}}{\u{a}}}(2018)}]{garrahan18a}%
  \BibitemOpen
  \bibfield  {author} {\bibinfo {author} {\bibfnamefont {J.~P.}\ \bibnamefont
  {Garrahan}}\ and\ \bibinfo {author} {\bibfnamefont {M.}~\bibnamefont
  {Gu{\c{t}}{\u{a}}}},\ }\bibfield  {title} {\emph {\bibinfo {title} {Catching
  and reversing quantum jumps and thermodynamics of quantum trajectories},}\
  }\href@noop {} {\bibfield  {journal} {\bibinfo  {journal} {Phys. Rev. A}\
  }\textbf {\bibinfo {volume} {98}},\ \bibinfo {pages} {052137} (\bibinfo
  {year} {2018})}\BibitemShut {NoStop}%
\bibitem [{Note1()}]{Note1}%
  \BibitemOpen
  \bibinfo {note} {These unravelings are not uniquely defined, since different
  jump operators can be chosen that yield the same master equation (by changing
  the Hamiltonian accordingly). These different unravelings would correspond to
  different detection schemes, such as photon counting or homodyne detection,
  that differ on the way the system is monitored~\cite
  {bartolo17a}.}\BibitemShut {Stop}%
\bibitem [{\citenamefont {Griessner}\ \emph {et~al.}(2006)\citenamefont
  {Griessner}, \citenamefont {Daley}, \citenamefont {Clark}, \citenamefont
  {Jaksch},\ and\ \citenamefont {Zoller}}]{griessner06a}%
  \BibitemOpen
  \bibfield  {author} {\bibinfo {author} {\bibfnamefont {A.}~\bibnamefont
  {Griessner}}, \bibinfo {author} {\bibfnamefont {A.}~\bibnamefont {Daley}},
  \bibinfo {author} {\bibfnamefont {S.}~\bibnamefont {Clark}}, \bibinfo
  {author} {\bibfnamefont {D.}~\bibnamefont {Jaksch}}, \ and\ \bibinfo {author}
  {\bibfnamefont {P.}~\bibnamefont {Zoller}},\ }\bibfield  {title} {\emph
  {\bibinfo {title} {Dark-state cooling of atoms by superfluid immersion},}\
  }\href@noop {} {\bibfield  {journal} {\bibinfo  {journal} {Phys. Rev. Lett.}\
  }\textbf {\bibinfo {volume} {97}},\ \bibinfo {pages} {220403} (\bibinfo
  {year} {2006})}\BibitemShut {NoStop}%
\bibitem [{\citenamefont {Aspect}\ \emph {et~al.}(1988)\citenamefont {Aspect},
  \citenamefont {Arimondo}, \citenamefont {Kaiser}, \citenamefont
  {Vansteenkiste},\ and\ \citenamefont {Cohen-Tannoudji}}]{aspect88a}%
  \BibitemOpen
  \bibfield  {author} {\bibinfo {author} {\bibfnamefont {A.}~\bibnamefont
  {Aspect}}, \bibinfo {author} {\bibfnamefont {E.}~\bibnamefont {Arimondo}},
  \bibinfo {author} {\bibfnamefont {R.~e.~a.}\ \bibnamefont {Kaiser}}, \bibinfo
  {author} {\bibfnamefont {N.}~\bibnamefont {Vansteenkiste}}, \ and\ \bibinfo
  {author} {\bibfnamefont {C.}~\bibnamefont {Cohen-Tannoudji}},\ }\bibfield
  {title} {\emph {\bibinfo {title} {Laser cooling below the one-photon recoil
  energy by velocity-selective coherent population trapping},}\ }\href@noop {}
  {\bibfield  {journal} {\bibinfo  {journal} {Phys. Rev. Lett.}\ }\textbf
  {\bibinfo {volume} {61}},\ \bibinfo {pages} {826} (\bibinfo {year}
  {1988})}\BibitemShut {NoStop}%
\bibitem [{\citenamefont {Touchette}(2009)}]{touchette09a}%
  \BibitemOpen
  \bibfield  {author} {\bibinfo {author} {\bibfnamefont {H.}~\bibnamefont
  {Touchette}},\ }\bibfield  {title} {\emph {\bibinfo {title} {The large
  deviation approach to statistical mechanics},}\ }\href@noop {} {\bibfield
  {journal} {\bibinfo  {journal} {Phys. Rep.}\ }\textbf {\bibinfo {volume}
  {478}},\ \bibinfo {pages} {1} (\bibinfo {year} {2009})}\BibitemShut {NoStop}%
\bibitem [{\citenamefont {Macieszczak}\ \emph
  {et~al.}(2016{\natexlab{b}})\citenamefont {Macieszczak}, \citenamefont
  {Gu{\c{t}}{\u{a}}}, \citenamefont {Lesanovsky},\ and\ \citenamefont
  {Garrahan}}]{macieszczak16b}%
  \BibitemOpen
  \bibfield  {author} {\bibinfo {author} {\bibfnamefont {K.}~\bibnamefont
  {Macieszczak}}, \bibinfo {author} {\bibfnamefont {M.}~\bibnamefont
  {Gu{\c{t}}{\u{a}}}}, \bibinfo {author} {\bibfnamefont {I.}~\bibnamefont
  {Lesanovsky}}, \ and\ \bibinfo {author} {\bibfnamefont {J.~P.}\ \bibnamefont
  {Garrahan}},\ }\bibfield  {title} {\emph {\bibinfo {title} {Dynamical phase
  transitions as a resource for quantum enhanced metrology},}\ }\href@noop {}
  {\bibfield  {journal} {\bibinfo  {journal} {Phys. Rev. A}\ }\textbf {\bibinfo
  {volume} {93}},\ \bibinfo {pages} {022103} (\bibinfo {year}
  {2016}{\natexlab{b}})}\BibitemShut {NoStop}%
\bibitem [{\citenamefont {Gardiner}\ and\ \citenamefont
  {Zoller}(2000)}]{gardiner_book00a}%
  \BibitemOpen
  \bibfield  {author} {\bibinfo {author} {\bibfnamefont {G.~W.}\ \bibnamefont
  {Gardiner}}\ and\ \bibinfo {author} {\bibfnamefont {P.}~\bibnamefont
  {Zoller}},\ }\href@noop {} {\emph {\bibinfo {title} {Quantum Noise}}},\
  \bibinfo {edition} {2nd}\ ed.\ (\bibinfo  {publisher} {Springer-Verlag,
  Berlin},\ \bibinfo {year} {2000})\BibitemShut {NoStop}%
\bibitem [{\citenamefont {del Valle}\ \emph {et~al.}(2009)\citenamefont {del
  Valle}, \citenamefont {Laussy},\ and\ \citenamefont {Tejedor}}]{delvalle09b}%
  \BibitemOpen
  \bibfield  {author} {\bibinfo {author} {\bibfnamefont {E.}~\bibnamefont {del
  Valle}}, \bibinfo {author} {\bibfnamefont {F.~P.}\ \bibnamefont {Laussy}}, \
  and\ \bibinfo {author} {\bibfnamefont {C.}~\bibnamefont {Tejedor}},\
  }\bibfield  {title} {\emph {\bibinfo {title} {Quantum regression formula and
  luminescence spectra of two coupled modes under incoherent continuous
  pumping},}\ }\href@noop {} {\bibfield  {journal} {\bibinfo  {journal} {AIP
  Conf. Proc.}\ }\textbf {\bibinfo {volume} {1147}},\ \bibinfo {pages} {238}
  (\bibinfo {year} {2009})}\BibitemShut {NoStop}%
\bibitem [{\citenamefont {del Valle}\ \emph {et~al.}(2012)\citenamefont {del
  Valle}, \citenamefont {Gonzalez-Tudela}, \citenamefont {Laussy},
  \citenamefont {Tejedor},\ and\ \citenamefont {Hartmann}}]{delvalle12a}%
  \BibitemOpen
  \bibfield  {author} {\bibinfo {author} {\bibfnamefont {E.}~\bibnamefont {del
  Valle}}, \bibinfo {author} {\bibfnamefont {A.}~\bibnamefont
  {Gonzalez-Tudela}}, \bibinfo {author} {\bibfnamefont {F.~P.}\ \bibnamefont
  {Laussy}}, \bibinfo {author} {\bibfnamefont {C.}~\bibnamefont {Tejedor}}, \
  and\ \bibinfo {author} {\bibfnamefont {M.~J.}\ \bibnamefont {Hartmann}},\
  }\bibfield  {title} {\emph {\bibinfo {title} {Theory of Frequency-Filtered
  and Time-Resolved $N$-Photon Correlations},}\ }\href@noop {} {\bibfield
  {journal} {\bibinfo  {journal} {Phys. Rev. Lett.}\ }\textbf {\bibinfo
  {volume} {109}},\ \bibinfo {pages} {183601} (\bibinfo {year}
  {2012})}\BibitemShut {NoStop}%
\bibitem [{\citenamefont {Ridolfo}\ \emph {et~al.}(2013)\citenamefont
  {Ridolfo}, \citenamefont {del Valle},\ and\ \citenamefont
  {Hartmann}}]{ridolfo13a}%
  \BibitemOpen
  \bibfield  {author} {\bibinfo {author} {\bibfnamefont {A.}~\bibnamefont
  {Ridolfo}}, \bibinfo {author} {\bibfnamefont {E.}~\bibnamefont {del Valle}},
  \ and\ \bibinfo {author} {\bibfnamefont {M.~J.}\ \bibnamefont {Hartmann}},\
  }\bibfield  {title} {\emph {\bibinfo {title} {Photon correlations from
  ultrastrong optical nonlinearities},}\ }\href@noop {} {\bibfield  {journal}
  {\bibinfo  {journal} {Phys. Rev. A}\ }\textbf {\bibinfo {volume} {88}},\
  \bibinfo {pages} {063812} (\bibinfo {year} {2013})}\BibitemShut {NoStop}%
\bibitem [{\citenamefont {Carmichael}\ and\ \citenamefont
  {Walls}(1977)}]{carmichael77a}%
  \BibitemOpen
  \bibfield  {author} {\bibinfo {author} {\bibfnamefont {H.}~\bibnamefont
  {Carmichael}}\ and\ \bibinfo {author} {\bibfnamefont {D.}~\bibnamefont
  {Walls}},\ }\bibfield  {title} {\emph {\bibinfo {title} {Hysteresis in the
  spectrum for cooperative resonance fluorescence},}\ }\href@noop {} {\bibfield
   {journal} {\bibinfo  {journal} {J. Phys. B.: At. Mol. Phys.}\ }\textbf
  {\bibinfo {volume} {10}},\ \bibinfo {pages} {L685} (\bibinfo {year}
  {1977})}\BibitemShut {NoStop}%
\bibitem [{\citenamefont {Cohen-Tannoudji}\ and\ \citenamefont
  {Reynaud}(1979)}]{cohentannoudji79a}%
  \BibitemOpen
  \bibfield  {author} {\bibinfo {author} {\bibfnamefont {C.}~\bibnamefont
  {Cohen-Tannoudji}}\ and\ \bibinfo {author} {\bibfnamefont {S.}~\bibnamefont
  {Reynaud}},\ }\bibfield  {title} {\emph {\bibinfo {title} {Atoms in Strong
  Light-Fields: Photon Antibunching in Single Atom Fluorescence},}\ }\href@noop
  {} {\bibfield  {journal} {\bibinfo  {journal} {Phil. Trans. R. Soc. Lond. A}\
  }\textbf {\bibinfo {volume} {293}},\ \bibinfo {pages} {223} (\bibinfo {year}
  {1979})}\BibitemShut {NoStop}%
\bibitem [{\citenamefont {Schrama}\ \emph {et~al.}(1992)\citenamefont
  {Schrama}, \citenamefont {Nienhuis}, \citenamefont {Dijkerman}, \citenamefont
  {Steijsiger},\ and\ \citenamefont {Heideman}}]{schrama92a}%
  \BibitemOpen
  \bibfield  {author} {\bibinfo {author} {\bibfnamefont {C.~A.}\ \bibnamefont
  {Schrama}}, \bibinfo {author} {\bibfnamefont {G.}~\bibnamefont {Nienhuis}},
  \bibinfo {author} {\bibfnamefont {H.~A.}\ \bibnamefont {Dijkerman}}, \bibinfo
  {author} {\bibfnamefont {C.}~\bibnamefont {Steijsiger}}, \ and\ \bibinfo
  {author} {\bibfnamefont {H.~G.~M.}\ \bibnamefont {Heideman}},\ }\bibfield
  {title} {\emph {\bibinfo {title} {Intensity correlations between the
  components of the resonance fluorescence triplet},}\ }\href@noop {}
  {\bibfield  {journal} {\bibinfo  {journal} {Phys. Rev. A}\ }\textbf {\bibinfo
  {volume} {45}},\ \bibinfo {pages} {8045} (\bibinfo {year}
  {1992})}\BibitemShut {NoStop}%
\bibitem [{\citenamefont {Nienhuis}(1993)}]{nienhuis93a}%
  \BibitemOpen
  \bibfield  {author} {\bibinfo {author} {\bibfnamefont {G.}~\bibnamefont
  {Nienhuis}},\ }\bibfield  {title} {\emph {\bibinfo {title} {Spectral
  correlations in resonance fluorescence},}\ }\href@noop {} {\bibfield
  {journal} {\bibinfo  {journal} {Phys. Rev. A}\ }\textbf {\bibinfo {volume}
  {47}},\ \bibinfo {pages} {510} (\bibinfo {year} {1993})}\BibitemShut
  {NoStop}%
\bibitem [{\citenamefont {Ulhaq}\ \emph {et~al.}(2012)\citenamefont {Ulhaq},
  \citenamefont {Weiler}, \citenamefont {Ulrich}, \citenamefont {Ro{\ss}bach},
  \citenamefont {Jetter},\ and\ \citenamefont {Michler}}]{ulhaq12a}%
  \BibitemOpen
  \bibfield  {author} {\bibinfo {author} {\bibfnamefont {A.}~\bibnamefont
  {Ulhaq}}, \bibinfo {author} {\bibfnamefont {S.}~\bibnamefont {Weiler}},
  \bibinfo {author} {\bibfnamefont {S.~M.}\ \bibnamefont {Ulrich}}, \bibinfo
  {author} {\bibfnamefont {R.}~\bibnamefont {Ro{\ss}bach}}, \bibinfo {author}
  {\bibfnamefont {M.}~\bibnamefont {Jetter}}, \ and\ \bibinfo {author}
  {\bibfnamefont {P.}~\bibnamefont {Michler}},\ }\bibfield  {title} {\emph
  {\bibinfo {title} {Cascaded single-photon emission from the {Mollow} triplet
  sidebands of a quantum dot},}\ }\href@noop {} {\bibfield  {journal} {\bibinfo
   {journal} {Nat. Photon.}\ }\textbf {\bibinfo {volume} {6}},\ \bibinfo
  {pages} {238} (\bibinfo {year} {2012})}\BibitemShut {NoStop}%
\bibitem [{\citenamefont {Silva}\ \emph {et~al.}(2016)\citenamefont {Silva},
  \citenamefont {S{\'a}nchez~Mu{\~n}oz}, \citenamefont {Ballarini},
  \citenamefont {Gonz{\'a}lez-Tudela}, \citenamefont {De~Giorgi}, \citenamefont
  {Gigli}, \citenamefont {West}, \citenamefont {Pfeiffer}, \citenamefont
  {Del~Valle}, \citenamefont {Sanvitto} \emph {et~al.}}]{silva16a}%
  \BibitemOpen
  \bibfield  {author} {\bibinfo {author} {\bibfnamefont {B.}~\bibnamefont
  {Silva}}, \bibinfo {author} {\bibfnamefont {C.}~\bibnamefont
  {S{\'a}nchez~Mu{\~n}oz}}, \bibinfo {author} {\bibfnamefont {D.}~\bibnamefont
  {Ballarini}}, \bibinfo {author} {\bibfnamefont {A.}~\bibnamefont
  {Gonz{\'a}lez-Tudela}}, \bibinfo {author} {\bibfnamefont {M.}~\bibnamefont
  {De~Giorgi}}, \bibinfo {author} {\bibfnamefont {G.}~\bibnamefont {Gigli}},
  \bibinfo {author} {\bibfnamefont {K.}~\bibnamefont {West}}, \bibinfo {author}
  {\bibfnamefont {L.}~\bibnamefont {Pfeiffer}}, \bibinfo {author}
  {\bibfnamefont {E.}~\bibnamefont {Del~Valle}}, \bibinfo {author}
  {\bibfnamefont {D.}~\bibnamefont {Sanvitto}},  \emph {et~al.},\ }\bibfield
  {title} {\emph {\bibinfo {title} {The colored {Hanbury Brown--Twiss}
  effect},}\ }\href@noop {} {\bibfield  {journal} {\bibinfo  {journal} {Sci.
  Rep.}\ }\textbf {\bibinfo {volume} {6}},\ \bibinfo {pages} {37980} (\bibinfo
  {year} {2016})}\BibitemShut {NoStop}%
\bibitem [{\citenamefont {Peiris}\ \emph {et~al.}(2015)\citenamefont {Peiris},
  \citenamefont {Petrak}, \citenamefont {Konthasinghe}, \citenamefont {Yu},
  \citenamefont {Niu},\ and\ \citenamefont {Muller}}]{peiris15a}%
  \BibitemOpen
  \bibfield  {author} {\bibinfo {author} {\bibfnamefont {M.}~\bibnamefont
  {Peiris}}, \bibinfo {author} {\bibfnamefont {B.}~\bibnamefont {Petrak}},
  \bibinfo {author} {\bibfnamefont {K.}~\bibnamefont {Konthasinghe}}, \bibinfo
  {author} {\bibfnamefont {Y.}~\bibnamefont {Yu}}, \bibinfo {author}
  {\bibfnamefont {Z.~C.}\ \bibnamefont {Niu}}, \ and\ \bibinfo {author}
  {\bibfnamefont {A.}~\bibnamefont {Muller}},\ }\bibfield  {title} {\emph
  {\bibinfo {title} {Two-color photon correlations of the light scattered by a
  quantum dot},}\ }\href@noop {} {\bibfield  {journal} {\bibinfo  {journal}
  {Phys. Rev. B}\ }\textbf {\bibinfo {volume} {91}},\ \bibinfo {pages} {195125}
  (\bibinfo {year} {2015})}\BibitemShut {NoStop}%
\bibitem [{\citenamefont {Bounouar}\ \emph {et~al.}(2017)\citenamefont
  {Bounouar}, \citenamefont {Strau\ss{}}, \citenamefont {Carmele},
  \citenamefont {Schnauber}, \citenamefont {Thoma}, \citenamefont {Gschrey},
  \citenamefont {Schulze}, \citenamefont {Strittmatter}, \citenamefont {Rodt},
  \citenamefont {Knorr},\ and\ \citenamefont {Reitzenstein}}]{bounouar17a}%
  \BibitemOpen
  \bibfield  {author} {\bibinfo {author} {\bibfnamefont {S.}~\bibnamefont
  {Bounouar}}, \bibinfo {author} {\bibfnamefont {M.}~\bibnamefont
  {Strau\ss{}}}, \bibinfo {author} {\bibfnamefont {A.}~\bibnamefont {Carmele}},
  \bibinfo {author} {\bibfnamefont {P.}~\bibnamefont {Schnauber}}, \bibinfo
  {author} {\bibfnamefont {A.}~\bibnamefont {Thoma}}, \bibinfo {author}
  {\bibfnamefont {M.}~\bibnamefont {Gschrey}}, \bibinfo {author} {\bibfnamefont
  {J.-H.}\ \bibnamefont {Schulze}}, \bibinfo {author} {\bibfnamefont
  {A.}~\bibnamefont {Strittmatter}}, \bibinfo {author} {\bibfnamefont
  {S.}~\bibnamefont {Rodt}}, \bibinfo {author} {\bibfnamefont {A.}~\bibnamefont
  {Knorr}}, \ and\ \bibinfo {author} {\bibfnamefont {S.}~\bibnamefont
  {Reitzenstein}},\ }\bibfield  {title} {\emph {\bibinfo {title}
  {Path-Controlled Time Reordering of Paired Photons in a Dressed Three-Level
  Cascade},}\ }\href@noop {} {\bibfield  {journal} {\bibinfo  {journal} {Phys.
  Rev. Lett.}\ }\textbf {\bibinfo {volume} {118}},\ \bibinfo {pages} {233601}
  (\bibinfo {year} {2017})}\BibitemShut {NoStop}%
\bibitem [{\citenamefont {Fern\'andez-Lorenzo}\ and\ \citenamefont
  {Porras}(2017)}]{fernandezlorenzo17a}%
  \BibitemOpen
  \bibfield  {author} {\bibinfo {author} {\bibfnamefont {S.}~\bibnamefont
  {Fern\'andez-Lorenzo}}\ and\ \bibinfo {author} {\bibfnamefont
  {D.}~\bibnamefont {Porras}},\ }\bibfield  {title} {\emph {\bibinfo {title}
  {Quantum sensing close to a dissipative phase transition: Symmetry breaking
  and criticality as metrological resources},}\ }\href@noop {} {\bibfield
  {journal} {\bibinfo  {journal} {Phys. Rev. A}\ }\textbf {\bibinfo {volume}
  {96}},\ \bibinfo {pages} {013817} (\bibinfo {year} {2017})}\BibitemShut
  {NoStop}%
\bibitem [{\citenamefont {Gammelmark}\ and\ \citenamefont
  {M{\o}lmer}(2014)}]{gammelmark14a}%
  \BibitemOpen
  \bibfield  {author} {\bibinfo {author} {\bibfnamefont {S.}~\bibnamefont
  {Gammelmark}}\ and\ \bibinfo {author} {\bibfnamefont {K.}~\bibnamefont
  {M{\o}lmer}},\ }\bibfield  {title} {\emph {\bibinfo {title} {Fisher
  information and the quantum Cram{\'e}r-Rao sensitivity limit of continuous
  measurements},}\ }\href@noop {} {\bibfield  {journal} {\bibinfo  {journal}
  {Phys. Rev. Lett.}\ }\textbf {\bibinfo {volume} {112}},\ \bibinfo {pages}
  {170401} (\bibinfo {year} {2014})}\BibitemShut {NoStop}%
\bibitem [{\citenamefont {Kiilerich}\ and\ \citenamefont
  {M{\o}lmer}(2014)}]{kiilerich14a}%
  \BibitemOpen
  \bibfield  {author} {\bibinfo {author} {\bibfnamefont {A.~H.}\ \bibnamefont
  {Kiilerich}}\ and\ \bibinfo {author} {\bibfnamefont {K.}~\bibnamefont
  {M{\o}lmer}},\ }\bibfield  {title} {\emph {\bibinfo {title} {Estimation of
  atomic interaction parameters by photon counting},}\ }\href@noop {}
  {\bibfield  {journal} {\bibinfo  {journal} {Phys. Rev. A}\ }\textbf {\bibinfo
  {volume} {89}},\ \bibinfo {pages} {052110} (\bibinfo {year}
  {2014})}\BibitemShut {NoStop}%
\bibitem [{\citenamefont {Kiilerich}\ and\ \citenamefont
  {M{\o}lmer}(2016)}]{kiilerich16a}%
  \BibitemOpen
  \bibfield  {author} {\bibinfo {author} {\bibfnamefont {A.~H.}\ \bibnamefont
  {Kiilerich}}\ and\ \bibinfo {author} {\bibfnamefont {K.}~\bibnamefont
  {M{\o}lmer}},\ }\bibfield  {title} {\emph {\bibinfo {title} {Bayesian
  parameter estimation by continuous homodyne detection},}\ }\href@noop {}
  {\bibfield  {journal} {\bibinfo  {journal} {Phys. Rev. A}\ }\textbf {\bibinfo
  {volume} {94}},\ \bibinfo {pages} {032103} (\bibinfo {year}
  {2016})}\BibitemShut {NoStop}%
\bibitem [{\citenamefont {Bartolo}\ \emph {et~al.}(2017)\citenamefont
  {Bartolo}, \citenamefont {Minganti}, \citenamefont {Lolli},\ and\
  \citenamefont {Ciuti}}]{bartolo17a}%
  \BibitemOpen
  \bibfield  {author} {\bibinfo {author} {\bibfnamefont {N.}~\bibnamefont
  {Bartolo}}, \bibinfo {author} {\bibfnamefont {F.}~\bibnamefont {Minganti}},
  \bibinfo {author} {\bibfnamefont {J.}~\bibnamefont {Lolli}}, \ and\ \bibinfo
  {author} {\bibfnamefont {C.}~\bibnamefont {Ciuti}},\ }\bibfield  {title}
  {\emph {\bibinfo {title} {Homodyne versus photon-counting quantum
  trajectories for dissipative Kerr resonators with two-photon driving},}\
  }\href@noop {} {\bibfield  {journal} {\bibinfo  {journal} {The European
  Physical Journal Special Topics}\ }\textbf {\bibinfo {volume} {226}},\
  \bibinfo {pages} {2705} (\bibinfo {year} {2017})}\BibitemShut {NoStop}%
\bibitem [{\citenamefont {Strobel}\ \emph {et~al.}(2014)\citenamefont
  {Strobel}, \citenamefont {Muessel}, \citenamefont {Linnemann}, \citenamefont
  {Zibold}, \citenamefont {Hume}, \citenamefont {Pezz{\`e}}, \citenamefont
  {Smerzi},\ and\ \citenamefont {Oberthaler}}]{strobel14a}%
  \BibitemOpen
  \bibfield  {author} {\bibinfo {author} {\bibfnamefont {H.}~\bibnamefont
  {Strobel}}, \bibinfo {author} {\bibfnamefont {W.}~\bibnamefont {Muessel}},
  \bibinfo {author} {\bibfnamefont {D.}~\bibnamefont {Linnemann}}, \bibinfo
  {author} {\bibfnamefont {T.}~\bibnamefont {Zibold}}, \bibinfo {author}
  {\bibfnamefont {D.~B.}\ \bibnamefont {Hume}}, \bibinfo {author}
  {\bibfnamefont {L.}~\bibnamefont {Pezz{\`e}}}, \bibinfo {author}
  {\bibfnamefont {A.}~\bibnamefont {Smerzi}}, \ and\ \bibinfo {author}
  {\bibfnamefont {M.~K.}\ \bibnamefont {Oberthaler}},\ }\bibfield  {title}
  {\emph {\bibinfo {title} {Fisher information and entanglement of non-Gaussian
  spin states},}\ }\href@noop {} {\bibfield  {journal} {\bibinfo  {journal}
  {Science}\ }\textbf {\bibinfo {volume} {345}},\ \bibinfo {pages} {424}
  (\bibinfo {year} {2014})}\BibitemShut {NoStop}%
\bibitem [{\citenamefont {Holstein}\ and\ \citenamefont
  {Primakoff}(1940)}]{holstein40a}%
  \BibitemOpen
  \bibfield  {author} {\bibinfo {author} {\bibfnamefont {T.}~\bibnamefont
  {Holstein}}\ and\ \bibinfo {author} {\bibfnamefont {H.}~\bibnamefont
  {Primakoff}},\ }\bibfield  {title} {\emph {\bibinfo {title} {Field Dependence
  of the Intrinsic Domain Magnetization of a Ferromagnet},}\ }\href@noop {}
  {\bibfield  {journal} {\bibinfo  {journal} {Phys. Rev.}\ }\textbf {\bibinfo
  {volume} {58}},\ \bibinfo {pages} {1098} (\bibinfo {year}
  {1940})}\BibitemShut {NoStop}%
\bibitem [{\citenamefont {Srinivas}\ and\ \citenamefont
  {Davies}(1981)}]{srinivas81a}%
  \BibitemOpen
  \bibfield  {author} {\bibinfo {author} {\bibfnamefont {M.~D.}\ \bibnamefont
  {Srinivas}}\ and\ \bibinfo {author} {\bibfnamefont {E.~B.}\ \bibnamefont
  {Davies}},\ }\bibfield  {title} {\emph {\bibinfo {title} {Photon counting
  probabilities in quantum optics},}\ }\href@noop {} {\bibfield  {journal}
  {\bibinfo  {journal} {Opt. Acta}\ }\textbf {\bibinfo {volume} {28}},\
  \bibinfo {pages} {981} (\bibinfo {year} {1981})}\BibitemShut {NoStop}%
\end{thebibliography}%

\section*{APPENDIX I: MEAN FIELD EQUATIONS}
\label{sec:appendix1}
The study of mean-field equations provides insight into the system dynamics and the different dissipative phases in the thermodynamic limit, $J\rightarrow \infty$. In that case, writing the commutator between the normalized angular momentum operators $s_i\equiv S_i/J$, $i\in\{x,y,z\}$, yields a value $[s_i,s_j]=i\epsilon^{ijk}s_k/J$ (with $\epsilon^{ijk}$ the Levy-Civita symbol) that tends to zero. One thus obtains the set of equations:
\begin{subequations}
\label{eq:sxsysz_meanfield}
\begin{align}
\dot s_x &= (\Gamma_--\Gamma_+)s_x s_z\label{eq:sxdot},\\\
\dot s_y &=-\Omega s_z +(\Gamma_--\Gamma_+)s_y s_z,\label{eq:sydot}\\
\dot s_z &= \Omega s_y - (\Gamma_--\Gamma_+)(s_x^2+s_y^2),\label{eq:szdot}
\end{align}
\end{subequations}
where $\Gamma_\pm$ are given by \cref{eq:gammaplus,eq:gammaminus}. At the level of description of the mean-field equations, the role of the squeezing angle $\theta$ is therefore to renormalize the decay rate $\Gamma$ by the factor $(\cos^2\theta-\sin^2\theta)$, since $\Gamma_--\Gamma_+=\Gamma(\cos^2\theta-\sin^2\theta)$. Given that these equations conserve the total norm $\mathcal N = s_x^2+s_y^2+s_z^2$, we can write them as a reduced set of dynamical equations in terms of the polar angles $\{\Theta\in[0,\pi],\Phi\in[-\pi,\pi]\}$, related to the cartesian coordinates as:
\begin{subequations}
 \label{eq:mz-meanfield}
\begin{eqnarray}
 s_x &=& \sin\Theta\cos\Phi,\\
 s_y &=& \sin\Theta\sin\Phi,\\
 s_z &=& -\cos\Theta.
\end{eqnarray}
\end{subequations}
Note that our definition differs from the standard one by the sign of the last equation, which means that the angle $\Theta$ is defined with respect to the \emph{negative} $z$-axis in order to make the lowest eigenstate of $S_z$ correspond to $\Theta=0$. The dynamical equations for the spherical angles are:
\begin{subequations}
\label{eq:angles_MF_dot}
\begin{align}
\dot\Theta &= \Omega\sin\Phi-(\Gamma_--\Gamma_+)\sin\Theta\label{eq:dottheta}\\
\dot\Phi &= \Omega\cos\Phi\cot\Theta\label{eq:dotPhi}
\end{align}
\end{subequations}
The previous equations define a vector field of derivatives on the Bloch sphere; these field lines are sketched in Fig.~\ref{fig:1}(b) for different values of $(\Omega/\Gamma,\theta)$, together with the spin Wigner function~\cite{agarwal81a,dowling94a} of the exact steady state on a finite system ($N=50$). 
A mean field approach does not necessarily offer a faithful description of the dynamics~\cite{mendoza16a}; in our model, it assumes a classical, point-like state on the Bloch sphere, therefore failing to describe spin fluctuations. Despite this, it is interesting to notice that, in a finite system, the shape of the fluctuations in the Bloch sphere actually bears some similarities with the vector field of derivatives predicted by the mean-field~\cite{strobel14a}. This is observed in Fig.~\ref{fig:1}, where it is clearly seen that the asymmetry in the density of field lines at both sides of the steady-state (moving along the meridian) is replicated as an asymmetry in the corresponding Wigner function.

We move now into analysing the steady solutions of these dynamical equations. Regarding the angle $\Phi$, Eq.~\eqref{eq:dotPhi} always has a stationary solution at $\Phi=\pm\pi/2$. It is instructive to consider the dynamics of $\Theta$ for $\Phi=\pi/2$, which reduces to:
\begin{equation}
\dot\Theta = \Omega - (\Gamma_--\Gamma_+)\sin\Theta.
\label{eq:dotTheta-effective}
\end{equation}
One can picture this as a dynamical equation for a pendulum, driven by the first term and damped by the second. The steady state solution $\Theta_0$ is determined by setting \eqref{eq:dotTheta-effective} to zero, which, from Eq.~\eqref{eq:mz-meanfield}, yields the magnetization $M\equiv s_z(t\rightarrow\infty)$ given by Eq.~\eqref{eq:M-mean-field}. In general, the stationary solutions of ~\cref{eq:sxdot,eq:sydot,eq:szdot} read:
\begin{subequations}
\label{eq:sxsysz_mf_ss}
\begin{align}
\langle s_z \rangle  &=M\label{eq:sz_mf},\\
\langle s_x \rangle &=  0 ,\\
\langle s_y \rangle &= \sqrt{1-M^2},\label{eq:sy_mf}
\end{align}
\end{subequations}

 There are two situations in which these solutions do not hold. 
\begin{enumerate}
\item At the point $\Gamma_--\Gamma_+=0$, where $S_x$ becomes a strong symmetry, Eq.~\eqref{eq:dotTheta-effective} does not have a stationary solution except for the trivial case $\Omega=0$. In particular, looking back at~\cref{eq:sxdot,eq:sydot,eq:szdot}, we see that at this point the evolution corresponds to a circular motion on a plane of constant $s_x$, with $s_y=(1-s_x^2)\cos(\Omega t)$, $s_z = (1-s_x^2)\sin(\Omega t)$.
%
\item  ~At the critical value
\begin{equation}
\Omega_c = \Gamma_--\Gamma_+=\Gamma(\cos^2\theta-\sin^2\theta),
\end{equation}
we have $M=0$, which means that the energy supplied by $\Omega$ is enough to reach the equator of the Bloch sphere, where the drag is maximum.
\end{enumerate}

Therefore, for values $\Omega>\Omega_c$, the pendulum is able to go beyond the equator, with a driving that now is large enough for it to engage in a perpetual oscillation across the Bloch sphere. This is reflected on the fact that Eq.~\eqref{eq:dotTheta-effective} has no stationary solution and in the unphysical imaginary value of $M$ predicted by Eq.~\eqref{eq:M-mean-field} for $\Omega>\Omega_c$. The emergence of initial-state-dependent closed trajectories at $\Omega>\Omega_c$ is represented on points (iv) and (viii) of Fig.~\ref{fig:1}(b). This transition to a phase with time-periodic steady states corresponds, in the case $\theta=0$, to the well-studied second order DPT of collective resonance flourescence~\cite{puri79a,lawande81a,gonzaleztudela13b}, it is related to the existence of steady states with imaginary eigenvalues~\cite{buca19a} and it was the subject of a recent work~\cite{iemini18a,tucker18a} where similar models have been used to describe dissipative time crystals.

In general, we observe that the role of the squeezed decay parametrized by $\theta$ is to lower the value of critical driving towards $\Omega_c\rightarrow 0 $ as $\theta\rightarrow \pi/4$ (and $\Gamma_-\rightarrow \Gamma_+$).  
Note that such an apparent non-ergodic dynamics does not survive in the full quantum solution for a finite system, which does reach stationarity on a time that, however, diverges with the system size (as predicted by the eigenvalue equation Eq.~\eqref{eq:exact-eigenvalues}). The stationary oscillations predicted by the mean-field equations are, therefore, the thermodynamic limit of a transient phenomena.

\section*{APPENDIX II: SPIN OBSERVABLES}
\label{sec:appendix2}
\subsection{Holstein-Primakoff approximation}
Hewe we use a Holstein-Primakoff (HP) approximation~\cite{holstein40a} to obtain analytical expressions for spin mean values and fluctuations, which can be linked to the Liouvillian gap in the ferromagnetic phase.  The exact HP transformation writes the angular momentum operator in terms of a bosonic mode with annihilation operator $b$:
\begin{eqnarray}
S_- &=& (\sqrt{2 J - b^\dagger b})b \nonumber\\
S_z &=& b^\dagger b - J
\label{eq:HP}
\end{eqnarray}
The HP approximation, consisting of a truncated series expansion of the square root in Eq.~\eqref{eq:HP}, is based on the premise that the upper levels of the finite ladder of eigenstates of $S_z$ are not occupied. Therefore, the nonlinear features that distinguish such a finite ladder from the infinite one of an harmonic oscillator are negligible, and $S_-$ is accurately described by the bosonic operator $b$. We will use the equations \eqref{eq:HP} for $\theta < \pi/4$ (where we know they are a better description since the system tends to be polarized towards the negative $z$ direction) and assume the same result applies for $\theta>\pi/4$ by flipping the spin and changing parameters $\Gamma_- \leftrightarrow \Gamma_-+$.

Following the approach outlined in  Ref.~\cite{kessler12a}, we  use a displaced operator:
\begin{equation}
b \rightarrow b + \sqrt{J}\beta
\end{equation}
that accounts for the mean polarization of the system. Using the renormalized operators $s_-\equiv S_-/J$ and $s_z\equiv S_z/J$, the corresponding HP expression expanded in terms of $\epsilon = 1/\sqrt{J}$ reads:
\begin{equation}
s_- = \sqrt{k}\sqrt{1-\epsilon\frac{\beta b^\dagger + \beta^* b}{k}-\epsilon^2\frac{b^\dagger b}{k}}(\beta + \epsilon b) = \sum_i \epsilon^i s^{(i)}_-.
\end{equation}
with $k = 2-|\beta|^2$. Up to first order in $\epsilon$, we have:
\begin{subequations}
\label{eq:sminusExpansion}
\begin{align}
s_-^{(0)} &= \sqrt{k} \beta,\\
s_-^{(1)} &= \frac{1}{2\sqrt{k}}[(2k-|\beta|^2)b-\beta^2 b^\dagger].
\end{align}
\end{subequations}
For the $s_z$ operator, we have $s_z = \sum_i \epsilon^i s_z^{(i)}$, with:
\begin{subequations}
\label{eq:szExpansion}
\begin{align}
s_z^{(0)} &= |\beta|^2-1, \\\
s_z^{(1)} &= \beta b^\dagger + \beta^* b.
\end{align}
\end{subequations}

It is useful to expand equation~\eqref{eq:master-equation} as:
\begin{multline}
\dot{\rho} = -i\left[ \Omega S_x,\rho \right] + \frac{\Gamma_-}{2J}\mathcal{L}_{S_-}\rho + \frac{\Gamma_+}{2J}\mathcal{L}_{S_+}\rho
\\+ \frac{\chi}{2J}(2 S_-\rho S_- -\{S_-^2,\rho\} + 2 S_+\rho S_+ - \{S_+^2,\rho \}),
\end{multline}
where $\Gamma_\pm$ are defined by \cref{eq:gammaplus,eq:gammaminus},  $\chi\equiv\Gamma\sin\theta\cos\theta$, and we defined the Lindblad operators $\mathcal L_O\{\rho\} \equiv 2O\rho O^\dagger-O^\dagger O \rho - \rho O^\dagger O$.
Then, we obtain
\begin{align}
\frac{1}{J}\dot{\rho} &= -i\left[\Omega \, s_x,\rho \right] + \frac{ \Gamma_-}{2}\mathcal{L}_{s_-}\{\rho\}+\frac{  \Gamma_+}{2}\mathcal{L}_{s_+ }\{\rho\} \nonumber\\
&+\frac{  \chi}{2 }(2 s_-\rho s_- -\{s_-^2,\rho\} + 2 s_+\rho s_+ - \{s_+^2,\rho \})
\nonumber\\
&= \left[ \mathcal{L}^{(0)} + \epsilon \mathcal{L}^{(1)} + \epsilon^2 \mathcal{L}^{(2)} + \mathcal{O}(\epsilon^3)\right]\rho.
\label{eq:me-expansion}
\end{align}
From Eq.~\eqref{eq:szExpansion} we immediately obtain  $\mathcal{L}^{(0)} = 0$. To all orders in the expansion, the Hamiltonian term describing coherent driving can be grouped together with a term coming from the dissipative part, in the following form:
\begin{align}
-\frac{i}{2}\left[s_{+}^{(n)}\left(\Omega-is_{-}^{(0)}(\Gamma_+-\Gamma_-)\right)+\mathrm{h.c.},\rho\right].
\label{eq:L-1st}
  \end{align}
We can therefore simplify the dynamics by eliminating the driving terms to all orders if we choose a proper value for the displacement $\beta$, such that
\begin{equation}
\Omega-is_{-}^{(0)}(\Gamma_+-\Gamma_-)=\Omega-i\sqrt{2-|\beta|^2}\beta(\Gamma_+-\Gamma_-)=0.
\end{equation}
This equation has three solutions that, written in terms of $r$ and $\phi$ as $\beta_i=r_i e^{i\phi_i}$, read:
\begin{subequations}
\begin{eqnarray}
r_1&=&\sqrt{1+M}\quad \phi_1=-\pi/2,\\
r_2&=&\sqrt{1-M} \quad \phi_2=-\pi/2,\\
r_3&=&\sqrt{1+Q}\quad \phi_3=\pi,
\end{eqnarray}
\end{subequations}
where $M$ is given by Eq.~\eqref{eq:M-mean-field} and we defined:
\begin{equation}
Q\equiv \sqrt{1+\left(\frac{\Omega}{\Gamma_+-\Gamma_-}\right)^2}.
\end{equation}

The first two solutions only exist only when $r_1$ and $r_2$ are real; we can identify the point at which these solutions cease to exist as the critical point where the phase transition occurs and the HP approximation is not well suited to describe the new phase. The critical lines $\Omega_c(\theta)$ that we get in this way coincide with the mean field result, Eq.~\eqref{eq:Omega_c}, since determining $\beta$ is essentially analogous to determining the steady-state mean-field solution.

We proceed now to demonstrate that $\beta=\beta_1$ is the only valid choice for the displacement by analysing the dynamics of the bosonic mode. Since all the terms of the form \eqref{eq:L-1st} are cancelled, we have $\mathcal L^{(1)}=0$. We define $A  \equiv (2k-|\beta|^2)/(2\sqrt{k})$ and $B\equiv-\beta^2/(2\sqrt{k})$, so that $s_-^{(1)} = A b + B b^\dagger$, and expand the density matrix $\rho(t) = \sum_n \epsilon^n \rho^{(n)}(t)$. By equating powers of $\epsilon$, Eq.~\eqref{eq:me-expansion} yields  a master equation for the lowest order density matrix, $\rho^{(0)}(t)$:
\begin{multline}
\dot\rho^{(0)}(t) =  \mathcal L^{(2)}\rho^{(0)}(t) = \frac{\gamma_-}{2}\mathcal{L}_b\{\rho^{(0)}\} +  \frac{\gamma_+}{2}  \mathcal{L}_{b^\dagger }\{\rho^{(0)}\}   \\
+\frac{\eta}{2}(2b\rho^{(0)} b - \{bb,\rho^{(0)}\} + 2 b^\dagger \rho^{(0)} b^\dagger - \{b^\dagger b^\dagger,\rho^{(0)}\}),
\label{eq:b-correlators}
\end{multline}
where
$\gamma_- \equiv  \Gamma_- A^2 +  \Gamma_+ B^2 + 2\chi AB$, $\gamma_+ \equiv  \Gamma_+ A^2 +  \Gamma_- B^2 + 2\chi AB$, $ \eta \equiv AB ( \Gamma_- +  \Gamma_+) + \chi(A^2+B^2)$ are all real quantities (since $\beta=-i r_1$ is purely imaginary).
The dynamics for $\langle b \rangle$ and $\langle b^\dagger \rangle$ is given by the equation $\dot{\mathbf{v}}=\mathcal W \mathbf{v}$, with $\mathbf{v}=(\langle b\rangle,\langle b^\dagger \rangle)^\mathrm{T}$ and 
\begin{equation}
\mathcal W = \frac{1}{2}\begin{pmatrix}\gamma_+ - \gamma_- & 0\\
0 & \gamma_+ - \gamma_-\end{pmatrix}.
\end{equation}
The eigenvalues of $\mathcal W$ describe the energy excitation spectrum of the Liouvillian~\cite{kessler12a} with highest real part, to the lowest order in $\epsilon$. We therefore find that the gap in the Liouvillian $\lambda=(\gamma_+-\gamma_-)/2$ is purely real:
\begin{equation}
\lambda = \frac{\Gamma_+ -\Gamma_-}{2}(A^2-B^2) =-(\Gamma_--\Gamma_+)(1-|\beta|^{2})
\end{equation}
From the three values of $\beta_i=r_i e^{i\phi_i}$ we get:
\begin{subequations}
\label{eq:lambda1}
\begin{eqnarray}
\lambda_1 = (\Gamma_--\Gamma_+)M,\\
\lambda_2 = -(\Gamma_--\Gamma_+)M,\\
\lambda_3 = (\Gamma_--\Gamma_+)Q.
\end{eqnarray}
\end{subequations}
\begin{figure}[t!]
\begin{center}
\includegraphics[width=1\columnwidth]{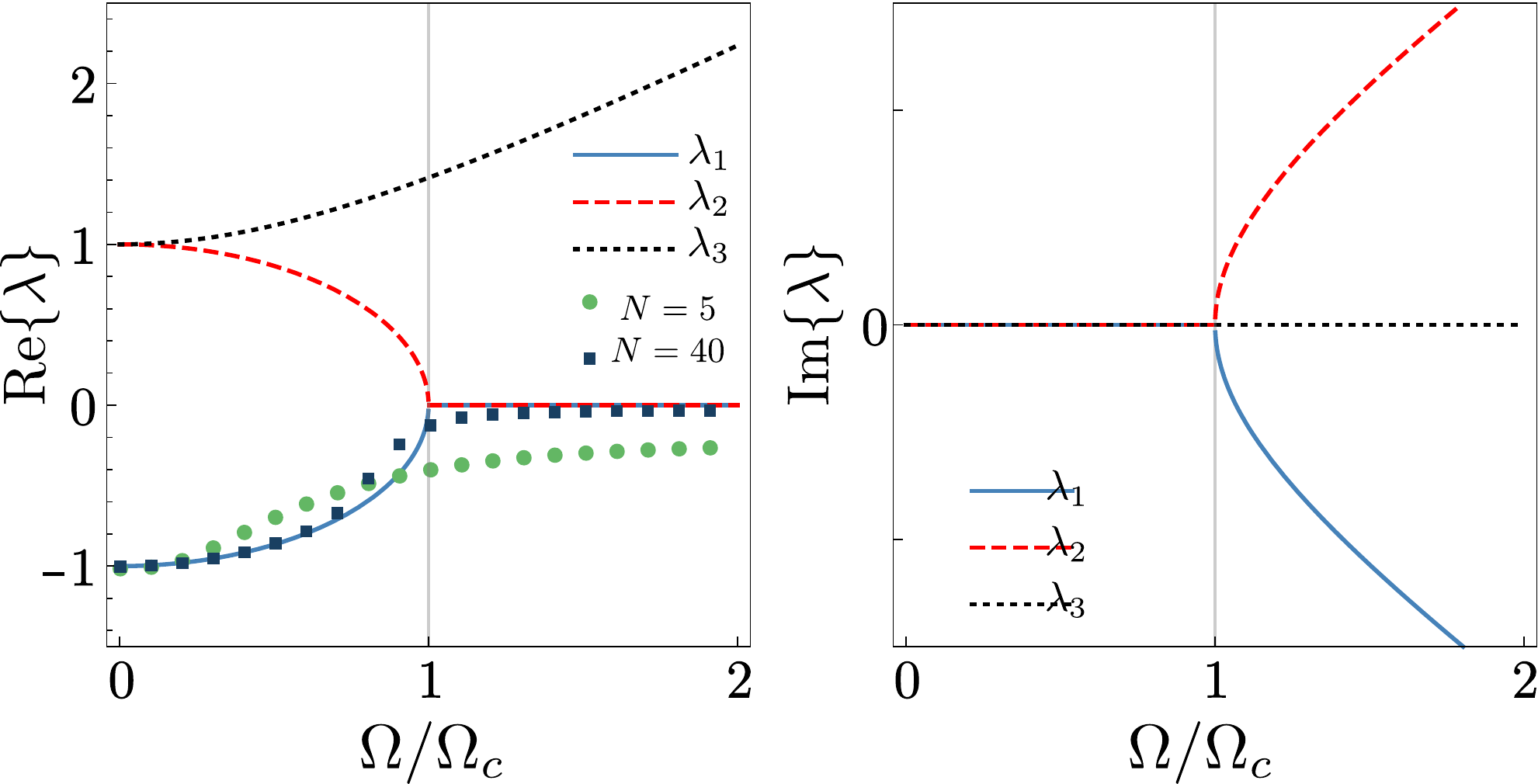}
\end{center}
\caption{Eigenvalues as a function of the normalized driving amplitude $\Omega/(\Gamma_--\Gamma_+)$, assuming $\Gamma_-> \Gamma_+$. Lines: analytical solutions given by Eq.~\eqref{eq:lambda1}. Markers: numerical solutions for finite systems. Below $\Omega/(\Gamma_--\Gamma_+)=1$, the only valid solution is $\lambda_1$. }
\label{fig:3}
\end{figure}
These three solutions are shown in Fig.~\ref{fig:3}. Only $\lambda_1$ has a negative real part in the region $\Omega < \Gamma_--\Gamma_+$ where these solutions are valid, and therefore the only valid choice of displacement is
\begin{equation}
\beta = e^{-i\pi/2}\sqrt{1+M}.
\label{eq:beta-good}
\end{equation}
The other choices give $\gamma_+>\gamma_-$, which clearly yield unstable equations of motion for the bosonic mode, since the effective pumping is larger than the losses and observables diverge; this is related to the instability of the corresponding steady mean-field solutions. The point where the gap closes $\gamma_+=\gamma_-$ is therefore associated with this instability in the equations of motion of the bosonic mode; this indicates that fluctuations in the spin become comparable to $J$ and indicates the onset of the dissipative phase transition.

\subsection{Spin polarization}

We can now compute spin observables in the ferromagnetic phase, where the HP expansion holds. 
In order to expand spin mean values $\langle s_{z/\pm} \rangle $  in powers of $\epsilon$, we must take into account both the HP expansions [Eqs.~\eqref{eq:sminusExpansion} and \eqref{eq:szExpansion}] and the expansion of $\rho(t)$. Doing so, one  obtains, to order $\epsilon^2$:
\begin{multline}
\langle  s_{z/\pm}(t)\rangle = \mathrm{Tr}[s_{z/\pm}^{(0)} \rho^{(0)}] + \epsilon\left\{ \mathrm{Tr}[s_{z/\pm}^{(1)} \rho^{(0)}] +\mathrm{Tr}[s_{z/\pm}^{(0)} \rho^{(1)}] \right\}\\
+\epsilon^2  \left\{\mathrm{Tr}[s_{z/\pm}^{(2)} \rho^{(0)}]   +   \mathrm{Tr}[s_{z/\pm}^{(1)} \rho^{(1)}]  +   \mathrm{Tr}[s_{z/\pm}^{(0)} \rho^{(2)}]  \right\}
+\mathcal O(\epsilon^3),
\label{eq:mean-spin-expansion}
\end{multline}
where we omitted the time dependence of the $\rho^{(n)}(t)$ for simplicity. 
Noting  that $s_{z/\pm}^{(0)}$ is a c-number, and that, by definition, $\mathrm{Tr}[\rho^{(1)}]=\mathrm{Tr}[\rho^{(2)}]=0$, the terms $\mathrm{Tr}[s_{z/\pm}^{(0)} \rho^{(1)}] $ and $\mathrm{Tr}[s_{z/\pm}^{(0)} \rho^{(2)}]$ in Eq.~\eqref{eq:mean-spin-expansion} are equal to zero. 

However, there are non-vanishing terms proportional to $\epsilon^2$ that depend on $\rho^{(1)}$. Since the effective master equation  $\dot\rho^{(1)}(t)=\mathcal L^{(2)}\rho^{(1)}+\mathcal L^{(3)}\rho^{(0)}$ is no longer quadratic, these terms prevent us to obtain a closed expression for $\langle s_{z/\pm}\rangle$ at order  $\epsilon^2=1/J$. 

Let us define the correlators to zeroth order in $\rho$ as $\langle O\rangle_0\equiv \mathrm{Tr}[O\rho_0]$. In order to evaluate the first-order terms $\langle s_{z/\pm}^{(1)}\rangle_0 $ in Eq.~\eqref{eq:mean-spin-expansion}, we must use Eq.~\eqref{eq:b-correlators} to obtain correlators of the form $\langle b\rangle_0$. In general, the dynamics of any arbitrary correlator $\langle {b^\dagger}^m b^n\rangle_0 $ will be given by:
\begin{multline}
\frac{d\langle {b^\dagger}^m b^n\rangle_0}{dt} = \frac{\gamma_+-\gamma_-}{2}(n+m)\langle {b^\dagger}^m b^n\rangle_0 \\  + \gamma_+ m n \langle {b^\dagger}^{m-1} b^{n-1}\rangle_0 - \frac{\eta}{2}m(m-1)\langle {b^\dagger}^{m-2} b^{n}\rangle_0 \\ - \frac{\eta}{2}n(n-1)\langle {b^\dagger}^{m} b^{n-2}\rangle_0.
\label{eq:regmat}
\end{multline}
In particular, we are interested in the stationary limit $t\rightarrow\infty$, where the density matrix fulfils $\mathcal L^{(2)}\rho^{(0)}=0$ (in the following, the notation $\langle \phantom{b}\rangle_0$ and $\rho^{(n)}$ will refer to stationary values).  We obtain steady state values of the correlators by setting the derivatives of Eq.~\eqref{eq:regmat} to zero.  This way, we get, for the case $m=0$, $n=1$:
\begin{equation}
\langle b \rangle_0 = 0.
\end{equation}
Since $\langle s_{z}^{(1)}\rangle_0 $ and $\langle s_{\pm}^{(1)}\rangle_0$ are  proportional to $\langle b\rangle_0$ and $\langle b^\dagger \rangle_0$, we find that they are all zero and, therefore, conclude that $\langle s_{z/\pm}\rangle$ has no first-order dependence on $\epsilon$. Therefore, using Eqs.~\eqref{eq:szExpansion}, ~\eqref{eq:sminusExpansion} and \eqref{eq:beta-good}, we find that the stationary expectation values $\langle s_z\rangle$, $\langle s_x\rangle$ and $\langle s_y\rangle$ are given, with corrections to \emph{second order} in $\epsilon$, by the zeroth-order terms $ \mathrm{Tr}[s_{z/\pm}^{(0)} \rho^{(0)}] $, which coincide with the solutions of the mean-field equations~\eqref{eq:sxsysz_mf_ss}:
\begin{subequations}
\begin{align}
\langle s_z \rangle  &=M=-R+\mathcal O (\epsilon^2) \label{eq:sz_appendix}\\
\langle s_x \rangle &=  0  + \mathcal{O}(\epsilon^2)\label{eq:sx_appendix}\\
\langle s_y \rangle &= \frac{\Omega}{\Gamma_+-\Gamma_-}+ \mathcal{O}(\epsilon^2)\label{eq:sy_appendix}
\end{align}
\end{subequations}
\vspace{1pt}
\subsection{Spin fluctuations: spin squeezing}
Our lack of an analytical expression of $\rho^{(n)}$ for $n>0$ prevents us from obtaining closed-form expressions for the second-order corrections to the mean spin. However, it is possible to get expressions for the fluctuations $\Delta s_{z/\pm}^2$ to second order, which is the lowest in their expansion. In particular, it is easy to prove that:
\begin{equation}
\Delta{ s_{z/\pm}}^2= \langle {s^{(1)}_{z/\pm}}^2\rangle_0 + \mathcal O (\epsilon^3).
\label{eq:fluctuations}
\end{equation}

The  mean spin direction in the thermodynamic limit, obtained from \cref{eq:sz_appendix,eq:sx_appendix,eq:sy_appendix}, can be written as:
\begin{equation}
\mathbf{u}_\mathrm{m}=\frac{\langle s_x \rangle \mathbf{u}_x+\langle s_y \rangle \mathbf{u}_y+\langle s_z \rangle \mathbf{u}_z}{\sqrt{\langle \mathbf{s}^2\rangle}}= \sqrt{1-M^2}\mathbf{u}_y+M\mathbf{u}_z
\end{equation}
We are interested in the squeezing along some direction in the plane perpendicular to $\mathbf u_m$, $\mathbf u_\bot(\varphi) \equiv \cos(\varphi)\mathbf{u}_x-\sin(\varphi)[-M \mathbf{u}_y+\sqrt{1-M^2}\mathbf{u}_z]$;  this direction to be determined by finding the  $\varphi$ that maximizes the squeezing.  
As we prove in Appendix~\ref{sec:A-squeezing}, $\mathbf u_x$ is always the preferential direction of squeezing.   In order to compute $\xi_\bot$,  it is useful to obtain, from the solution of Eq.~\eqref{eq:b-correlators}, the expression for the mean quadratic correlators:
\begin{subequations}
\begin{eqnarray}
\langle b^\dagger b\rangle &=& \frac{\gamma_+}{\gamma_--\gamma_+ },\\
\langle b^2 \rangle = \langle {b^\dagger}^2 \rangle &=& \frac{\eta}{\gamma_+-\gamma_-}.
\end{eqnarray}
\label{eq:bpbb2}
\end{subequations}
Using these, we  can write the following expression for the variance:
\begin{equation}
\Delta s_x^2 = \frac{k}{2J}\left(\langle b^\dagger b\rangle -\langle b^2\rangle + \frac{1}{2} \right)+\mathcal{O}(\epsilon^3),
\label{eq:sbot}
\end{equation}
and from there, obtain the  expression for the spin squeezing:
\begin{equation}
\xi_\bot^2 = \frac{N(\Delta S_x)^2}{\langle \mathbf S \rangle^2} = k\left(\frac{\gamma_+-\eta}{\gamma_--\gamma_+}+\frac{1}{2}\right)+\mathcal O(\epsilon).
\end{equation}
which can be rewritten in the form shown in Eq.~\eqref{eq:spin_squeezing}.

\subsection{Preferential direction of squeezing}
\label{sec:A-squeezing}
To complete our previous discussion, we demonstrate here that $\mathbf u_x$ is the direction with minimum fluctuations finding the angle $\varphi$ that minimizes spin fluctuations along the general direction $\mathbf u_\bot(\varphi)$.  
We define the a short notation for the following quantities with the properties of sine and cosines, $c\equiv \cos(\varphi)$, $s\equiv \sin(\varphi)$, $\tilde c \equiv M$ and $\tilde s \equiv \sqrt{1-M^2}$, and define the covariance $\cov[X,Y]\equiv\langle(X-\langle X\rangle)(Y-\langle Y \rangle) \rangle$. Then, we get, for the fluctuations along a general direction perpendicular to the mean spin: 
\begin{widetext}
\begin{multline}
\frac{(\Delta S_\bot)^2}{J^2}=
\frac{1}{J^2}\left\{
 s^2 \left[\tilde c^2\frac{2\cov[S_+,S_-]-(\Delta S_+)^2-(\Delta S_-)^2-2\langle S_z\rangle}{4}+\tilde s^2 (\Delta S_z)^2 \right.\right.\\
 \left.\left.
 -i\tilde s \tilde c\left( (\Delta S_+ S_z)^2-(\Delta S_- S_z)^2+\frac{\langle S_+\rangle+\langle S_-\rangle}{2}\right) \right] +
c^2 \left[ \frac{2\cov[S_+,S_-]+(\Delta S_+)^2+(\Delta S_-)^2-2\langle S_z\rangle}{4}   \right] 
\right.\\
\left. 
+\right.\\
\left.
+
sc\left[i\tilde c\frac{(\Delta S_+)^2-(\Delta S_-)^2}{2}-\tilde s \,\left(\cov[S_+,S_z]+\cov[S_-,S_z] +\frac{\langle S_+\rangle -\langle S_-\rangle}{2}  \right) \right]
 \right\}
\end{multline}
\end{widetext}
that we can express, grouping the coefficients of $s^2$, $c^2$ and $sc$ into three parameters $\kappa$, $\lambda$ and $\mu$ respectively, as:
\begin{multline}
\frac{(\Delta S_\bot)^2}{J^2}=
\frac{1}{J}\left[\kappa \sin(\varphi)^2 + \lambda \cos(\varphi)^2 + \mu \cos(\varphi)\sin(\varphi)\right]\\
=
\frac{1}{2J}\left[(\lambda-\kappa)\cos(2\varphi)+ \mu \sin(2\varphi)+\kappa+\lambda \right].
\end{multline}
To find the angle $\varphi$ that minimizes $(\Delta S_\bot)^2$, we take the derivative with respect to $\varphi$ and make it equal to zero, giving the following solution for $\varphi$:
\begin{equation}
2\varphi = \arctan\left(\frac{\mu}{\lambda-\kappa}\right).
\end{equation}
This function is usually  treated as a single-valued function by restricting the domain of $\tan(x)$ to $x\in [-\pi/2,\pi/2]$. We know from numerical calculations that indeed $\varphi \approx 0$, so we use this single-valued definition of $\arctan(x)$. In that case, we can use the properties:
\begin{subequations}
\begin{eqnarray}
\cos[\arctan(x)] &=& \frac{1}{\sqrt{x^2+1}},\\
\sin[\arctan(x)] &=& \frac{x}{\sqrt{x^2+1}},
\end{eqnarray}
\end{subequations} 
and then get:
\begin{equation}
\frac{(\Delta S_\bot)^2}{J^2} =\frac{1}{2J}\left[\kappa+\lambda+(\lambda-\kappa)\sqrt{1+\left(\frac{\mu}{\lambda-\kappa} \right)^2} \right].
\label{eq:sbot2kappa}
\end{equation}
We are now left to compute the values of $\kappa$, $\lambda$ and $\mu$. To do so, let us observe that, to order $\epsilon^2$:
\begin{equation}
\frac{\langle S_-^2\rangle}{J^2}=\langle {s_-^{(0)}}^2\rangle+\frac{1}{J}\left[\langle {s_-^{(1)}}^2\rangle +\langle s_-^{(0)} s_-^{(2)}\rangle+\langle s_-^{(2)}s_-^{(0)} \rangle\right]
\end{equation}
and, since $s_-^{(0)}$ is a c-number, we have that $(\Delta S_-)^2/J^2=\langle {s_-^{(1)}}^2\rangle/J$. By following the same argument to express the rest of variances and covariances present in the equation in terms of the $s_{\pm,z}^{(n)}$, we can write down the following values of $\kappa$, $\lambda$ and $\mu$, to zero order in $\epsilon$:
\begin{widetext}
\begin{eqnarray}
\kappa &=& \tilde c^2 \frac{2\langle s_+^{(1)}s_-^{(1)} \rangle-\langle {s_+^{(1)}}^2 \rangle-\langle {s_-^{(1)}}^2 \rangle-2\langle s_z^{(0)}\rangle}{4}
+\tilde s^2 \langle {s_z^{(1)}}^2 \rangle 
-i\tilde s \tilde c
\left[
\langle s_+^{(1)}s_z^{(1)} \rangle - \langle s_-^{(1)}s_z^{(1)} \rangle+\frac{\langle s_+^{(0)}\rangle+\langle s_-^{(0)}\rangle}{2} 
\right], \\
\lambda &=& \frac{2\langle s_+^{(1)}s_-^{(1)} \rangle+\langle {s_+^{(1)}}^2 \rangle+\langle {s_-^{(1)}}^2 \rangle-2\langle s_z^{(0)}\rangle}{4},\\
\mu &=& i\tilde c\frac{\langle {s_+^{(1)}}^2 \rangle-\langle {s_-^{(1)}}^2 \rangle}{2}-\tilde s \left(\langle s_+^{(1)} s_z^{(1)} \rangle + \langle s_-^{(1)} s_z^{(1)} \rangle +\frac{\langle s_+^{(0)}\rangle-\langle s_-^{(0)}\rangle}{2}\right) .
\end{eqnarray}
\end{widetext}

Taking into account that $\langle b^2\rangle = \langle {b^\dagger}^2\rangle$, and $s_-^{(1)}=A b + B b^\dagger$,  we can write the expressions of the correlators appearing in the equations:
\begin{subequations}
\begin{eqnarray}
\langle {s_\pm^{(1)}}^2\rangle &=&\langle b^2\rangle (A^2+B^2) +2AB\langle b^\dagger b \rangle + AB \\
\langle s_+^{(1)} s_-^{(1)}\rangle &=& \langle b^\dagger b\rangle (A^2+B^2) + \langle b^2\rangle 2AB + B^2 \\
\langle {s_z^{(1)}}^2\rangle &=& |\beta|^2 \left[2(\langle b^\dagger b\rangle-\langle b^2\rangle) + 1\right]  \\
\langle s_+^{(1)}s_z^{(1)}\rangle &=&  i|\beta|(A - B)(\langle b^\dagger b \rangle-\langle b^2\rangle) + B\beta  \\
\label{eq:spsz1}
\langle s_-^{(1)}s_z^{(1)}\rangle &=& i|\beta|(A- B)(\langle b^2\rangle-\langle b^\dagger b \rangle)  + A\beta 
\label{eq:smsz1}
\end{eqnarray}
\end{subequations}
We know $\langle s_-^{(0)}\rangle = \sqrt{k}{\beta}=- \langle s_+^{(0)}\rangle$,  and from Eq.~\eqref{eq:spsz1} and \eqref{eq:smsz1} we have that $\langle s_+^{(1)} s_z^{(1)} \rangle + \langle s_-^{(1)} s_z^{(1)} \rangle = \beta(A+B)=-i|\beta|\sqrt{k}=-i\tilde s$. Also, $\langle {s_+^{(1)}}^2 \rangle = \langle {s_-^{(1)}}^2 \rangle$. It is then easy to see that
\begin{equation}
\mu = 0 \rightarrow \varphi = 0
\end{equation}
proving that, in the thermodynamic limit, $\mathbf u_x$ is always the preferential direction for squeezing.

\vspace{10pt}
\section*{APPENDIX III: PROBABILITY AMPLITUDES OF GENERAL MONTE CARLO TRAJECTORIES}
\label{sec:appendix3}
In this section demonstrate Eq.~\eqref{eq:S-freezing} of the main text. By expanding the wavefunction in eigenstates $|m\rangle$ of the strong symmetry, we find, for a trajectory with jumps at times $(t_1,\ldots,t_n)<t$:
\begin{widetext}
\begin{multline}
|\psi(t)\rangle \propto e^{-i\tilde H(t-t_n)}A|\psi(T_n)\rangle \propto \sum_m e^{-i\tilde H(t-t_n)}m\, c_m(t_n)|m\rangle \propto \sum_m e^{-i\tilde H(t-t_n)}m e^{-i\tilde H(t_n-t_{n-1})}\, m c_m(t_{n-1})|m\rangle \\
\propto \ldots \propto  e^{-i \tilde H t}m^n\, c_m(0)|m\rangle
\end{multline}
\end{widetext}
From here, taking into account that $\tilde H = H-i\Gamma A^\dagger A /(2J) $, the probability to find the $|\psi(t)\rangle$ in an eigenstate $|m\rangle$ simply reads:
\begin{equation}
p(m; t,n) = |\langle m|\psi(t)\rangle|^2 =  \frac{1}{\mathcal N_{t,n}} e^{-\Gamma |m|^2 t/J}|m|^{2n }|c_m(0)|^2
\label{eq:p_MC_original}
\end{equation}
with $\mathcal N_{t,n}$ a normalization constant. Defining a rate $\alpha=n J/(\Gamma t)$, we can rewrite Eq.~\eqref{eq:p_MC_original} as
\begin{equation}
p(m; t,\alpha) = \frac{1}{\mathcal N_{t,\alpha}} \left(e^{- |m|^2}|m|^{2\alpha}\right)^{t\Gamma/J}|c_m(0)|^2.
\label{eq:p_MC_2}
\end{equation}
\vspace{10pt}

\section*{APPENDIX IV: EXACT EXPRESSION FOR THE ACTIVITY DISTRIBUTION}\label{sec:appendix4}
In this section, we demonstrate Eq.~\eqref{eq:pn} of the main text.
 Defining the quantum-jump superoperator~$\mathcal J\{\cdot\}\equiv L\{\cdot\}L^\dagger$ and the no-jump part of the Liouvillian $\mathcal S = \mathcal L - \mathcal J$, the probability for the system to experience $K$ quantum jumps on a time $T$, starting at the state $\rho(0)$, is given by~\cite{srinivas81a,zoller87a}:
\begin{widetext}
\begin{equation}
p_T(K)=\int_0^T dt_K \int_0^{t_K} dt_{K-1}\ldots\int_0^{t_2} dt_1 \mathrm{Tr}\left[ e^{\mathcal S(T-t_K)}\mathcal J e^{\mathcal S(t_K-t_{K-1})}\cdots \mathcal J e^{\mathcal S t_1}\rho(0) \right].
\label{eq:photon-counting}
\end{equation}
\end{widetext}
From~\cref{eq:strong-symmetry1,eq:strong-symmetry2}, we find:
\begin{subequations}
\begin{align}
e^{\mathcal S t}|m\rangle\langle m|&=e^{-\Gamma |m|^2 t/J}|m\rangle\langle m|\\
\mathcal J |m\rangle \langle m| &= \frac{\Gamma}{J}|m|^2|m\rangle \langle m|
\end{align}
\end{subequations}
and therefore:
\begin{multline}
p_T(K)=\int_0^T dt_K ...\int_0^{t_2} dt_1 \sum_m c_m \left(\frac{\Gamma}{J}|m|^2 \right)^K e^{-\Gamma |m|^2 T/J} \\
= \sum_m \frac{1}{K!}\left(\frac{T\Gamma |m|^2}{J} \right)^K e^{-\Gamma |m|^2 T/J}c_m.
\label{eq:pn-appendix}
\end{multline}

\end{document}